\documentclass{egpubl}
\usepackage{eg2026}
 
\Tutorial               % uncomment for Tutorial contribution
\usepackage[T1]{fontenc}
\usepackage{dfadobe}  

\usepackage{cite}  % comment out for biblatex with backend=biber
\BibtexOrBiblatex
\electronicVersion
\PrintedOrElectronic
\ifpdf \usepackage[pdftex]{graphicx} \pdfcompresslevel=9
\else \usepackage[dvips]{graphicx} \fi

\usepackage{egweblnk} 
\usepackage{amssymb}
\usepackage{amsmath}
\usepackage{mathtools}
\usepackage{algpseudocode}
\usepackage{algorithm}
\usepackage[table]{xcolor}

\usepackage{tocloft}
\renewcommand{\cftsecaftersnum}[1]{}% Remove \hfil after number

\usepackage{booktabs}
\usepackage{tikz}
\usetikzlibrary{arrows.meta,calc}

\title[Simulation Methods for Multiphysics Phenomena in Visual Computing]%
      {Simulation Methods for Multiphysics Phenomena\\ in Visual Computing}

\author[Löschner et al.]
{\parbox{\textwidth}{\centering 
		Fabian Löschner \orcid{0000-0001-6818-2953},		
		Stefan Rhys Jeske \orcid{0000-0003-3920-7765}, 
		José Antonio Fernández-Fernández \orcid{0000-0003-4651-7542}, 
		and
		Jan Bender \orcid{0000-0002-1908-4027}
	}
	\\
	{\parbox{\textwidth}{\centering 
			RWTH Aachen University, Germany
		} 
	}
}

\usepackage{color}
\definecolor{flcolor}{rgb}{0.7,0.1,0.1}
\definecolor{jbcolor}{rgb}{0.0,0.5,0.0}
\definecolor{sjcolor}{rgb}{0.1,0.1,0.7}
\definecolor{jfcolor}{rgb}{0.7,0.1,0.7}

\usepackage{bm}
\usepackage{bbold}
\renewcommand{\vec}[1]{\ensuremath{\boldsymbol{\mathbf{#1}}}}
\newcommand{\matr}[1]{\ensuremath{\mathbf{#1}}}          % matrix
\newcommand{\identity}{\mathbb{1}}
\newcommand{\Real}{\mathbb{R}}
\newcommand{\xnext}{\vec x}
\newcommand{\xprev}{\vec x^{\mathrm{prev}}}
\newcommand{\vnext}{\vec v}
\newcommand{\vprev}{\vec v^{\mathrm{prev}}}

\newcommand{\doublesumpotelem}{\sum_{p \in \mathcal{P}} \; \sum_{e \in \mathcal{T}_p}}
\newcommand{\potelem}{\phi_p(\vec \xnext_{e})}
\begin{document}

% uncomment for using teaser
\teaser{
    \includegraphics[trim={0 2.0cm 0 0.6cm}, clip, width=0.495\linewidth, height=125pt]{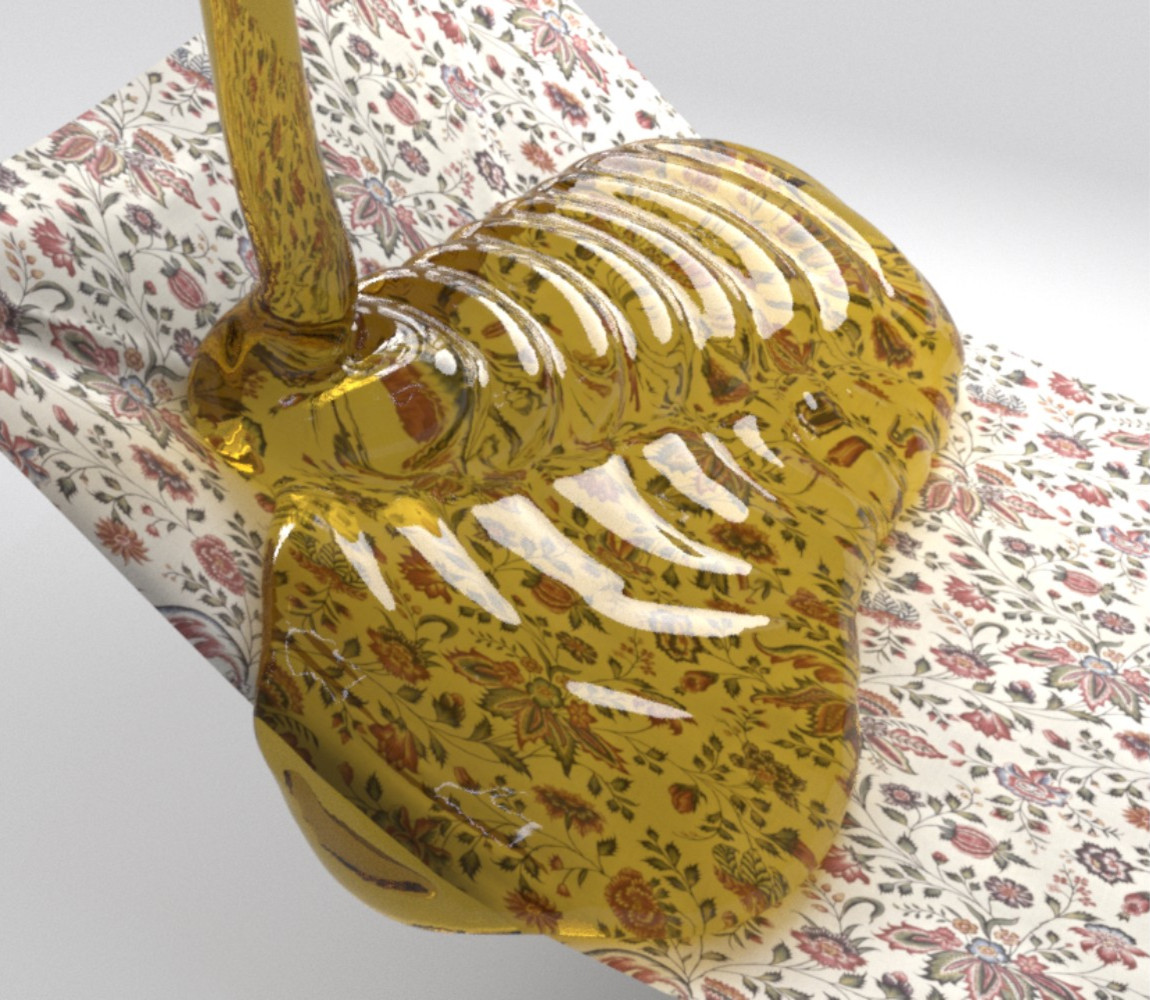}
    \hfill
    \includegraphics[trim={2.2cm 1.0cm 0.6cm 1.7cm}, clip, width=0.495\linewidth, height=125pt]{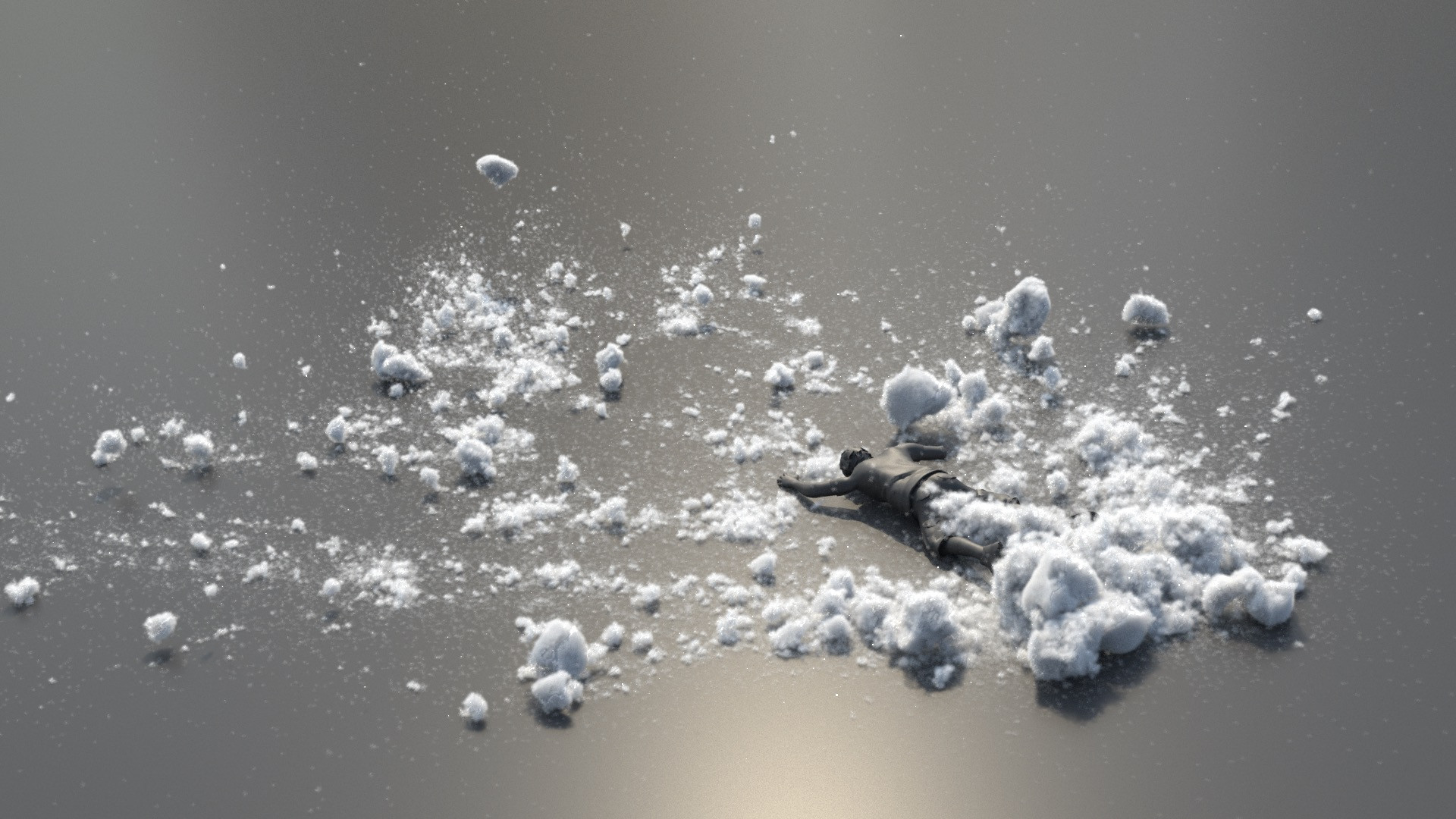} \vspace{0.075cm} \\
    \includegraphics[trim={6cm 5cm 4.6125cm 3.0cm}, clip, width=0.495\linewidth, height=125pt]{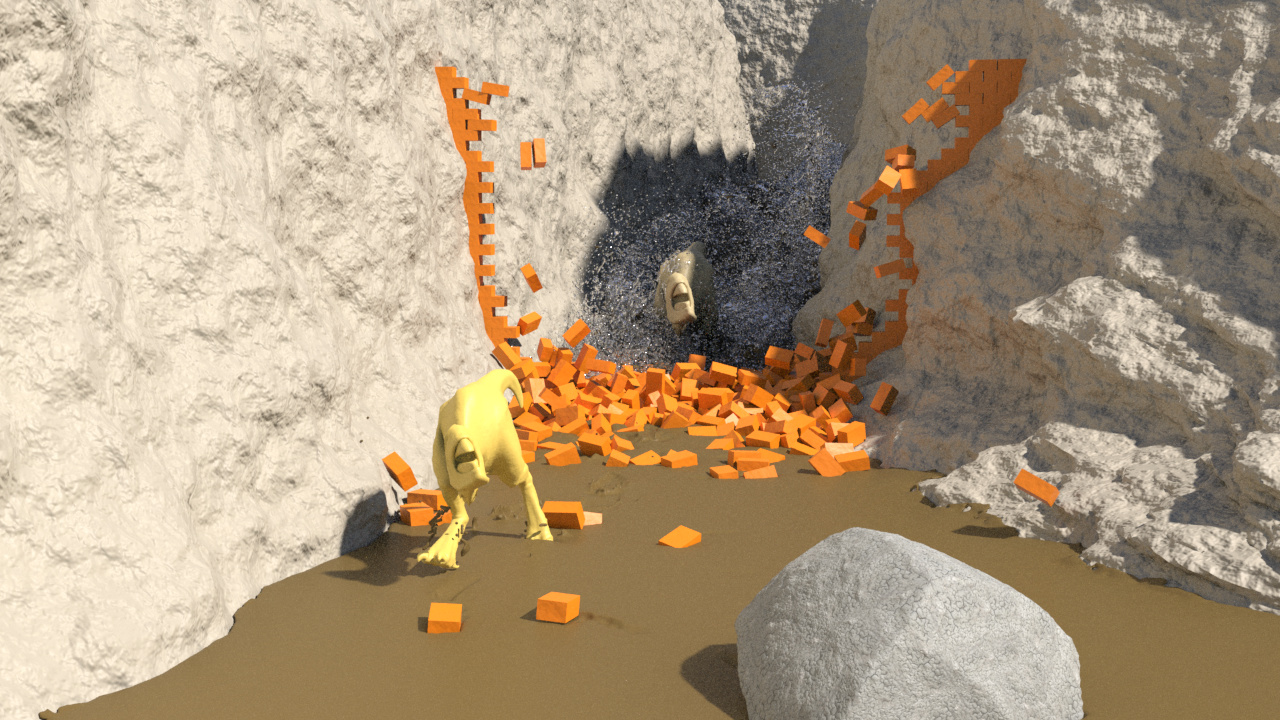}
    \hfill
    \includegraphics[trim={0 2.0cm 0 2.0cm}, clip, width=0.495\linewidth, height=125pt]{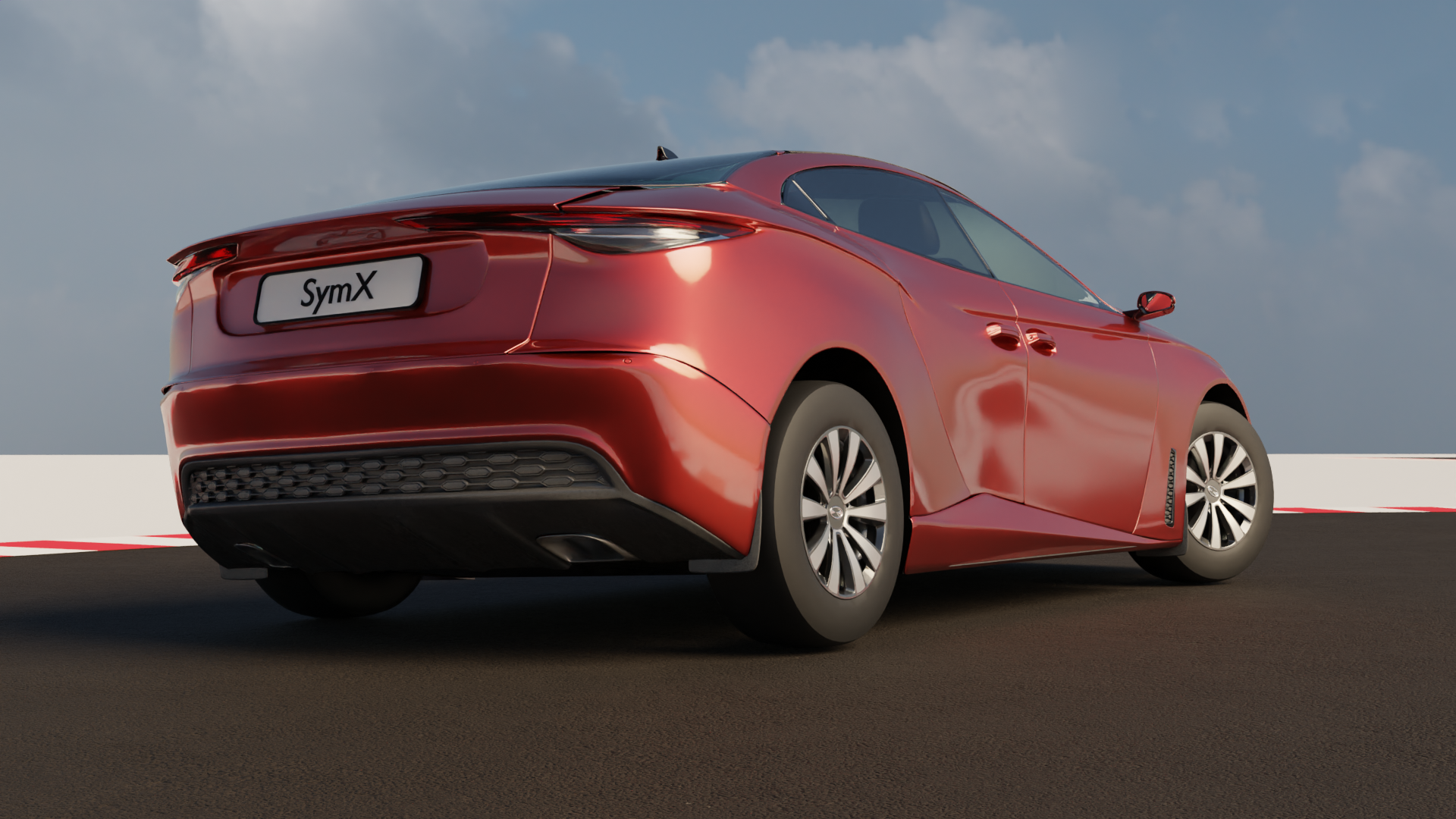} \null
    \centering
    \caption{\label{fig:teaser}Complex multiphysics simulations of different materials interacting with each other. \textbf{Top:} (Left) Honey is poured onto a stretched, elastic fabric~\cite{li_dynamic_2024}. (Right) A snowball fractures into clumps after impact with an obstacle~\cite{yu_xpbi_2024}. \textbf{Bottom:} (Left) Two creatures run from a flood wave through highly viscous mud, breaking through a wall of rigid bodies~\cite{gisslerInterlinkedSPHPressure2019}.
    (Right) Drifting around a corner, a rigid car with deformable tires as well as suspension and steering modeled by motors and joints is simulated with frictional contact~\cite{FLW25}.}
}

\maketitle
%-------------------------------------------------------------------------
\begin{abstract}
Physics simulation is a cornerstone of many computer graphics applications, ranging from video games and virtual reality to visual effects and computational design. 
The number of techniques for physically-based modeling and animation has thus skyrocketed over the past few decades, facilitating the simulation of a wide variety of materials and physical phenomena.
These course notes provide an in-depth introduction to multiphysics simulation methods for computer graphics, covering the mathematical foundations, key algorithms, and practical considerations behind the most widely used approaches.
We focus on methods developed by the computer graphics community for simulating various physical phenomena and materials -- including rigid and deformable bodies, fluids, and granular materials -- as well as the interactions between them.
For each method, we present the underlying mathematical framework with detailed derivations and discuss how different materials and coupling strategies fit into the formulation.
A selection of software frameworks that offer out-of-the-box multiphysics modeling capabilities is also presented.
Finally, we touch on emerging trends in physics-based animation, including machine learning-based methods which have become increasingly popular in recent years.

%\printccsdesc   
\end{abstract}  

{% Redefine \tableofcontents to flow within the two-column layout
\makeatletter
\renewcommand\tableofcontents{%
    \section*{Contents}%
    \@starttoc{toc}%
}
\makeatother
\tableofcontents
}

\section{Introduction}

Physics-based simulation is a key component of computer graphics, powering applications from video games and virtual reality to visual effects and computational design.
In this tutorial, we provide an in-depth introduction to multiphysics simulation techniques developed by the graphics community, covering the simultaneous simulation of interacting physical phenomena such as rigid and deformable bodies, fluids, and granular materials (see Figure~\ref{fig:teaser}).
For each method, we present the underlying mathematical framework with detailed derivations and equations, discuss how different materials and physical phenomena are represented, and show how models can be coupled.

Work in the field can be classified into two fundamental modeling approaches -- \emph{unified models} and \emph{coupling techniques}.
Unified models treat multiphysics systems with a single discretization across different material types and a monolithic mathematical formulation, which facilitates strong coupling and simplifies implementation. However, this approach may not provide the most suitable model for every material or phenomenon in a given simulation~\cite{tu_unified_2024}.
Coupling techniques take a modular design perspective, permitting the use of specialized simulation methods for different materials and phenomena. For example, mesh-free Lagrangian descriptions are beneficial for tracking large deformations, whereas Eulerian representations excel at computing spatial derivatives and enforcing global constraints such as incompressibility. The modular approach enables best-in-class models but can introduce challenges such as additional boundary handling costs and weaker inter-model coupling.

We organize the material covered in this tutorial into four main categories based on the general modeling approach.

\begin{itemize}
    \item Section~\ref{sec:unified:energy-based} presents energy-based models in which materials and physical phenomena are modeled using scalar energy potentials, permitting formulations as unconstrained optimization problems.
    \item Section~\ref{sec:unified:constraint-based} introduces constraint-based multiphysics models that use the concept of constraints as the fundamental building block for physically-based animation.
    \item Section~\ref{sec:point-based} presents Lagrangian methods using mesh-free point-based discretizations.
    \item Section~\ref{sec:eulerian_hybrid} discusses Eulerian and hybrid simulation methods involving grid-based discretizations.
\end{itemize}

The distinction between energy-based and constraint-based methods is not always clear cut.
For example, constraint-based methods may also rely on strain energies to model deformables while energy-based formulations may account for (soft) constraints using penalty energies.
Broadly speaking, we consider a model to be energy-based if its formulation permits application in unconstrained optimization, covering strain energies, barrier potentials, penalty functions, and incremental potentials such as Projective Dynamics.
Simulation methods that directly support hard equality or inequality constraints, such as Position Based Dynamics, are covered in Section~\ref{sec:unified:constraint-based}.

Section~\ref{sec:frameworks} discusses a selection of multiphysics frameworks that expose combined simulation capabilities in a single platform.
Section~\ref{sec:data-driven} presents emerging trends in physics-based simulation before we conclude in Section~\ref{sec:conclusion}.

\section{Energy-Based Multiphysics Modeling}
\label{sec:unified:energy-based}
\begin{figure*}
    \centering
    \includegraphics[trim={190 0 325 40},clip,height=3.845cm]{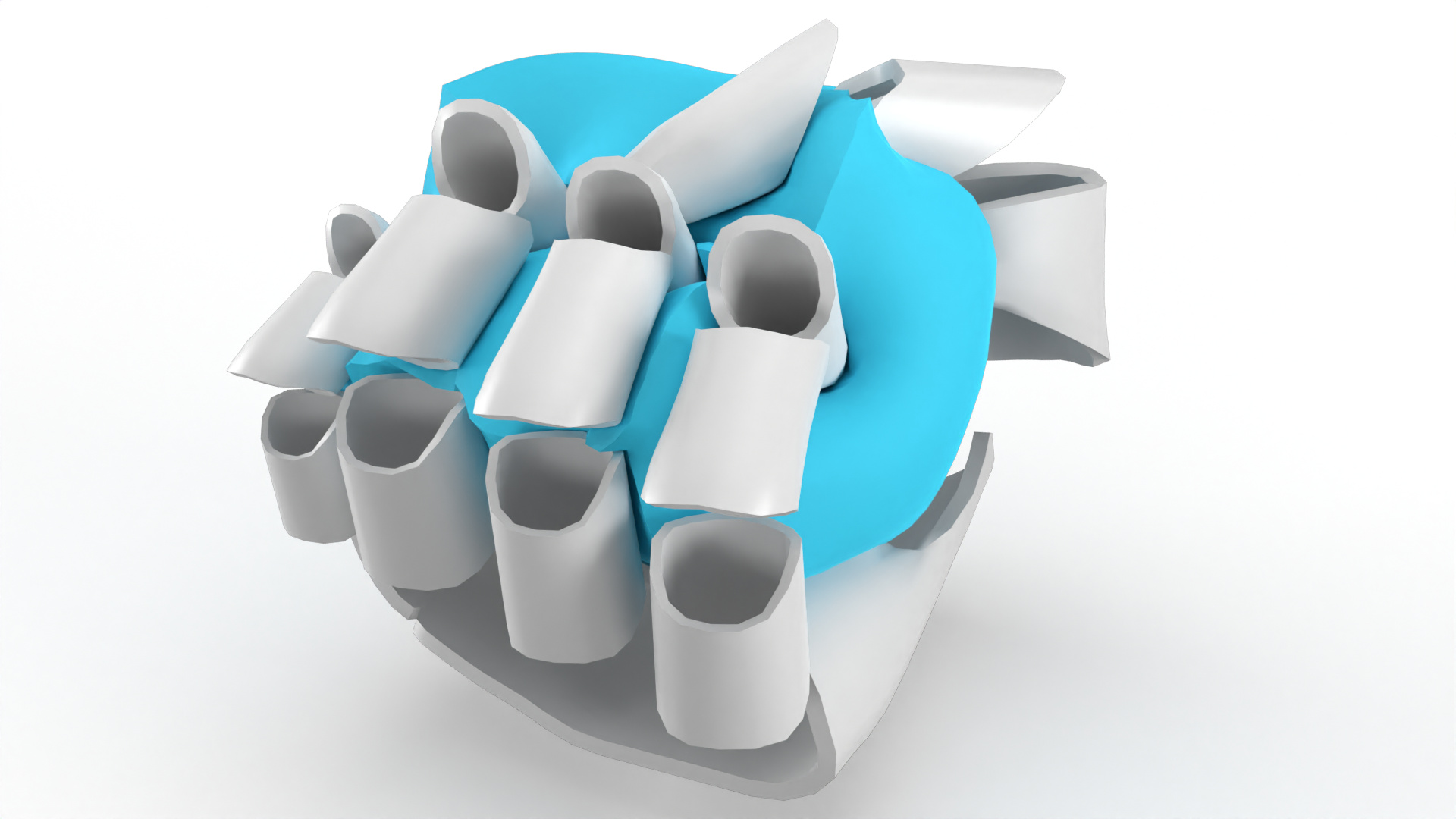}
    \hfill
    \includegraphics[trim={255 150 280 175},clip,height=3.845cm]{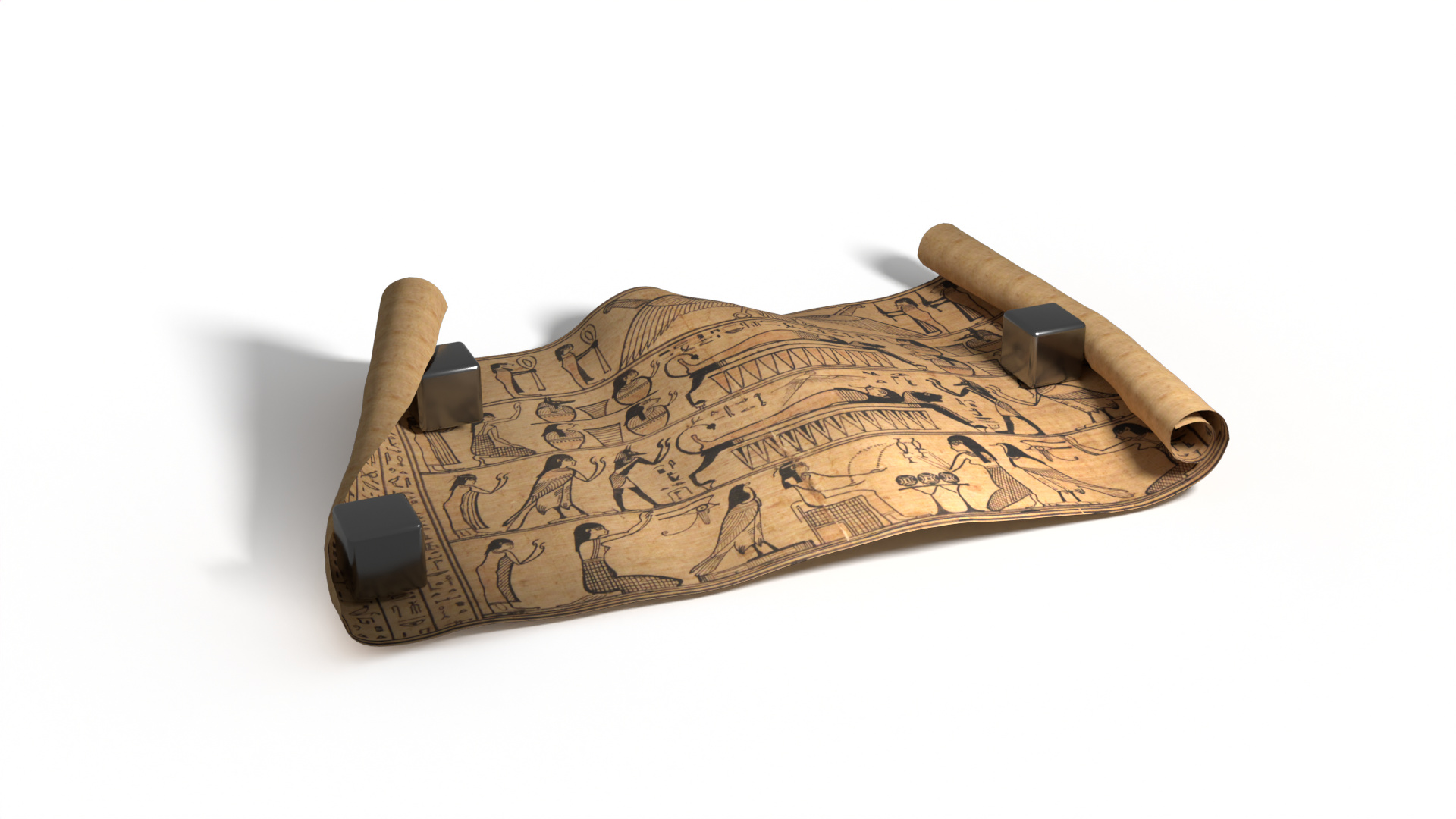}
    \hfill
    \includegraphics[trim={290 0 290 75},clip,height=3.845cm]{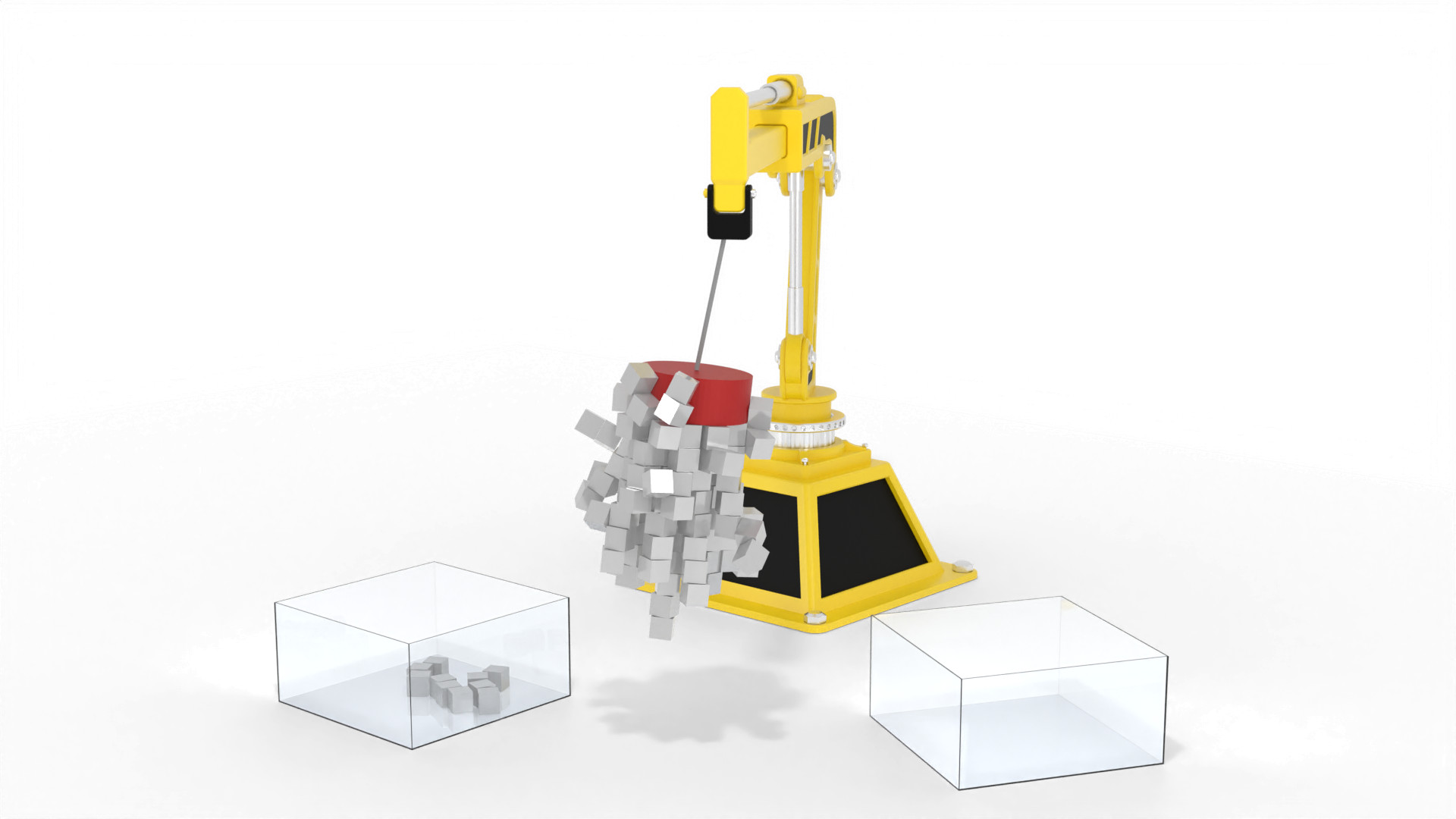}
    \caption{\label{fig:energy-based} A selection of energy-based multiphysics simulations with strong coupling. \textbf{Left:} A robotic hand modeled by rigid bodies, joints and motors compresses an elastic cube~\cite{FLL+24}. \textbf{Middle:} A scroll modeled as a Cosserat shell is pushed open by rigid bodies through frictional contact~\cite{LFJB24}. \textbf{Right:} A crane modeled as a multibody system lifts metallic cubes using magnetic forces that are strongly coupled with frictional contact~\cite{WFJB24}.}
\end{figure*}

Instead of approaching multiphysics systems with a unified discretization, in this section we focus on a unified formulation in terms of scalar energy potentials.
If all physical systems can be represented only in terms of such energies, we obtain a strongly coupled, fully implicit formulation that can be solved as a monolithic optimization problem.
Each physical effect---elasticity, contact, gravity, and others---contributes an additive term to a single global objective, the \emph{incremental potential}, whose minimizer solves the discrete equations of motion.
Casting time integration as optimization rather than root-finding has important practical advantages: it enables globalization techniques such as backtracking line search that make Newton-type solvers substantially more robust, and the additive structure provides a natural interface for multiphysics coupling, since all effects are assembled into and solved through the same objective.
This optimization-based time integration approach has a long history~\cite{RO99} and is well established in computer graphics~\cite{KYT+06,MTGG11,liu_fast_2013,GSS+15}.
In recent years, it gained significantly in popularity, for example with the introduction of barrier methods for handling of deformables coupled with frictional contact~\cite{LFS+20,LKJ21} and multibody dynamics~\cite{FLS+21,CLL+22}.
Notably, many constraints that traditionally require explicit enforcement, such as contact non-penetration and joint limits, can be incorporated directly into the unconstrained optimization through penalty or barrier energies.
Although an energy-based formulation is well suited for these models, dissipative effects such as damping and friction require adaptations, since they do not derive from a potential in the unknown positions alone, and coupling with fluids or granular media usually has to fall back to operator splitting or weak coupling.
Still, energy-based and optimization-based approaches can be applied to a wide range of multiphysics applications as shown in Figure~\ref{fig:energy-based}.

We want to highlight that a formulation as an unconstrained optimization problem is possible for many phenomena and has advantages in terms of complexity and robustness of the numerical methods (see, e.g.,~\cite{LFS+20}). 
However, more complex simulators that seek to combine different physical systems may still choose to combine such energy-based models with hard constraints and apply constrained optimization, for example, to enforce incompressibility in solid-fluid coupling (see, e.g.,~\cite{TB22}).

In the following, we first lay out the mathematical foundations of optimization-based time integration and its numerical building blocks, including Newton's method, derivative assembly, penalty and barrier constraints, line search, convergence criteria, and linear solves (see Section~\ref{sec:energy-based-foundations}).
We then present recent works on concrete physical systems and effects as well as their coupling (see Section~\ref{sec:energy-based-materials}).
Finally, we explore related methods that are built upon optimization-based integration but aim to improve performance for interactive applications in comparison to a full Newton method, notably Vertex Block Descent (see Section~\ref{sec:VBD}) and Projective Dynamics (see Section~\ref{sec:projective-dynamics}).

\subsection{Foundations of Energy-Based Simulation}
\label{sec:energy-based-foundations}

% -------------------------------------------------------------------------------------------------------------
In this section, we introduce the foundations of energy-based modeling and optimization-based solvers.
This viewpoint is particularly appealing in multiphysics settings because many different physical effects can be expressed in a \emph{unified} way through scalar potentials.
Elasticity, gravity, contact, pressure, and other effects can often be modeled by additive energy terms.
These contributions are then assembled into a single global objective whose stationary points correspond to the discrete equations of motion.
All effects contribute directly to the same objective, and can therefore be solved simultaneously within one nonlinear optimization problem.

\subsubsection{Optimization Time Integration}
\label{sec:opt-time-int}
To cast dynamics as an optimization problem, we begin by writing time integration itself in variational form.
Consider the equations of motion
\begin{equation} \label{eq:eom}
    \matr M \dot{\vec v} = \sum_i \vec f_i(\vec x, \vec v),
\end{equation}
where $\matr M$ is the global mass matrix, $\dot{\vec v}$ is the acceleration, and $\vec f_i$ are the forces acting on the system.
These forces may be internal or external, and may in general depend on both positions $\vec x$ and velocities $\vec v$.

Let $\Delta t$ denote the time step, and let $(\xprev, \vprev)$ and $(\xnext, \vnext)$ denote the state at the beginning and end of the step, respectively.
For simplicity, we first consider Euclidean degrees of freedom, such as nodal positions in a particle system or a deformable body discretization.
Using Backward Euler, we write
\begin{equation}
    \dot{\vec v} \approx \frac{\vnext - \vprev}{\Delta t},
    \qquad
    \vnext \approx \frac{\xnext - \xprev}{\Delta t}.
\end{equation}
For non-Euclidean configuration spaces, such as rigid body rotations, the same variational viewpoint still applies, but the kinematic update must be written in a form consistent with the underlying representation.
Substituting these expressions into Eq.~\eqref{eq:eom}, and assuming for now that the forces are evaluated at the end of the step, yields
\begin{equation} \label{eq:eom_discrete}
    \matr M \frac{\xnext - \xprev - \Delta t \vprev}{\Delta t^2}
    - \sum_i \vec f_i(\xnext, \vnext) = \vec 0.
\end{equation}

A state $\vec x$ satisfying Eq.~\eqref{eq:eom_discrete} is a state that satisfies the discrete balance of forces for the current time step.
One may attempt to solve this nonlinear system directly, for instance using Newton's method.
However, solving a nonlinear root-finding problem directly can be fragile:
Newton steps may overshoot, enter invalid intermediate states, or fail to converge when started far from the solution.
These difficulties are particularly pronounced in stiff and strongly nonlinear simulation problems.

For a prominent subset of such systems, however, the same step can be reformulated as a nonlinear optimization problem.
Assume that the forces are conservative, so that each force can be written as
\begin{equation}
    \vec f_i(\vec x) = -\nabla \phi_i(\vec x),
\end{equation}
for some potential energy $\phi_i$.
Then the Backward Euler step is equivalent to minimizing the incremental potential
\begin{equation} \label{eq:be_potential}
    E(\vec x)
    =
    \frac{1}{2\Delta t^2}
    (\vec x - \tilde{\vec x})^T \matr M (\vec x - \tilde{\vec x})
    + \sum_i \phi_i(\vec x),
\end{equation}
where $\tilde{\vec x} = \xprev + \Delta t \vprev$ is the inertial position.
This equivalence follows from the stationarity condition $\nabla E(\xnext)=\vec 0$, which recovers Eq.~\eqref{eq:eom_discrete} for conservative forces.

The time step update can therefore be written as
\begin{align}
    \xnext &= \text{arg\,min}_{\vec x} E(\vec x), \label{eq:step_x} \\
    \vnext &= \frac{\xnext - \xprev}{\Delta t}. \label{eq:step_v}
\end{align}

This \emph{variational} form is very useful in practice.
Rather than solving a nonlinear root-finding problem directly, we can minimize a scalar objective.
This immediately enables the use of globalization techniques such as backtracking line search or trust-region strategies~\cite{NW06}, which make Newton-like methods substantially more robust.
In particular, line search enforces sufficient decrease of the objective and prevents overly aggressive steps, which is crucial in the presence of strong nonlinearities such as large deformations or contact.

The energy-based viewpoint also provides a natural interface for multiphysics simulation.
Each physical effect contributes an additive term to the same objective, and the total gradient and Hessian are simply the sums of the corresponding per-effect contributions.
This leads to a conceptually simple and implementation-friendly framework in which elasticity, inertia, contact, and other effects are handled in a uniform manner.

Unfortunately, not all physically relevant forces can be incorporated directly into this framework.
Dissipative or more general non-conservative effects do not, in general, derive from a scalar potential in the unknown state variables.
This includes many common models for friction, damping, and plastic flow, and is often reflected by a force Jacobian that is not symmetric.
Nevertheless, approaches have been proposed to bring such effects to optimization-based solvers, for example through lagged or split treatments of selected terms.

Although Eq.~\eqref{eq:be_potential} was derived for Backward Euler, the variational viewpoint is not limited to this integrator.
Incremental potential formulations have also been developed for other implicit schemes, including implicit Newmark~\cite{KMOW00,LFS+20}, the implicit midpoint method~\cite{dinev_stabilizing_2018}, and higher-order schemes such as the trapezoidal rule, BDF2, and TR-BDF2~\cite{BOFN18,CLL+22}.

% -------------------------------------------------------------------------------------------------------------
\subsubsection{Newton's Method}
\label{sec:newtons-method}
We will use Newton's method to minimize the incremental potential $E(\vec x)$ due to the robustness and fast local convergence it provides.
It makes direct use of first- and second-order derivatives and shows great robustness and fast local convergence, which is particularly effective for the stiff nonlinear problems that arise in practice.

At a current iterate $\vec x$, Newton's method builds the local quadratic model
\begin{equation}
    m_k(\Delta \vec x)
    =
    E(\vec x)
    +
    \nabla E(\vec x)^T \Delta \vec x
    +
    \frac{1}{2}
    \Delta \vec x^T
    \nabla^2 E(\vec x)
    \Delta \vec x,
\end{equation}
and chooses the search direction $\Delta \vec x$ as its minimizer.
This yields the linear system
\begin{equation} \label{eq:newton_system}
    \nabla^2 E(\vec x) \, \Delta \vec x
    =
    - \nabla E(\vec x).
\end{equation}
The iterate is then updated as
\begin{equation}
    \vec x^{(k+1)} = \vec x + \alpha \Delta \vec x,
\end{equation}
where $\alpha \in (0,1]$ is typically chosen by line search.

From the viewpoint of mechanics, the gradient $\nabla E(\vec x)$ is the residual force of the current state.
A stationary point satisfies $\nabla E(\vec x)=\vec 0$, which is precisely the condition of discrete force balance.
The Hessian $\nabla^2 E(\vec x)$ is the local stiffness, that is, the linearization of the residual forces with respect to the degrees of freedom.
Newton's method can therefore be interpreted as repeatedly solving a linearized equilibrium problem around the current iterate.
It converges in one iteration for a quadratic objective, and typically converges quadratically once sufficiently close to a regular minimizer.

The quality of Newton's method depends strongly on the regularity of the objective.
At a minimum, $C^1$ continuity is required so that the gradient is well defined.
In practice, $C^2$ continuity is desirable, since Newton's method relies on the Hessian to build a local quadratic model.
If the objective is discontinuous, optimization progress may stall entirely, and line search may get stuck at the discontinuity.
The construction of smooth and well-behaved energies is a central theme in simulation.

Through this document, we will visit important aspects of this optimization process: the assembly of gradients and Hessians, Hessian projection to positive definiteness, line search conditions, convergence criteria, and inexact linear solves.
The overall procedure is summarized in Algorithm~\ref{alg:newton}.

\begin{algorithm}[tb]
	\caption{Newton's method for optimization of the Backward Euler incremental potential.}
	\label{alg:newton}
	\begin{algorithmic}[1]
	\State $\vec x \gets \tilde{\vec x} = \xprev + \Delta t \vprev$ \Comment{initialize with inertial position}
	\While{not converged}
		\State assemble gradient $\nabla E(\vec x)$ and Hessian $\nabla^2 E(\vec x)$ \Comment{Sec.~\ref{sec:derivatives-and-assembly}}
		\State solve $\nabla^2 E(\vec x)\, \Delta \vec x = -\nabla E(\vec x)$ \Comment{Eq.~\eqref{eq:newton_system}}
		\State find step size $\alpha$ by line search \Comment{Sec.~\ref{sec:line-search}}
		\State $\vec x \gets \vec x + \alpha \Delta \vec x$
	\EndWhile
	\State $\vnext \gets (\xnext - \xprev) / \Delta t$
	\end{algorithmic}
\end{algorithm}

% -------------------------------------------------------------------------------------------------------------
\subsubsection{Global Derivatives and Assembly}
\label{sec:derivatives-and-assembly}
The global potential is usually an aggregation of different potential energies \emph{sources}, each of which is itself a sum of many \emph{local} contributions:
\begin{equation} \label{eq:global_energy_sum}
    E(\xnext) = \doublesumpotelem \potelem.
\end{equation}
Here, $\mathcal{P}$ denotes the different classes of potentials, such as inertia, elasticity, or contact, and $\mathcal{T}_p$ is the corresponding set of local contributions.
The vector $\xnext_e$ contains the subset of degrees of freedom associated with contribution $e$.
Since each local contribution only depends on a small part of the full unknown vector $\xnext$, we may write
\begin{equation} \label{eq:selector}
    \xnext_e = \matr P_e \xnext,
\end{equation}
where $\matr P_e$ is a selection matrix that extracts the relevant entries from the global vector.

For example, in a tetrahedral discretization of an elastic body, the total strain energy is the sum of the strain energies of the individual tetrahedra,
\begin{equation}
    \phi_{\text{strain}}(\xnext)
    =
    \sum_{e \in \mathcal{T}_{\text{strain}}}
    \phi_{\text{strain}}(\xnext_e),
\end{equation}
where $\mathcal{T}_{\text{strain}}$ is the set of tetrahedra and $\xnext_e$ contains the degrees of freedom of tetrahedron $e$.

A single simulation problem may also contain multiple sets of degrees of freedom.
This is especially common in multiphysics settings, where, for example, rigid and deformable bodies may be solved together, even though they may have different representations of state.
To keep the notation simple, we will continue to write the global unknown as $\xnext$, even though it refers to the stacked vector of all degrees of freedom
\begin{equation}
    \xnext = \left( \xnext_0, \xnext_1, \dots, \xnext_n \right)^T.
\end{equation}

Although the optimization problem is global, each local contribution only ``sees'' a few unknowns.
The global coupling arises because different local terms may depend on overlapping degrees of freedom:
for example, two neighboring tetrahedra share nodes, and those same nodes may also appear in contact primitives during a collision.
This overlap is reflected in the derivatives: local force contributions accumulate in the global gradient, and local stiffness contributions accumulate in the global Hessian.
Consequently, physical effects such as elasticity and contact are solved in a fully coupled manner through the same global minimization problem.

\subsubsection{Local-to-global derivatives.}
\label{sec:loca-global-derivatives}
Since the global potential is an aggregation, its derivatives are the sums of the derivatives of the individual local contributions:
\begin{align}
    \nabla E(\xnext)   &= \doublesumpotelem \nabla \potelem,\\
    \nabla^2 E(\xnext) &= \doublesumpotelem \nabla^2 \potelem.
\end{align}
However, these expressions must be interpreted carefully:
each local gradient and Hessian act only on the subset of degrees of freedom associated with their corresponding contribution.
For example, a linear tetrahedron with four nodes contributes a local gradient of size $12 \times 1$ and a local Hessian of size $12 \times 12$.
Another tetrahedron produces derivatives of the same size, but with respect to a different subset of vertices.
The process of inserting these local derivatives into the global gradient and Hessian according to their local-to-global mapping is called \emph{assembly}.

Using the selection matrix notation from Eq.~\eqref{eq:selector}, this mapping can be written compactly through the chain rule as
\begin{align}
    \nabla E(\xnext)   &= \doublesumpotelem \matr P_e^T \, \nabla \potelem,\\
    \nabla^2 E(\xnext) &= \doublesumpotelem \matr P_e^T \, \nabla^2 \potelem \, \matr P_e.
\end{align}
Local element derivatives are scattered into the global derivatives.
Since most local potentials only involve a small number of unknowns, the global Hessian is sparse, that is, most of its entries are zero.
Moreover, the sparsity pattern is not always fixed: dynamic contact, changing constraints, or adaptive remeshing may modify the connectivity of the problem over time.

When multiple sets of degrees of freedom are present in multiphysics simulation, the global derivatives naturally acquire a \emph{super-block} structure:
\begin{align}
    \def\arraystretch{1.3}
    \nabla E(\xnext) &=
    \begin{pmatrix}
    \nabla_{\vec x_1} E \\
    \nabla_{\vec x_2} E \\
    \vdots \\
    \nabla_{\vec x_n} E
    \end{pmatrix},
    \, \\
    \nabla^2 E(\xnext) &=
    \begin{pmatrix}
    \nabla^2_{\vec x_1 \vec x_1} E &
    \nabla^2_{\vec x_1 \vec x_2} E &
    \dots &
    \nabla^2_{\vec x_1 \vec x_n} E \\
    \nabla^2_{\vec x_2 \vec x_1} E &
    \nabla^2_{\vec x_2 \vec x_2} E &
    \dots &
    \nabla^2_{\vec x_2 \vec x_n} E \\
    \vdots & \vdots & \ddots & \vdots \\
    \nabla^2_{\vec x_n \vec x_1} E &
    \nabla^2_{\vec x_n \vec x_2} E &
    \dots &
    \nabla^2_{\vec x_n \vec x_n} E
    \end{pmatrix}.
\end{align}
Contributions that depend only on one set of degrees of freedom describe energy variations internal to a single subsystem.
These include, for example, inertial or elastic terms, as well as contact terms internal to one subsystem.
Off-diagonal Hessian blocks, on the other hand, describe couplings between different subsystems, such as contact, attachments, or fluid--structure interaction.

% -------------------------------------------------------------------------------------------------------------
\subsubsection{Penalty Constraints}
\label{sec:penalty}
Within an energy-based formulation, constraints are often incorporated through additional \emph{penalty} or \emph{barrier} terms.
Common examples include contact, strain-limiting, joint constraints and joint limits in rigid-body systems, and prescribed boundary conditions.
This preserves the unified optimization structure introduced above: constraints contribute to the same global potential and are handled by the same Newton solve.

Let $C(\vec x)$ be a scalar constraint function such that $C(\vec x)=0$ corresponds to exact satisfaction of the constraint.
A penalty energy can then be defined as
\begin{equation}
    \phi_c(\vec x) = \frac{k_c}{2}\, C(\vec x)^2,
\end{equation}
where $k_c > 0$ is the penalty stiffness.
Its gradient produces a restoring force that increases with the amount of violation,
\begin{equation}
    \vec f_c = -\nabla \phi_c(\vec x) = -k_c\, C(\vec x)\, \nabla C(\vec x).
\end{equation}
Thus, the larger the constraint violation, the stronger the corrective force.

For inequality constraints, however, corrective forces should act only when the constraint is violated.
If the feasible set is defined by $C(\vec x) \geq 0$, the corresponding penalty may be written as
\begin{equation}
    \phi_c(\vec x) =
    \begin{cases}
        0, & C(\vec x) \geq 0,\\[4pt]
        -\frac{k_c}{3}\, C(\vec x)^3, & C(\vec x) < 0.
    \end{cases}
\end{equation}
A cubic term is used here because it provides a $C^2$-continuous transition at activation, which is generally beneficial for the convergence of Newton's method.

In some cases, the constraint defines an additional \emph{hard limit} of the admissible state, as in contact with non-penetration guarantees.
For such constraints, logarithmic barrier energies are often used, since they can produce arbitrarily large corrective forces as the constraint approaches the boundary of admissibility, at the price of increased nonlinearity.

% -------------------------------------------------------------------------------------------------------------
\subsubsection{Line Search}
\label{sec:line-search}

Line search is central to the success of Newton's method in complex multiphysics problems, particularly in the presence of contact.
Newton's method provides a search direction $\Delta \vec x$ based solely on local information at the current iterate $\vec x$.
This direction may therefore be unaware of any nearby but currently inactive stiff barrier constraint, and may point directly across it.
In such a case, the full Newton step is clearly unacceptable.
More generally, far from the minimizer, the local quadratic model may simply be too inaccurate for the full step to be reliable.
In all such situations, taking the full step may lead to overshooting or even complete solver failure.

For this reason, Newton's method is almost always combined with a line search, which replaces the full update by
\begin{equation}
    \vec x^{(k+1)} = \vec x + \alpha \Delta \vec x,
\end{equation}
where the step length $\alpha \in (0,1]$ is chosen to satisfy suitable acceptance conditions.

The first requirement is usually \emph{sufficient decrease}.
Given a descent direction $\Delta \vec x$, one seeks a step length such that
\begin{equation} \label{eq:armijo}
    E(\vec x + \alpha \Delta \vec x)
    \leq
    E(\vec x) + c_1 \alpha \nabla E(\vec x)^T \Delta \vec x,
\end{equation}
where $c_1 \in (0,1)$ is a small constant.
This is the classical \emph{Armijo condition}~\cite{NW06}.
It ensures that the accepted step decreases the objective by an amount commensurate with the predicted local descent, and therefore prevents overly aggressive updates.

A standard strategy to enforce this condition is \emph{backtracking}.
One starts with $\alpha = 1$, corresponding to the full Newton step, and repeatedly reduces it, typically by a constant factor $\rho \in (0,1)$, until the acceptance condition is satisfied.
This mechanism retracts a step from regions of very large penalty or barrier growth, such as a step deep into a contact potential.

In physical simulation, however, decrease of the objective is not the only concern.
The iterates must also remain \emph{admissible}.
Depending on the application, this may exclude inverted finite elements, penetrations, or other invalid configurations.
In contact simulation, this is commonly obtained using continuous collision detection (CCD) to truncate the step in order to keep it admissible.
A simpler alternative is to test admissibility during backtracking itself, for example by testing for intersection-freeness, although this does not prevent tunnelling.

Finally, for very stiff problems, line search may become numerically delicate near convergence, since the change in total energy can approach floating-point accuracy.
In this regime, direct evaluation of the Armijo condition may become noisy and reject otherwise acceptable steps.
A standard remedy is to complement the direct energy comparison by a local Taylor approximation; see \cite{LLF+23}.

% -------------------------------------------------------------------------------------------------------------
\subsubsection{Convergence and Inexactness}
\label{sec:convergence}
In computer graphics applications, Newton's method is rarely driven to very high accuracy.
Instead, one usually accepts an \emph{inexact} solution once the remaining error is sufficiently small for the purpose of the simulation.
This is an important design choice: solving too accurately increases computational cost with little visible benefit, while terminating too early may introduce excessive numerical dissipation, poor force balance, and mechanically implausible behavior.
Moreover, the convergence criterion itself should remain meaningful across disparate materials and interactions, especially in multiphysics settings.

Validity-preserving mechanisms such as IPC make early termination substantially safer.
A search terminated before full convergence may still exhibit large force imbalances, but it remains collision-free and is therefore generally admissible as the starting point for the next time step.

Common stopping criteria include the following:
\begin{itemize}
    \item \textbf{Residual norm.}
    Terminate when
    \[
        \|\nabla E(\xnext)\| < \varepsilon_{\mathrm{res}}.
    \]
    This is the standard criterion in nonlinear optimization, since the gradient measures the remaining force imbalance.

    \item \textbf{Newton step norm.}
    Terminate when
    \[
        \|\Delta \xnext\| < \varepsilon_{\mathrm{step}}.
    \]
    This measures the size of the update and therefore how much the current iterate is still changing geometrically.

    \item \textbf{Fixed iterations.}
    Terminate after a prescribed number of Newton iterations.
\end{itemize}

From a mechanical viewpoint, the residual norm is the most direct measure of convergence, since it quantifies the remaining force imbalance.
However, for inexact solves, it is often difficult to choose a coarse residual tolerance that both correlates with visual outcome and remains meaningful across very different stiffness scales.
Consider, for example, a dryer machine rotating a piece of cloth inside.
A residual force of a given magnitude may correspond to a negligible geometric error in the stiff and heavy rigid components, but to a large visible deformation in the cloth.

In this setting, the Newton step norm is often more practical because it measures the size of the correction being applied.
In fact, the Newton step is quite literally the residual weighted by the inverse of mass and stiffness (see Eq.~\refeq{eq:newton_system}).
This relative scaling of the force residual to the local stiffness makes it more consistent across mixed-material and multiphysics settings.
For better temporal consistency, the tolerance is often scaled with the time step:
\[
    \|\Delta \xnext\| < \Delta t \, \varepsilon_{\mathrm{vel}}.
\]
Note that neither the residual nor the step norms are guaranteed to decrease monotonically during a Newton search.
Nevertheless, it is usually reasonable to expect the accepted steps to become progressively smaller as the iterates approach stationarity.

Finally, while terminating after a fixed number of iterations provides predictable runtime, it is not tied to any direct measure of convergence or mechanical accuracy.

The appropriate choice ultimately depends on the goals of the application, but both the stopping criterion and its tolerance should be chosen deliberately.
Stopping too late wastes work, while stopping too early reduces accuracy.
Consistency across models, discretizations, time steps, and parameter choices should therefore be considered carefully.

% -------------------------------------------------------------------------------------------------------------
\subsubsection{Projection to Positive Definite}
\label{sec:projection-PD}
A central requirement for Newton's method is that the search direction be a \emph{descent direction}.
Recall the Newton system
\begin{equation}
    \nabla^2 E \, \Delta \vec x
    =
    - \nabla E.
\end{equation}
If the Hessian is symmetric positive definite (SPD) at the current iterate, that is,
\begin{equation}
    \vec r^T \nabla^2 E \, \vec r > 0
    \qquad
    \forall \vec r \neq \vec 0,
    \label{eq:spd}
\end{equation}
then the local quadratic model is strictly convex and the Newton direction is guaranteed to be a descent direction.
Indeed, multiplying the Newton equation by $\Delta \vec x^T$ gives
\begin{equation}
    \Delta \vec x^T \nabla E(\vec x)
    =
    -\Delta \vec x^T \nabla^2 E(\vec x) \Delta \vec x,
\end{equation}
implying
\begin{equation}
    \Delta \vec x^T \nabla E(\vec x) < 0.
\end{equation}
Thus, the Newton step is aligned with the negative gradient and can be used safely within a line search.

In practice, however, Hessians arising in physical simulation are not always positive definite and therefore will generally not guarantee descend directions.
For this reason, many simulation methods replace the true Hessian by a positive definite approximation, thereby guaranteeing descent while remaining as faithful as possible to the original model.
This process is commonly referred to as \emph{projection to positive definite}.

A standard approach is to filter, for example by clamping, the negative eigenvalues of the element Hessians prior to assembly.
Consider the eigendecomposition of the Hessian of element $e$,
\begin{equation}
    \vec{H}^{e}=\vec{Q}^{e}\,\vec{D}^{e}\bigl(\vec{Q}^{e}\bigr)^T,
\end{equation}
where the columns of $\vec{Q}^{e}$ are the eigenvectors of $\vec{H}^{e}$ and $\vec{D}^{e}$ is a diagonal matrix containing the corresponding eigenvalues.
Applying clamping, the projected element Hessian is
\begin{equation}
    \widehat{\vec{H}}^{e}=\vec{Q}^{e}\,\widehat{\vec{D}}^{e}\bigl(\vec{Q}^{e}\bigr)^T,
\end{equation}
with
\begin{equation}
    \widehat{D}^{e}_{ii}=\max(D^{e}_{ii},\varepsilon),
\end{equation}
for some $\varepsilon>0$.
Since a sum of SPD matrices is SPD, the resulting assembled global matrix is also SPD.
This approach is commonly referred to as Projected Newton~\cite{teran05}.

In some cases, however, projecting all element Hessians at every iteration introduces unnecessary distortion of the global Hessian and adds computational overhead.
This is particularly relevant in problems with natural regularization, for example in dynamics, where the inertial term already contributes positive curvature.
In such settings, it can be beneficial to apply projection only when needed.
In particular, Project-On-Demand Newton~\cite{LLF+23} projects local Hessians only when the assembled global Hessian is detected to be indefinite, and Progressively Projected Newton~\cite{PPN} selectively projects contributions in regions more likely to induce indefiniteness.

Beyond clamping negative eigenvalues, other spectral modifications have also been explored.
A common alternative is \emph{eigenvalue mirroring}, in which negative eigenvalues are reflected to positive values by an absolute-value operation rather than clamped to a floor~\cite{eigenvalue_mirroring}.
Related approaches interpolate between clamping and mirroring in order to balance robustness and fidelity to the original curvature~\cite{eigenvalue_blending,mirroring_trustregion}.
These variants have shown strong results in simulations of near-incompressible materials undergoing extreme deformation.

A related but conceptually distinct approach is to avoid the true Hessian altogether and instead build a positive semidefinite approximation from the outset.
The most common example is the Gauss--Newton method, which arises when the objective is written as a sum of squared residuals.
Its Hessian approximation discards second-order residual terms and retains only a matrix of the form $\matr J^T \matr J$, which is positive semidefinite by construction.

% -------------------------------------------------------------------------------------------------------------
\subsubsection{Linear System Solves}
\label{sec:linear-system-solves}
Once the gradient and Hessian have been assembled, each Newton iteration requires the solution of the linear system in Eq.~\eqref{eq:newton_system}.
In large-scale simulation, the assembly of the Hessian and the solution of the linear system often account for a large fraction of the total runtime.
Equally important, the linear solve is usually \emph{inexact}: in practice, one does not seek a highly accurate solution of the Newton system at every iteration, but rather a level of accuracy commensurate with the current quality of the quadratic model.

As discussed in Section~\ref{sec:derivatives-and-assembly}, the assembled Hessian is typically very sparse, and its sparsity pattern may change over time due to contact activation, remeshing, or changing constraints.
Practical implementations therefore rely on sparse formats that support efficient assembly and matrix--vector products.
Popular choices include Blocked Compressed Row Storage (BCRS), where each nonzero stores a small dense block such as $3\times 3$ in three-dimensional nodal discretizations, and ELL-type formats, which are attractive for their regular memory access and parallel efficiency.
In all cases, assembly and sparse linear algebra are usually memory-bound, making data layout and access critical for performance.

When the Hessian has been made positive definite, as discussed in the previous section, one may use solvers designed for SPD systems.
Direct factorizations such as Cholesky or LDLT are robust and accurate, and are often attractive for small problems or for simulations so stiff that iterative methods become ineffective.
When iterative solvers are applicable, the standard choice is the Preconditioned Conjugate Gradient (PCG) method, equipped with a good preconditioner.
At a minimum, diagonal or block-diagonal preconditioners should be used; stronger alternatives such as incomplete factorizations may be worthwhile when their setup cost is justified.

\paragraph*{Inexact Newton Solves}
\label{sec:inexact-newton-solves}
A key practical advantage of iterative linear solvers is that they support inexact Newton methods naturally.
Far from the minimizer, solving the Newton system to very tight tolerances is usually wasteful, since the quadratic model is itself the limiting factor for progress.
It is then preferable to compute a larger number of cheaper Newton steps.
Near the solution, however, more accurate solves are desirable in order to recover fast local convergence.

This strategy is commonly described in terms of a \emph{forcing sequence}, which tightens the accuracy of the inner linear solve as the outer Newton iteration progresses.
A common practical surrogate is to terminate CG once its relative residual has been reduced by a fixed factor, e.g.
\begin{equation}
    \|\vec r_k\| \leq \eta \, \|\vec r_0\|,
\end{equation}
with $\eta \approx 10^{-4}$.
This is simple and effective in practice, and often provides the desired behavior across Newton iterations.
Strictly speaking, however, it is not a forcing sequence in the classical sense, where the inner tolerance is tied explicitly to the nonlinear Newton residual rather than fixed a priori.
Nevertheless, this relative-residual criterion is widely used and it provides a good balance between efficiency and convergence in simulation.

In summary, the linear system solve should not be viewed as an isolated numerical subroutine.
It is tightly coupled to sparse representation, assembly, Hessian definiteness, preconditioning, and the inexactness strategy of the outer Newton method.
A good implementation balances all of these aspects simultaneously.

% -------------------------------------------------------------------------------------------------------------
\subsubsection{Differentiability}
The optimization-based simulation framework described above requires a certain degree of differentiability, which is likewise needed in many inverse design and parameter identification applications.
When the underlying models and solvers are sufficiently smooth, one can compute sensitivities of the simulation output with respect to its parametrization.
This makes it possible not only to simulate a system for fixed parameters, but also to adjust those parameters so that the simulation matches a desired outcome.
Here we briefly state the main technique used to compute such sensitivities, and refer the reader to the course notes by Coros et al.~\cite{diff_sim_course} for a more complete treatment.

Let $\vec \theta \in \mathbb{R}^N$ denote a vector of parameters, and let $\vec x(\vec \theta)$ denote the converged simulation state obtained by minimizing the incremental potential.
We define a loss $\mathcal{L}(\vec x(\vec \theta), \vec \theta) \in \mathbb{R}$ and seek to solve
\begin{equation}
    \vec \theta = \text{argmin}_{\vec \theta} \mathcal{L}(\vec x(\vec \theta), \vec \theta).
\end{equation}

To use gradient-based optimization methods, we need the total derivative $\frac{\mathrm{d}\mathcal{L}}{\mathrm{d}\vec \theta}$.
Differentiating every operation inside Newton's method is possible in principle, for example by automatic differentiation, but is often impractical for large-scale problems.
Instead, one typically applies the \emph{Implicit Function Theorem}.

We have already established that the solution of the nonlinear mechanical problem satisfies
\begin{equation}
    \nabla_{\vec x} E(\vec x,\vec \theta) = \vec 0.
\end{equation}
Then the total derivative of the loss can be written as
\begin{align}
    \label{eq:adjoint}
    \frac{\mathrm{d}\mathcal{L}}{\mathrm{d}\vec \theta}
    &=
    \frac{\partial \mathcal{L}}{\partial \vec \theta}
    +
    \vec \lambda^T
    \frac{\partial^2 E}{\partial \vec x \, \partial \vec \theta},
    \quad
    \text{for}
    \quad
    \left(\frac{\partial^2 E}{\partial \vec x^2}\right)^T \vec \lambda
    =
    -
    \left(\frac{\partial \mathcal{L}}{\partial \vec x}\right)^T,
\end{align}
where $\vec \lambda$ is the adjoint state.
This expression connects the simulation degrees of freedom $\vec x$ and the parameters $\vec \theta$ through mixed derivatives of the physical model and the loss.
Importantly, evaluating Eq.~\eqref{eq:adjoint} requires only one additional linear solve after the forward simulation has converged, regardless of how many Newton iterations were needed.

Finally, Eq.~\eqref{eq:adjoint} describes the sensitivity of the solution of a single nonlinear solve.
In dynamic simulation, however, we must solve a sequence of such problems.
The corresponding sensitivities must then be propagated through that sequence, which leads to a sequence of adjoint computations, typically applied in reverse time order.

% -------------------------------------------------------------------------------------------------------------
\subsubsection{Frameworks}
We have seen that optimization time integrators provide a convenient way to unify complex multiphysics problems through a simple mathematical interface: a global scalar objective and a collection of degrees of freedom.
Their implementation, however, can still be highly involved.
A complete system may require many local potentials, several coupled sets of degrees of freedom, first- and second-order derivatives, sparse assembly, positive-definite projection, line search, and robust, efficient linear solves.
Implementing, optimizing, testing, and integrating all these components into a correct and efficient solver is a substantial engineering task.
Several frameworks have therefore been proposed to facilitate the construction of such systems by taking over large parts of the differentiation and solution process from a concise problem definition.

Representative examples include TinyAD~\cite{TinyAD}, SymX~\cite{FLW25}, and the mesh-based simulation framework of Herholz et al.~\cite{Herholz24}.
These may be viewed as lower-level mathematical engines on top of which a multiphysics solver can be built.
TinyAD is a lightweight C++ library for first- and second-order differentiation of sparse mesh-based problems, and is particularly attractive for its simplicity and ease of integration.
It was introduced in the context of geometry processing and provides simple solver facilities while relying heavily on the Eigen library.
SymX targets energy-based simulation more directly, and provides a more extensive infrastructure for the ingredients discussed throughout this tutorial.
The authors report strong performance on complex nonlinear and multiphysics problems, including higher-order FEM, rigid bodies, and frictional contact.
Herholz et al.~\cite{Herholz24} pursue a similar goal, although their system is not publicly released.

At a higher level, there are also full multiphysics solvers available.
STARK~\cite{FLL+24} is built on top of SymX, thereby removing differentiation from the manual implementation pipeline.
It provides ready-to-use models for FEM, rigid bodies, and frictional contact.
As such, it can serve as a useful starting point for researchers and practitioners who wish to develop new models within an already integrated simulation environment.

Alternatively, the IPC Toolkit~\cite{ipc_toolkit} provides a modular C++ implementation of the core IPC building blocks that will be introduced in the following section, including barrier potentials, friction, and continuous collision detection, and can be integrated into existing simulation pipelines.
PolyFEM~\cite{polyfem} is a finite element library that supports a wide range of element types and formulations, including higher-order bases and mixed methods, and can be combined with IPC Toolkit.

\subsection{Energy-Based Materials and Phenomena}
\label{sec:energy-based-materials}

Any materials or physical effects that admit a formulation in terms of an energy or scalar potential can be easily combined to obtain a strongly coupled multiphysics simulator by summing their contributions and any interaction potentials in the global incremental potential from Equation~\eqref{eq:be_potential}.
As a result, these models benefit from the numerical methods and their stability as outlined before.
In the following, we give a brief overview of recent works that use this approach to model specific effects or to perform coupled simulations.

\subsubsection{Elastic Deformables} %
The simulation of elastic deformables naturally fits into an optimization-based framework.
Continuous material models for deformable bodies are typically defined in terms of a scalar strain energy density function $\Psi(\matr F)$ depending on the local deformation gradient $\matr F$.
The total strain energy of a body occupying reference domain $\Omega_0$ is obtained by integrating this density over the volume,
\begin{equation} \label{eq:strain_energy_continuous}
    \phi_{\text{strain}}(\vec x) = \int_{\Omega_0} \Psi\!\bigl(\matr F(\vec x, \vec X)\bigr) \, \mathrm{d}\vec X,
\end{equation}
where $\vec X$ denotes material (reference) coordinates.
This strain energy can be added directly as one of the potential terms $\phi_i$ in the incremental potential of Equation~\eqref{eq:be_potential}.

To evaluate this integral numerically, a spatial discretization is required.
Using the Finite Element Method (FEM) with a Lagrangian view, the domain is partitioned into elements (e.g.\ tetrahedra in 3D), and the integral is approximated using numerical quadrature,
\begin{equation} \label{eq:strain_energy_discrete}
    \phi_{\text{strain}}(\vec x) \approx \sum_{e \in \mathcal{T}_{\text{strain}}} \sum_{q=1}^{Q_e} w_{eq} \, \Psi\!\bigl(\matr F(\vec x_e, \vec X_{eq})\bigr),
\end{equation}
where $Q_e$ is the number of quadrature points in element $e$, $w_{eq}$ are the corresponding quadrature weights, $\vec X_{eq}$ are the quadrature point locations in the reference configuration, and $\vec x_e = \matr P_e \vec x$ are the nodal positions of element $e$ extracted via the selection operator from Eq.~\eqref{eq:selector}.
The deformation gradient $\matr F(\vec x_e, \vec X_{eq})$ at each quadrature point is computed from the element's nodal positions using the shape function derivatives evaluated at $\vec X_{eq}$.
For higher-order elements with non-constant deformation gradients, multiple quadrature points per element are necessary to accurately capture the variation of the strain energy density across the element.

Different element geometries can be used in practice, such as tetrahedra, hexahedra or even polyhedra~\cite{WBG07,BHS22}.
However, in graphics the most common choice is linear tetrahedra, where the shape functions are linear and the deformation gradient is therefore constant within each element.
A single quadrature point per element (with weight $w_e = V_e^0$, the reference volume) is then sufficient, and Eq.~\eqref{eq:strain_energy_discrete} simplifies to
\begin{equation}
    \phi_{\text{strain}}(\vec x) \approx \sum_{e \in \mathcal{T}_{\text{strain}}} V_e^0 \, \Psi\!\bigl(\matr F_e(\vec x_e)\bigr).
\end{equation}
For a linear tetrahedron with nodes $\vec x_0, \dots, \vec x_3$, the constant deformation gradient is given by
\begin{equation} \label{eq:deformation_gradient}
    \matr F_e
    =
    \begin{pmatrix} \vec x_1 - \vec x_0 & \vec x_2 - \vec x_0 & \vec x_3 - \vec x_0 \end{pmatrix}
    \matr D_e^{-1},
\end{equation}
where $\matr D_e = \begin{pmatrix} \vec X_1 - \vec X_0 & \vec X_2 - \vec X_0 & \vec X_3 - \vec X_0 \end{pmatrix}$ is formed from the reference positions and its inverse can be precomputed.
In either case, each element contribution takes the role of a local potential in the assembly from Eq.~\eqref{eq:global_energy_sum}, and contributes its gradient and Hessian to the global system as described in Section~\ref{sec:loca-global-derivatives}.

Commonly used examples of hyperelastic material models that can be formulated in this way include the Corotated material model~\cite{SHST12,MZS11}, the Neohookean and Stable Neohookean materials~\cite{SGK18} and the ARAP model~\cite{LCK22,SA07}.
Non-classical material models such as micropolar materials with rotational degrees of freedom can also be considered in this framework as shown by Löschner et al.~\cite{LFJ+23,LFJB24}.
While FEM is a sophisticated numerical method that supports higher-order discretizations~\cite{SHD+18,LLK+20,SHG+22}, linear elements were successfully used for very efficient methods and possibly interactive applications, for example the Fast Corotated FEM approach by Kugelstadt et al.~\cite{KKB18}.
For more details on deformable solids and their spatial discretization using FEM, we recommend the courses of Kim and Eberle~\cite{KE22} and Sifakis and Barbi\v{c}~\cite{SB12}.

\begin{algorithm}[tb]
	\caption{Incremental potential contact (IPC) with Newton's method.}
	\label{alg:ipc}
	\begin{algorithmic}[1]
	\State $\vec x \gets \tilde{\vec x} = \xprev + \Delta t \vprev$ \Comment{initialize with inertial position}
	\While{not converged}
		\State detect active contact set $\hat{\mathcal{C}}(\vec x) = \{k \in \mathcal{C} : d_k(\vec x) \leq \hat d\}$
		\State assemble $E(\vec x) = E_{\text{inertia}} + \phi_{\text{strain}} + \phi_{\text{contact}} + \phi_{\text{friction}}$
		\State assemble gradient $\nabla E(\vec x)$ and Hessian $\nabla^2 E(\vec x)$
		\State project $\nabla^2 E(\vec x)$ to positive definiteness
		\State solve $\nabla^2 E(\vec x)\, \Delta \vec x = -\nabla E(\vec x)$ \Comment{Eq.~\eqref{eq:newton_system}}
		\State $\alpha_{\text{CCD}} \gets$ largest step preserving intersection-freeness via CCD
		\State $\alpha \gets$ backtracking line search starting from $\alpha_{\text{CCD}}$ \Comment{Eq.~\eqref{eq:armijo}}
		\State $\vec x \gets \vec x + \alpha \Delta \vec x$
		\State update barrier stiffness $\kappa$ adaptively
	\EndWhile
	\State $\xnext \gets \vec x$
	\State $\vnext \gets (\xnext - \xprev) / \Delta t$
	\end{algorithmic}
\end{algorithm}

\subsubsection{Contact Potentials} %
The coupling of deformable solids between themselves and to other materials is typically realized through frictional contact.
In the context of unconstrained optimization-based integration, \emph{Incremental Potential Contact} (IPC)~\cite{LFS+20} gained significant popularity in recent years.
IPC is designed to enable intersection-free simulations and is based on a fully implicit and unconstrained formulation of contact.
The overall procedure is summarized in Algorithm~\ref{alg:ipc}.

The key idea is to define admissibility in terms of unsigned distances $d_k(\vec x)$ between non-adjacent surface primitive pairs $k \in \mathcal{C}$, where $\mathcal{C}$ is the set of all non-incident point-triangle and non-adjacent edge-edge pairs on the surface mesh.
A configuration is intersection-free if $d_k(\vec x) > 0$ for all $k \in \mathcal{C}$.
To enforce this within unconstrained optimization, IPC augments the incremental potential from Eq.~\eqref{eq:be_potential} with a sum of smoothly clamped barrier potentials (see Section~\ref{sec:penalty}) over all primitive pairs,
\begin{equation} \label{eq:ipc_barrier_potential}
    \phi_{\text{contact}}(\vec x) = \kappa \sum_{k \in \mathcal{C}} b\!\left(d_k(\vec x), \hat d\right),
\end{equation}
where $\kappa > 0$ is an adaptively chosen barrier stiffness, $\hat d > 0$ is a user-specified distance threshold controlling geometric accuracy of the contact, and $b(d, \hat d)$ is a $C^2$-continuous, smoothly clamped log-barrier defined as
\begin{equation}
    b(d, \hat d) =
    \begin{cases}
        -(d - \hat d)^2 \ln\!\left(\dfrac{d}{\hat d}\right), & 0 < d < \hat d,\\[6pt]
        0, & d \geq \hat d.
    \end{cases}
\end{equation}
This barrier is exactly zero for $d \geq \hat d$, providing a smooth clamping that allows efficient pruning of distant pairs, and tends to infinity as $d \to 0$, generating arbitrarily large repulsive forces to prevent intersection.
Because $b$ vanishes for distant primitives, only the active constraint set $\hat{\mathcal{C}}(\vec x) = \{k \in \mathcal{C} : d_k(\vec x) \leq \hat d\}$ contributes non-zero terms, making evaluation efficient despite the quadratic number of potential pairs.
Each barrier term plays the role of a local potential in the assembly from Eq.~\eqref{eq:global_energy_sum}, contributing its gradient and Hessian to the global system.
The resulting contact forces are given by the barrier derivatives $\lambda_k = -\kappa \, \partial b / \partial d_k$, which are guaranteed to be non-negative (repulsive) by construction.

To guarantee that the intersection-free invariant is maintained not just at the solution but throughout all Newton iterations, IPC augments the backtracking line search (see Section~\ref{sec:line-search}) with continuous collision detection (CCD): before accepting a step $\alpha \Delta \vec x$, CCD determines the largest feasible step size $\alpha_{\text{CCD}}$ along the search direction such that no primitive pair intersects along the linear trajectory from $\vec x$ to $\vec x + \alpha_{\text{CCD}} \Delta \vec x$.
The line search then backtracks starting from $\alpha_{\text{CCD}}$ to find sufficient energy decrease, ensuring that every iterate remains admissible.
Combined with a non-inverting elastic energy density such as neo-Hookean, this yields a trajectory that is intersection- and inversion-free at every solver iteration.

Further contributions of IPC include a smoothed and lagged friction potential (discussed below in the context of damping and friction) and an adaptive barrier stiffness update for $\kappa$ that improves conditioning.
Subsequent works aim to resolve spurious tangential forces in the original formulation~\cite{DLCT24}, ensure convergence under refinement~\cite{LFS+23} and extend it to high-order (curved) meshes~\cite{high_order_ipc}.
More recently, Huang et al.~\cite{HPF25} proposed the \emph{Geometric Contact Potential} (GPC), a barrier formulated as a surface integral that addresses several limitations of IPC simultaneously: it eliminates spurious contact forces in rest configurations by construction, ensures convergence under mesh refinement, supports both smooth and piecewise-smooth (sharp-featured) surfaces in a unified framework, and allows substantially larger barrier extents without artificial stiffness.

While the original IPC framework did not target interactive applications, recent works aim to fill this gap by proposing efficiency and robustness improvements such as better convergence using cubic barriers with dynamic stiffness~\cite{cubic_barriers}, improved conditioning using a barrier-augmented Lagrangian~\cite{GLY+24}, Preconditioned Conjugate Gradient solvers~\cite{SCBL24}, improved Hessian projection and filtering in a Gauss-Newton method~\cite{HCLK24}, localized CCD updates and improved Parallelization~\cite{LLJ+23} or integration into the Projective Dynamics framework~\cite{lan_penetration-free_2022} which all advertise fast GPU implementations.
Taking a fundamentally different approach to real-time performance, \textit{Offset Geometric Contact} (OGC)~\cite{CHL+25} bypasses IPC's expensive continuous collision detection entirely. By offsetting surface geometry to ensure orthogonal contact forces and restricting solver steps within safe, per-vertex conservative bounds, OGC provides a penetration-free guarantee that is exceptionally efficient for local solvers like VBD (see Section~\ref{sec:VBD}).

Alternatively, if the requirement of fully intersection-free configurations can be relaxed, various penalty-based collision energies as investigated by Shi and Kim~\cite{SK23} or by Macklin et al.~\cite{macklin_primaldual_2020} with accurate penetration-depths obtained by e.g. the method of Chen et al.~\cite{CDY23} can also be applied in the context of optimization-based integration.
These barrier and penalty energy-based approaches naturally fit in an optimization framework and provide strongly coupled contact.
For more details on contact and friction handling in computer graphics, we recommend the course by Andrews et al.~\cite{andrews_contact_2022}.

\subsubsection{Damping, Friction and Plasticity} %
So far, all forces we have considered are \emph{conservative}: they derive from a scalar potential energy that depends only on the configuration $\vec x$.
This is what makes the optimization-based formulation work: every force contributes a term to the incremental potential from Eq.~\eqref{eq:be_potential}, and minimizing the sum solves the equations of motion.
However, many important physical effects are \emph{dissipative}: damping removes kinetic energy, and friction opposes relative sliding motion.
Such dissipation is inherently path-dependent---the energy lost depends on the trajectory taken, not merely on the current state---so, unlike conservative forces, it cannot be described by a potential that assigns an energy to a single configuration.
Only once time integration fixes a discrete path across the step it becomes meaningful to ask whether a dissipative force can still be captured by a potential defined over that step.

A useful framework for modeling such effects is the \emph{dissipation potential} $R(\vec x,\vec v)$, a non-negative scalar function of positions and velocities whose negative velocity gradient yields the dissipative force, $\vec f_d = -\nabla_{\vec v} R(\vec x,\vec v)$ (see~\cite{BOFN18,strain_rate_damping}); Coulomb friction and Rayleigh damping can both be cast in this form.
After discretization, the velocity update $\vec v = (\vec x - \xprev)/\Delta t$ makes $\vec f_d$ a function of $\vec x$ alone, and a per-time-step potential $\phi_d(\vec x)$ with $\vec f_d = -\nabla_{\vec x}\phi_d$ requires this force to be \emph{integrable}, which in particular demands a symmetric Jacobian.
For some models it is: Rayleigh damping, for instance, admits an exact potential that is merely inconvenient to use directly (see below).
For others, such as Coulomb friction and plastic flow, the Jacobian is inherently non-symmetric and cannot equal the Hessian of any scalar energy; these require \emph{lagging} (freezing selected position-dependent quantities at a known state until the remaining force becomes integrable) or other specialized treatments.
When $R$ depends on velocities alone, integrability is automatic and the objective term $\phi_d = \Delta t\, R(\vec v)$ can be used directly.

\paragraph*{Rayleigh Damping}
Rayleigh damping is a widely used model that defines the damping force as a linear combination of mass- and stiffness-proportional terms,
\begin{equation} \label{eq:rayleigh_damping_force}
    \vec f_d = -(\alpha \matr M + \beta \matr K(\vec x))\,\vec v,
\end{equation}
where $\alpha \geq 0$ and $\beta \geq 0$ are the mass and stiffness damping coefficients, and $\matr K(\vec x) = \nabla^2 \phi_{\text{strain}}(\vec x)$ is the elastic stiffness matrix.
This corresponds to the dissipation potential
\begin{equation} \label{eq:rayleigh_dissipation}
    R(\vec x, \vec v) = \tfrac{1}{2}\,\vec v^T (\alpha \matr M + \beta \matr K(\vec x))\,\vec v.
\end{equation}
Here the stiffness matrix $\matr K$ depends on positions, so this dissipation potential depends on both $\vec x$ and $\vec v$.
Substituting the Backward Euler velocity update nonetheless yields an exact objective term as a function of $\vec x$ alone.
Retaining the full position dependence of $\matr K(\vec x)$ is nevertheless problematic in practice: since $\matr K = \nabla^2 \phi_{\text{strain}}$, its contribution to the Hessian of the incremental potential would require third derivatives of the elastic energy, and the damping matrix $\alpha \matr M + \beta \matr K(\vec x)$ is not guaranteed to be positive semi-definite, so an indefinite contribution could inject energy and cause spurious accelerations rather than dissipate them.
To avoid this, Gast et al.~\cite{GSS+15} \emph{lag} the position dependence: the stiffness matrix is evaluated at the known positions from the start of the time step, $\matr K(\xprev)$, yielding the lagged dissipation potential
\begin{equation} \label{eq:rayleigh_lagged}
    R^{\text{lag}}(\vec v) = \tfrac{1}{2}\,\vec v^T (\alpha \matr M + \beta \matr K(\xprev))\,\vec v.
\end{equation}
Because $R^{\text{lag}}$ depends only on velocities, and using the Backward Euler substitution $\vec v = (\vec x - \xprev) / \Delta t$, the corresponding objective term for the incremental potential becomes
\begin{equation} \label{eq:rayleigh_potential}
    \phi_{\text{damp}}(\vec x) = \frac{1}{2\,\Delta t}\, (\vec x - \xprev)^T (\alpha \matr M + \beta \matr K(\xprev))\,(\vec x - \xprev).
\end{equation}
For energy-based simulation of crowds, Karamouzas et al.~\cite{KSNG17} instead argue that a fully implicit formulation of Rayleigh damping is desirable and show that omitting the lagging still results in forces that are first-order accurate w.r.t.\ the time step size.
Brown et al.~\cite{BOFN18} extend this to Coulomb friction, strain rate damping and other models but demonstrate that this first-order approximation results in forces that are not invariant under rigid transformations.
They introduce a correction term to remedy the resulting angular momentum loss.
However, they apply a local-global optimization approach using ADMM which solves individual potentials locally (see Projective Dynamics in Section~\ref{sec:projective-dynamics}), and a fully implicit formulation for an unconstrained global optimization problem was not considered.
Alternatively, for strain rate-based damping, an energy can be formulated by directly discretizing the strain rate in time which was used to simulate viscous threads~\cite{viscous_threads} and sheets~\cite{viscous_sheets}.

\paragraph*{Coulomb Friction}
A major contribution of the IPC framework~\cite{LFS+20} is a variational treatment of Coulomb friction within unconstrained optimization.
The physical model is based on the Maximal Dissipation Principle (MDP), which states that frictional forces maximize the rate of dissipation in relative tangential motion subject to the Coulomb constraint.
For each active contact pair $k \in \hat{\mathcal{C}}$, a tangential sliding basis $\matr T_k(\vec x) \in \Real^{3n \times 2}$ is constructed in the plane orthogonal to the contact normal.
The local relative sliding displacement is then $\vec u_k = \matr T_k(\vec x)^T (\vec x - \xprev) \in \Real^2$, and the sliding velocity is $\vec v_k = \vec u_k / \Delta t$.
Maximizing the dissipation rate subject to the Coulomb constraint yields the friction force
\begin{equation} \label{eq:ipc_friction_force}
    \vec F_k(\vec x, \lambda) = \matr T_k(\vec x) \operatorname*{arg\,min}_{\boldsymbol\beta \in \Real^2} \boldsymbol\beta^T \vec v_k \quad \text{s.t.} \quad \|\boldsymbol\beta\| \leq \mu \lambda_k,
\end{equation}
where $\lambda_k = -\kappa\,\partial b / \partial d_k$ is the (non-negative) contact force magnitude from the barrier potential and $\mu$ is the friction coefficient.
This friction force is bimodal: when $\|\vec v_k\| > 0$ (sliding), we have $\vec F_k = -\mu\lambda_k \matr T_k \vec u_k / \|\vec u_k\|$; when $\|\vec v_k\| = 0$ (sticking), the friction direction is not uniquely defined.

Two challenges prevent direct inclusion of these friction forces in unconstrained optimization: the non-smoothness at the transition between sticking and sliding, and the absence of a well-defined dissipation potential whose spatial gradient generates friction forces.
To address the first issue, IPC introduces a $C^1$-smooth approximation of the friction magnitude function using a velocity threshold $\epsilon_v > 0$ below which contacts are treated as static,
\begin{equation} \label{eq:ipc_friction_smooth}
    f_1(y) =
    \begin{cases}
        -\dfrac{y^2}{\epsilon_v^2 \Delta t^2} + \dfrac{2y}{\epsilon_v \Delta t}, & y \in (0,\, \epsilon_v \Delta t),\\[6pt]
        1, & y \geq \epsilon_v \Delta t.
    \end{cases}
\end{equation}
This yields smooth friction forces $\vec F_k = -\mu\lambda_k \matr T_k\, f_1(\|\vec u_k\|)\, \vec u_k / \|\vec u_k\|$ that transition continuously from zero (sticking) to the full Coulomb magnitude (sliding).

The second issue is more fundamental: even with this smoothing, the friction force is not integrable.
Because the sliding basis $\matr T_k(\vec x)$ and the normal force magnitude $\lambda_k(\vec x)$ themselves depend on the positions, the Jacobian of the friction force with respect to $\vec x$ is non-symmetric, so no scalar potential can generate it directly and further adaptation is required.
For this purpose, IPC lags the tangential sliding basis $\matr T_k^n$ and contact force magnitude $\lambda_k^n$ at their values from the previous nonlinear solve (or previous time step).
With these lagged quantities held fixed, the friction contribution to the incremental potential can be written as a sum over all active contact pairs,
\begin{equation} \label{eq:ipc_friction_potential}
    \phi_{\text{friction}}(\vec x) = \sum_{k \in \hat{\mathcal{C}}} \mu \lambda_k^n\, D(\vec x, \matr T_k^n),
\end{equation}
where $D$ is a per-contact dissipation potential that integrates the smoothed friction magnitude and whose negative gradient with respect to $\vec x$ recovers the smoothed friction force with the lagged quantities.
This lagged potential can be added directly to the incremental potential from Eq.~\eqref{eq:be_potential} and resolved within the same Newton solve.
To improve the accuracy of the friction direction and magnitude, the lagged quantities $\matr T_k^n$ and $\lambda_k^n$ can be updated by solving successive minimizations within each time step.

Similar approximations were used in the penalty formulation by Macklin et al.~\cite{macklin_primaldual_2020}.
Outside of simple test cases, however, it is not recommended to update the lagged quantities during optimization as there are no convergence guarantees in general.
Larionov et al.~\cite{LLA+24} claim that this lagging can lead to inaccurate and time step dependent behavior, especially close to the slip threshold and demonstrate that force-based simulations with a fully implicit friction formulation can be beneficial for simulations where accurate frictional behavior is important.

\paragraph*{Plasticity}
For plasticity, Li et al.~\cite{LLJ22} present implicit augmented energy densities of finite strain von Mises and Drucker-Prager plasticity models as well as hardening and viscoelasticity models and evaluate them in MPM and FEM frameworks.
However, their approach is currently limited to the St.~Venant-Kirchhoff elastic material model, while the Drucker-Prager model and hardening model require additional fixed-point iterations around the optimization problem as their yield strain is fully implicit.

\subsubsection{Shells and Rods} %
In addition to volumetric deformables, codimensional deformables including cloth, shells, rods and hair are an important aspect of multimaterial simulations.
They can also often be described by energies that can be added as contributions to the incremental potential formulation.
The IPC method was also extended to provide strongly coupled frictional contact between codimensional and volumetric deformables~\cite{LKJ21}.

For cloth or shells, an established approach in computer graphics is to combine a membrane deformation model for in-plane forces with a bending energy for out-of-plane deformations.
The total shell energy then takes the form
\begin{equation} \label{eq:shell_total}
    \phi_{\text{shell}}(\vec x) = \phi_{\text{membrane}}(\vec x) + \phi_{\text{bend}}(\vec x),
\end{equation}
where $\phi_{\text{membrane}}$ is obtained by integrating a membrane energy density over the surface and $\phi_{\text{bend}}$ is a discrete bending energy.
This total shell energy enters directly as one of the potential terms $\phi_i$ in the incremental potential of Equation~\eqref{eq:be_potential}.
Common choices for the membrane part include the Baraff-Witkin model~\cite{Kim20,baraff_large_1998} or Koiter's model~\cite{CSvRV18}, while established bending energies include the Discrete Shells model~\cite{GHDS03,BMF03,TG13} and the quadratic and cubic shell variants for inextensible surfaces~\cite{BWH+06,GGWZ07}.
In the following, we briefly introduce the Baraff-Witkin and Discrete Shells models as representative examples.
Analytic eigenanalyses of such membrane and bending energies, which are required for Hessian projection to positive definiteness, were presented by Kim~\cite{Kim20}, Haomiao and Kim~\cite{WK23} and Wang et al.~\cite{WYW23}.

\paragraph*{Baraff-Witkin Model}
The Baraff-Witkin model~\cite{baraff_large_1998} is a widely used membrane energy for cloth that measures stretching and shearing separately along the warp and weft thread directions of the fabric.
As shown by Kim~\cite{Kim20}, it can be expressed as an anisotropic strain energy density using the invariants
\begin{equation}
    I_5(\vec a) = \vec a^T \matr F^T \matr F \, \vec a, \qquad I_6(\vec a, \vec b) = \vec a^T \matr F^T \matr F \, \vec b,
\end{equation}
where $\matr F$ is the deformation gradient and $\vec a$, $\vec b$ are directions in the 2D material space.
The stretching energy density is given by
\begin{equation} \label{eq:bw_stretch}
    \Psi_{\text{stretch}} = \frac{k_s}{2}\Bigl[\bigl(\sqrt{I_5(\vec a_u)} - b_u\bigr)^2 + \bigl(\sqrt{I_5(\vec a_v)} - b_v\bigr)^2\Bigr],
\end{equation}
where $\vec a_u$, $\vec a_v$ are orthogonal unit vectors along the warp and weft directions (the standard basis in material space), $k_s$ is the stretching stiffness, and $b_u$, $b_v$ are rest-state stretch parameters (typically~1).
Using $I_5$ with these specific directions measures stretch independently along each thread direction rather than in the principal strain directions, which is physically consistent with woven fabrics.
The shearing energy density penalizes changes of the angle between the two thread directions,
\begin{equation} \label{eq:bw_shear}
    \Psi_{\text{shear}} = \frac{k_{\text{shear}}}{2}\, I_6(\vec a_u, \vec a_v)^2,
\end{equation}
and plays a critical role in producing characteristic wrinkling behavior under extension.

\paragraph*{Discrete Shells Bending Model}
While membrane energies resist in-plane deformations, a separate bending energy is needed to penalize out-of-plane deformations (i.e.\ curvature changes) of the surface.
The Discrete Shells model~\cite{GHDS03,BMF03} defines a bending energy by measuring the change in dihedral angles along mesh edges.
The discrete flexural energy is defined as a sum over all interior edges of the triangle mesh,
\begin{equation} \label{eq:discrete_shells}
    \phi_{\text{bend}}(\vec x) = k_B \sum_{e \in \mathcal{E}} \frac{\lVert \bar{\vec e} \rVert}{\bar{h}_e} \bigl(\theta_e(\vec x) - \bar{\theta}_e\bigr)^2,
\end{equation}
where $k_B$ is the bending stiffness, $\theta_e$ is the dihedral angle at edge $e$ in the deformed configuration, $\bar{\theta}_e$ is its counterpart in the undeformed (rest) configuration, $\lVert \bar{\vec e} \rVert$ is the edge length in the rest configuration, and $\bar{h}_e$ is related to the heights of the two triangles sharing the edge.
By storing the rest angles $\bar{\theta}_e$, this model naturally supports curved rest configurations such as hats, bent paper, or pre-shaped garments.

\paragraph*{Strain Limiting}
In physical cloth, large in-plane deformations are strongly resisted, and many fabrics are nearly inextensible.
Achieving this behavior in simulation by simply using a very stiff membrane model leads to numerical difficulties: on practical mesh resolutions, such stiff membranes can introduce \emph{membrane locking}, a numerical artifact where the material appears artificially resistant to bending~\cite{LKJ21}.
A common workaround is to use a relatively soft membrane energy and impose separate \emph{strain limits} that prevent excessive stretching.

Strain limiting can be formulated as an energy potential within the optimization-based framework.
In the C-IPC approach~\cite{LKJ21}, the singular values $\sigma_i$ of the deformation gradient $\matr F$ of each element are constrained to lie below an upper bound $s$ (the strain limit), and this constraint is enforced through a $C^2$-continuous barrier, analogous in spirit to the contact barrier from Eq.~\eqref{eq:ipc_barrier_potential}.
A second parameter $\hat{s} < s$ defines an activation threshold below which no forces are generated.
To make the barrier independent of the specific choice of $s$ and $\hat{s}$, the strain is mapped to the normalized variable
\begin{equation}
    y = \frac{\hat{s} - \sigma_i^t}{s - \hat{s}},
\end{equation}
which is positive when the strain is below the threshold and approaches $-1$ as $\sigma_i^t \to s$.
The smoothly clamped barrier is then defined as
\begin{equation} \label{eq:strain_barrier}
    b(y) = \begin{cases} -y^2 \ln(1+y), & y < 0, \\ 0, & y \geq 0. \end{cases}
\end{equation}
This function is exactly zero when the strain remains below the activation threshold ($y \geq 0$) and diverges as $y \to -1$ (i.e.\ $\sigma_i^t \to s$), guaranteeing that the strain limit is never exceeded.
The resulting strain-limiting potential is integrated over the cloth volume to ensure mesh-resolution independent behavior,
\begin{equation} \label{eq:strain_limiting}
    \phi_{\text{SL}}(\vec x) = \kappa_s \sum_{t \in \mathcal{T}} \sum_{i} V^t \, b(\sigma_i^t),
\end{equation}
where $\kappa_s$ is the barrier stiffness and $V^t$ is the reference volume of element $t$.
Because the barriers are unilateral---they only activate when the strain threshold is exceeded---they do not over-constrain the system and thus avoid triggering locking.
Furthermore, since the strain-limiting potential is resolved simultaneously with all other forces in the optimization, it avoids the compression, jittering, and numerical softening artifacts that can arise from time-step splitting approaches where strain limits are enforced as a separate projection step.

\paragraph*{Continuum-based Shell Models}
Instead of manually combining stretch and bending energies, continuum-based shell models, such as Kirchhoff-Love shells, inherit their bending stiffness from an underlying volumetric material model~\cite{WB23}.
Although Kirchhoff-Love shells typically require either sophisticated $\mathcal{C}^1$ continuous~\cite{TWS06,WHP11} or non-conforming~\cite{KGBG09} discretizations, it is possible to apply results from discrete differential geometry~\cite{Wei12} to construct an efficient piecewise linear discretization~\cite{WB23}.
In contrast to Kirchhoff-Love shells which are rigid to transverse shearing, Cosserat shells are shear deformable and introduce additional degrees of freedom (directors) which can also be incorporated in an energy-based formulation~\cite{LFJB24}.
As an alternative to codimensional approaches, thick shells can also be simulated with volumetric prism elements and reduced integration to avoid locking~\cite{CXY+23}.

\paragraph*{Hair and Rods}
The main challenge of modeling hair and rods is to capture the bending and torsion of their one-dimensional geometry in three-dimensional space.
Common approaches include the shear-deformable Cosserat rods~\cite{ST07,kugelstadt_position_2016} as well as shear-rigid Kirchhoff rods~\cite{HB23} as used in the Discrete Elastic Rods model~\cite{BWR+08}.
These models are based on an orthonormal frame attached to the centerline of the geometry to measure bending and torsion.
Depending on the model, the orthonormal frame might be explicitly represented by additional degrees of freedom or is only implicitly used during derivation.
Shi et al.~\cite{SWP+23} recognized that established rod models are unsuitable for simulation of tightly coiled hair and proposed an energy-based model defined directly in terms of the vertex positions of the hair.

\subsubsection{Rigid Bodies and Multibody Systems} %
Rigid bodies and articulated multibody systems are ubiquitous in multiphysics simulations, appearing as tools, mechanisms, characters and environmental objects that interact with deformable materials through frictional contact.
Unlike deformable bodies, rigid bodies do not require a volumetric discretization, but their rotational degrees of freedom introduce challenges for optimization-based formulations that we discuss in the following.

\paragraph*{Rigid Body Dynamics}
To simulate rigid bodies in the context of optimization-based integration, Macklin et al.~\cite{macklin_primaldual_2020} present a formulation based on generalized positions (translation and orientation of the bodies) and a kinematic map~\cite{BET14} which maps linear and angular velocities to time derivatives of the generalized positions.
This enables rigid bodies and deformables to be solved in a unified way.
To define a corresponding potential, the kinematic map is approximated to be constant during a time step and gyroscopic forces are integrated explicitly.
Contacts are resolved using penalty energies.

Ferguson et al.~\cite{FLS+21} propose a formulation that avoids these approximations by parametrizing each rigid body $i$ with a translation vector $\vec q_i \in \Real^3$ and a rotation vector $\vec \theta_i \in \Real^3$, where the rotation matrix is recovered via Rodrigues' formula $\mathcal{R}(\vec \theta_i)$.
Because Rodrigues' formula maps arbitrary vectors to valid rotation matrices, orthogonality does not need to be imposed as a constraint, yielding a fully unconstrained optimization problem.
The incremental potential splits into a translational and a rotational part.
The translational potential is the standard inertial term,
\begin{equation} \label{eq:rb_translational}
    E_{\vec q}(\vec q) = \sum_{i} \frac{m_i}{2 \Delta t^2} \lVert \vec q_i - \tilde{\vec q}_i \rVert^2,
\end{equation}
where $\tilde{\vec q}_i = \vec q_i^{\,t} + \Delta t\, \dot{\vec q}_i^{\,t} + \Delta t^2 m_i^{-1} \vec f_i$ is the inertial position.
For the rotational part, following the simplified form of Romero et al.~\cite{RCO25}, the rotational inertial potential is
\begin{equation} \label{eq:rb_rotational}
    E_{\matr R}(\matr R) = -\frac{1}{\Delta t^2} \sum_{i} \operatorname{tr}\!\bigl(\matr R_i\, \matr J_i^l\, \tilde{\matr R}_i^T\bigr),
\end{equation}
where $\matr J_i^l$ is the local inertia tensor of body $i$ and $\tilde{\matr R}_i$ is the explicit rotation prediction.
Instead of using angular velocities, the prediction is expressed in terms of incremental rotation steps $\Delta \matr R_i^{t} = \matr R_i^{t}\, (\matr R_i^{t-1})^T$ as $\tilde{\matr R}_i = (2\identity - (\Delta \matr R_i^{t})^T)\, \matr R_i^{t}$, which avoids the need for explicit velocity variables.
The full time step is then obtained by minimizing the combined incremental potential
\begin{align} \label{eq:rb_full_potential}
    \operatorname*{arg\,min}_{\vec \theta, \vec q} \; &E_{\vec q}(\vec q) + E_{\matr R}(\mathcal{R}(\vec \theta)) \nonumber \\ 
    &+ \phi_{\text{contact}}(\vec x(\mathcal{R}(\vec \theta), \vec q)) + \phi_{\text{friction}}(\vec x(\mathcal{R}(\vec \theta), \vec q)),
\end{align}
where the contact and friction potentials from Eqs.~\eqref{eq:ipc_barrier_potential} and~\eqref{eq:ipc_friction_potential} are evaluated on the world-space surface positions $\vec x$ obtained from the rigid transformations.
This formulation generalizes the IPC method to guarantee intersection-free contact with strong coupling between rigid bodies and deformables.
Because rigid bodies undergo curved trajectories during a time step, CCD queries must account for these nonlinear paths, which is more involved than the linear CCD used for deformables~\cite{FLS+21}.
Romero et al.~\cite{RCO25} instead keep the rotation matrices as state variables and, following a retraction, let each Newton step optimize an incremental rotation vector in the tangent space at the identity, applied to the state through the full exponential map.
Because this increment is re-centered at zero in every iteration, the exponential map and its derivatives are only ever evaluated at the identity, where they reduce to exact linear and quadratic terms.
This avoids the complex and poorly conditioned derivatives of the Rodrigues formula required by Ferguson et al.~\cite{FLS+21}, yielding simpler, well-conditioned expressions and significantly improved convergence of the Newton solver.

\paragraph*{Affine Body Dynamics}
Reasoning that the idealized assumption of fully rigid bodies is not always necessary, Lan et al.~\cite{LKL+22} propose Affine Body Dynamics (ABD) which relaxes the rigid body constraint to an affine transformation penalized by a very stiff orthogonality potential.
Intuitively, an affine body can be understood as embedding the entire object inside a single linear finite element~\cite{LLK+20}: just as a linear tetrahedron maps its interior through an affine function with a spatially constant deformation gradient $\matr F$, ABD represents the whole body by one global affine map, so the deformation gradient is uniform over the body.
Each affine body $b$ is accordingly described by 12 degrees of freedom: a translation vector $\vec p_b \in \Real^3$ and a linear transformation matrix $\matr A_b \in \Real^{3 \times 3}$, stored as a generalized coordinate vector $\vec q_b = (\vec p_b^T, \vec a_1^T, \vec a_2^T, \vec a_3^T)^T \in \Real^{12}$, where $\vec a_i$ are the rows of $\matr A_b$.
A material point with rest position $\bar{\vec x}_k$ in the body frame is mapped to world coordinates via
\begin{equation} \label{eq:abd_kinematics}
    \vec x_k = \matr A_b \bar{\vec x}_k + \vec p_b = \matr J(\bar{\vec x}_k)\, \vec q_b, \qquad
    \matr J(\bar{\vec x}) = \begin{pmatrix} \identity & \identity \otimes \bar{\vec x}^T \end{pmatrix},
\end{equation}
where $\matr J$ is a constant Jacobian matrix.
This is directly analogous to the FEM shape function derivatives from Eq.~\eqref{eq:deformation_gradient}: the deformation gradient of the affine body is simply $\matr F = \matr A_b$, and because this single-element view has constant $\matr F$, all quantities that would normally be integrated over many elements reduce to a single evaluation per body.
Because the coordinates are affine (flat), velocities are linear in $\dot{\vec q}_b$ and the generalized mass matrix $\matr M_b = \int_{\Omega_b} \rho\, \matr J(\bar{\vec x})^T \matr J(\bar{\vec x})\, \mathrm{d}\Omega$ is constant, avoiding the nonlinear Coriolis terms that arise in standard rigid body dynamics.

To keep the body nearly rigid, a stiff orthogonality potential penalizes deviations of $\matr A_b$ from a rotation matrix,
\begin{equation} \label{eq:abd_orthogonality}
    \phi_{\perp}(\vec q_b) = \kappa\, V_b \lVert \matr A_b \matr A_b^T - \identity \rVert_F^2,
\end{equation}
where $\kappa$ is a large penalty stiffness and $V_b$ is the volume of the body.
This potential can be evaluated efficiently as a polynomial in the rows of $\matr A_b$ without requiring matrix decompositions.
The per-body incremental potential then takes the form
\begin{equation} \label{eq:abd_body_potential}
    E_b(\vec q_b) = \frac{1}{2\Delta t^2}(\vec q_b - \tilde{\vec q}_b)^T \matr M_b\, (\vec q_b - \tilde{\vec q}_b) + \phi_{\perp}(\vec q_b),
\end{equation}
where $\tilde{\vec q}_b = \vec q_b^{\,t} + \Delta t\, \dot{\vec q}_b^{\,t} + \Delta t^2 \matr M_b^{-1} \vec f_b$ is the inertial state, analogous to Eq.~\eqref{eq:be_potential}.
A key advantage of ABD is that the affine map produces linear point trajectories within each time step, which greatly simplifies continuous collision detection compared to the curved trajectories of true rigid bodies, significantly reducing its cost while still ensuring intersection-free simulations.

Recently, Huang et al.~\cite{HLL25} demonstrate how to simulate affine bodies and deformable bodies in a unified system and propose a hash-based two-level reduction strategy to improve performance of contact Hessian assembly for GPU implementations.
Previously, this assembly process could present a problematic bottleneck as the degrees of freedom of a single affine body might be involved in thousands of collisions.

\paragraph*{Multibody Systems}
Finally, rigid-, affine- or deformable bodies are often coupled with joints and motors to form multibody systems.
Chen et al.~\cite{CLL+22} propose a unified approach for modeling articulation joints within the framework of unconstrained optimization.
Their key insight is to categorize joint constraints into three types---linear equality, nonlinear equality and inequality constraints---each handled by a different mechanism that preserves the unconstrained optimization structure.

Joint constraints are formulated in terms of \emph{attachment points} defined via barycentric coordinates on the bodies' degrees of freedom, $\vec p^{\,i} = \sum_{j=1}^{N^i} \alpha_j \vec x_j^{\,i}$, which makes the approach applicable to rigid, affine and deformable bodies alike.
\emph{Linear equality constraints} of the form $\vec c_i^T \vec x = s_i$ are enforced exactly through a change of variables $\vec x = \matr U \vec z$ that reduces the system to a lower-dimensional space where the constraints are satisfied by construction.
Joints formulated in this way include point connections ($\vec p^{\,1} = \vec p^{\,2}$), hinges (two shared points defining a rotation axis), and sliding constraints (restricting displacement to be orthogonal to a plane normal).
\emph{Nonlinear equality constraints}, such as a distance constraint maintaining a fixed length $l$ between two points,
\begin{equation}
    c(\vec x) = \lVert \vec p_1 - \vec p_2 \rVert - l = 0,
\end{equation}
cannot be eliminated by a linear coordinate transformation and are instead enforced through a stiff quadratic penalty potential,
\begin{equation}
    \phi_{\text{eq}}(\vec x) = \frac{k}{2}\, c(\vec x)^2.
\end{equation}
Finally, \emph{inequality constraints} such as joint limits are formulated as $c(\vec x) \geq 0$ and enforced using barrier potentials,
\begin{equation} \label{eq:joint_barrier}
    \phi_{\text{ineq}}(\vec x) = \kappa\, b(c(\vec x)),
\end{equation}
where $b$ is a smoothly clamped log-barrier analogous to the contact barrier from Eq.~\eqref{eq:ipc_barrier_potential}.
This is a natural fit: just as the contact barrier prevents intersection by generating arbitrarily large repulsive forces as the distance approaches zero, the joint barrier prevents the constraint from being violated.
All inequality joints are formulated purely in terms of bounded distances.
For example, upper and lower distance bounds are expressed as $c(\vec x) = d_u - \lVert \vec p_1 - \vec p_2 \rVert \geq 0$ and $c(\vec x) = \lVert \vec p_1 - \vec p_2 \rVert - d_l \geq 0$, respectively, and rotation limits for hinges are converted from dihedral angle bounds to equivalent arc-distance constraints, avoiding trigonometric functions.
All of these joint potentials are added as additional terms to the global incremental potential from Eq.~\eqref{eq:be_potential} and are resolved simultaneously with elasticity, contact and friction.

Beyond traditional joint types, Chen et al.~\cite{CDGB19} introduce energy-based point, curve and surface connections to couple different codimensional and volumetric models which also supports non-manifold deformable bodies.

\subsubsection{Multiphysics Phenomena} %
Besides deformables and multibody systems, other physical effects can also be incorporated into optimization-based formulations.
For thin shells discretized with FEM, Chen et al.~\cite{CSvRV18} simulate the effects of heating and wetting, like curling paper or shape-changing pasta.

Another aspect that can be considered in multiphysics systems are forces due to magnetic effects.
For inverse design of magnetically actuated thin shells, Chen et al.~\cite{CNZ+22} couple thin shell deformations with a linear, ideal hard-magnetic potential.
Westhofen et al.~\cite{WFJB24} propose an incremental potential formulation that supports strong coupling of linear magnetic rigid bodies with frictional contact considering para- and diamagnetic effects.
They demonstrate much more robust simulations in comparison to previous weakly coupled approaches.

\subsubsection{Fluids and coupling}
While energy-based formulations could facilitate coupling between deformables and fluids, full strong coupling is hard to achieve.
To our knowledge, an energy-based, fully implicit discretization of fluid dynamics with e.g. SPH does not exist.
Instead, an operator splitting approach is a viable alternative, such as the approach proposed by Xie et al.~\cite{xieContactProxySplitting2023} for the coupling of SPH fluids and FEM deformables.
For fluids, they use a semi-implicit approach to formulate pressure and viscosity as conservative forces and to obtain corresponding potentials.
For the coupling, they introduce a contact proxy of the deformables into the SPH solver as well as a fluid contact proxy into the incremental potential of the subsequent solid and frictional contact solve.

Takahashi and Batty~\cite{TB22} proposed a monolithic solver for strongly coupled fluids, rigid bodies and deformable elastic solids with frictional contact based on constrained optimization.
Fluids are discretized with APIC~\cite{jiang_affine_2015} (see Section~\ref{sec:MPM}) using a variational formulation of viscosity~\cite{BB08,LBB17} reformulated as a compliant constraint in combination with a hard constraint for incompressibility.
Deformables are discretized with FEM and formulated using an incremental potential as shown in Equation~\eqref{eq:be_potential}.
Fluid-solid coupling is achieved through a cut-cell approach with volume fractions while contacts between solids and rigid bodies are formulated as hard constraints.
All these contributions are combined into a single constrained optimization problem with primal variables for fluid velocity, elastic displacement and rigid body velocities and dual variables or Lagrange multipliers in terms of viscous stress, pressure and contact force.
The system is solved using a custom Newton-type solver leveraging the structure of the problem which was improved from previous, less general monolithic solvers~\cite{TB20,TB21}.

A different approach for the implicit simulation of fluid dynamics are Lagrangian mesh-based discretizations using FEM.
Misztal et al.~\cite{MEB+14} used this approach to simulate immiscible multi-phase flow and perform an explicit advection step followed by an implicit solve for all other forces such as inertia, viscosity and surface tension.
For the discretization they also discretize ambient space around the fluid.
A similar formulation was used by Clausen et al.~\cite{CWSO13} but they also handle advection within the implicit solve.
They also support coupling with elastic and plastic materials through explicit collision forces and they only mesh the materials themselves, without ambient space.
This Lagrangian mesh-based approach, however, requires involved remeshing operations for merging, splitting and refinement.
Also Lagrangian in nature but working on simplicial complexes, i.e.~collections of points, lines, triangles and tetrahedra, Zhu et al.~\cite{zhuCodimensionalSurfaceTension2014} proposed a codimensional approach for the simulation of surface tension.
They extended this approach to simulate non-Newtonian fluids \cite{zhuCodimensionalNonNewtonianFluids2015}.

Coupling of MPM and FEM as introduced by Li et al.~\cite{li_dynamic_2024} is described in Section~\ref{sec:MPM}.

\subsection{Vertex Block Descent}
\label{sec:VBD}

Newton's method applied to an incremental potential demonstrates great stability, even for large time steps.
However, it is a relatively costly approach.
In computer graphics, there has been particular interest to parallelize solver algorithms and better leverage GPU hardware for interactive or learning applications.
Recently, Chen et al.~\cite{chen_vertex_2024} introduced \emph{Vertex Block Descent} (VBD) as a block coordinate descent algorithm~\cite{Wri15} for minimizing the variational energy. 
Their approach solves a per vertex optimization problem.
Compared to the global variational formulation of Backward Euler from Equation~\eqref{eq:be_potential}, VBD introduces local variational energies 
\begin{equation} \label{eq:vbd}
    E_i(\vec x) = \frac{m_i}{2 \Delta t^2} || \vec x_i - \tilde{\vec x}_i ||^2 + \sum_{j \in \mathcal E_i} \phi_j(\vec x) \,,
\end{equation}
where $\mathcal{E}_i$ is the set of all elements (e.g., FEM elements or constraints) that act on vertex $i$. Minimizing this local energy with Newton's method leads to a series of linear problems with $3 \times 3$ matrices that can be solved analytically.
The local solves are guaranteed to reduce the global variational energy - however, the method does not necessarily converge to the same solution as Newton's method.
They demonstrate impressive simulations of elastic deformables coupled with rigid bodies, frictional contact, damping and simple constraints.

Giles et al.~\cite{GDY25} addressed VBD's inability to enforce explicit hard constraints and its poor convergence when handling high stiffness ratios. By integrating an augmented Lagrangian formulation that progressively increases stiffness, their Augmented-VBD (AVBD) method enables the stable and efficient simulation of complex frictional contacts, articulated rigid bodies, and stiff systems interacting with soft bodies.
For contact resolution with VBD Chen et al.~\cite{CHL+25} introduced \textit{Offset Geometric Contact} (OGC), a penetration-free contact model that offsets surface faces along their normal directions to ensure orthogonal contact forces and eliminate stretching artifacts. By pairing OGC's per-vertex displacement bounds with the naturally small local steps of the VBD solver, their method avoids expensive continuous collision detection to deliver highly parallelized, real-time simulations.

The local view and vertex-centric reformulation in VBD gives significant opportunities for parallelization (e.g. on GPUs~\cite{warp2022}), and the authors report faster and more accurate results in comparison to XPBD~\cite{macklin_xpbd_2016}.
Concurrently, Chen et al.~\cite{chen_position-based_2024} presented a Position-Based Nonlinear Gauss-Seidel method which follows a similar idea of vertex-based Gauss-Seidel iterations and can be applied to quasistatic and dynamic simulations.
However, they focus mostly on isotropic hyperelastic materials.

\subsection{Projective Dynamics}
\label{sec:projective-dynamics}
A different optimization-based framework for the interactive simulation of various materials and phenomena is \emph{Projective Dynamics} (PD) as introduced by Bouaziz et al.~\cite{bouaziz_projective_2014}.
It follows the alternating local/global optimization scheme proposed for static geometry problems by Bouaziz et al.~\cite{shape_up} and for Hookean mass-spring systems by Liu et al.~\cite{liu_fast_2013} and generalizes it to a wider range of geometrically motivated constraints for dynamic simulations.
Originally, PD supported the simulation of elastic solids, shells and rods with contact and was later extended and generalized to support physical material models, rigid bodies, fluids, penetration-free frictional contact and more.
The main advantage of the method in comparison to a full Newton method is that updating a Hessian matrix involving the constraint's second derivatives is not needed and early iterations can be more efficient in reducing the objective.
This is achieved by restricting the general optimization-based framework described before to quadratic potentials that pull the degrees of freedom towards target positions obtained through non-linear local constraint projections.

\subsubsection{Foundations of Projective Dynamics}
Projective Dynamics is based on the Backward Euler incremental potential introduced in Equation~\eqref{eq:be_potential}, but uses constraint functions $\vec C_i$ to define potentials for individual physical effects.
To facilitate a local/global split of the solver, PD introduces auxiliary variables $\vec p_i$ that can be interpreted as target configurations satisfying each constraint such that $\vec C_i(\vec p_i) = 0$.
The global objective that has to be minimized in PD is then given by
\begin{equation} \label{eq:pd_global_energy}
    E_{\mathrm{PD}}(\vec x, \vec p) = \frac{1}{2 \Delta t^2} (\vec x - \tilde{\vec x})^T \matr M (\vec x - \tilde{\vec x}) + \sum_i d_i(\vec x, \vec p_i) \,,
\end{equation}
where $d_i$ are constraint specific functions measuring the distance between the current configuration $\vec x$ and the target configuration on the constraint manifold $\vec p_i$.
In particular, PD restricts this to quadratic functions of the form
\begin{equation} \label{eq:pd_constraint}
    d_i(\vec x, \vec p_i) = \frac{w_i}{2} || \matr A_i \matr S_i \vec x - \matr B_i \vec p_i||^2 \,,
\end{equation}
where $w_i$ is a nonnegative weight, $\matr A_i$ and $\matr B_i$ are constant matrices specific to the respective constraint type and $\matr S_i$ is a selection matrix mapping the global to the constraint specific degrees of freedom.
This ensures that the global potential in Equation~\eqref{eq:pd_global_energy} remains quadratic for fixed $\vec p_i$.

\paragraph*{Local/Global Solver}
In each iteration of the solver algorithm, PD first determines the target positions $\vec p_i$ for every constraint by performing a local \emph{constraint projection} for a fixed current configuration $\vec x$
\begin{equation} \label{eq:pd_projection}
    \arg\min_{\vec p_i} \frac{w_i}{2} || \matr A_i \matr S_i \vec x - \matr B_i \vec p_i||^2 \,, \quad \mathrm{s.t.}\, \vec C_i(\vec p_i) = 0 \,,
\end{equation}
which corresponds to finding the projection of $\vec x$ to the nearest point on the constraint manifold of $\vec C_i$ with respect to distance $d_i$ (e.g. finding the closest zero strain configuration of a deformed triangle).
These local projections capture the nonlinearity of the constraint forces and closed-form solutions exist for many constraint types.
The individual projections are independent and can be performed in parallel.
This is followed by a global solve for updated positions $\vec x$ using the fixed, previously obtained target configuration $\vec p$
\begin{equation}
    \min_{\vec x} E_{\mathrm{PD}}(\vec x, \vec p) \,,
\end{equation}
which can be done in a single linear solve as $E_{\mathrm{PD}}$ is quadratic in $\vec x$.
This global solve pulls the configuration of the system towards the target positions.
The linear solve can be accelerated by pre-factorizing the constant system matrix in case of fixed constraint sets.
To handle topology changes for contact, cutting or fluids efficient refactorization schemes~\cite{LLKC21,wang_optimized_2021} or iterative solvers~\cite{wang_chebyshev_2015,weiler_projective_2016,lan_penetration-free_2022} were proposed.

A broad range of geometrically motivated constraints fit in this framework, including deformation (Hookean or ARAP models), bending, joints and contact~\cite{bouaziz_projective_2014}.
However, many non-linear constitutive laws, such as the Neohookean model, as well as hard constraints, cannot be directly considered in this way.
This was addressed later in generalizations of PD to a Quasi-Newton~\cite{LBK17} or ADMM~\cite{OBLN17} method.
Although methods for simulating fluids or granular and plastic media based on PD were presented~\cite{weiler_projective_2016,he_projective_2018}, a coupled multiphysics simulation together with, e.g., deformables and frictional contact within the PD framework was not thoroughly explored, yet.

\paragraph*{Relation to (Extended) Position Based Dynamics}
Projective Dynamics and Extended Position Based Dynamics (XPBD) (see Section~\ref{sec:PBD}) both directly build on constraint functions and are targeted at interactive applications.
However, a fundamental difference is that PD is a \emph{primal} method solving for the physical system's degrees of freedom directly (i.e. positions or velocities), whereas XPBD is a \emph{dual} method that enforces constraints by first solving for Lagrange multipliers.
A practical consequence is that PD can easily deal with high mass ratios and may struggle with high stiffness ratios while the converse applies to XPBD~\cite{macklin_primaldual_2020}.
Similar observations can be made for the original PBD algorithm as it can be interpreted as a restriction of XPBD to constraints with infinite stiffness.

The additional global solve in PD ensures that a global compromise between all constraints and the implicit inertia potential is found, even considering incompatible constraints (corresponding to a least squares fit~\cite{shape_up}).
For PBD and XPBD, the solution of problems with incompatible constraints depends on the ordering of the local constraint projections.
In general, the original XPBD algorithm also does not converge to a state of force balance, mainly because a residual term is discarded in its derivation from the full equations of motion (see Section~\ref{sec:PBD}).
This was recently addressed by Chen et al.~\cite{chen_primal_2024}.

XPBD has the advantage that it can naturally handle hard constraints and also supports more general constraint functions whereas PD always requires a sensible split into a quadratic distance measure and a constraint manifold.
As noted in the following paragraph, this was also addressed in the literature but requires significant deviations from the original concept of PD~\cite{LBK17,OBLN17}.

\paragraph*{Generalizations and Enhancements}

Applying the Chebyshev Semi-Iterative Method was proposed by Wang~\cite{wang_chebyshev_2015} to improve the efficiency of the global solve in Projective Dynamics.
In combination with a Jacobi solver, this results in a highly efficient GPU implementation of PD which also works well with dynamically changing constraint connectivity due to the iterative methods.
However, Fratarcangeli et al.~\cite{FTP16} claim that the Chebyshev approach can introduce instabilities and instead propose a parallel Gauss-Seidel approach for PD and PBD.

Liu et al.~\cite{LBK17} show that the original PD algorithm can be interpreted as a Quasi-Newton method.
In this context they generalize PD to more hypereleastic materials which also necessitates the introduction of a line search to the method.
They improve convergence by enhancing their initial Hessian approximation from PD with L-BFGS updates.
Concurrently, Overby et al.~\cite{OBLN17} demonstrate that PD can also be interpreted as a special case of ADMM and used this insight to generalize PD to arbitrary conservative forces, hard constraints and to incorporate implicit friction models~\cite{BOFN18}.
However, they note that the convergence speed of the ADMM approach heavily depends on the choice of weights in the optimization algorithm.

A typical approach to further improve performance of PD is to combine it with subspace or reduced order modeling, such as Hyper-Reduced Projective Dynamics~\cite{brandt_reduced_2018}, Medial Elastics~\cite{LLF+20}, or other subspace approaches, e.g., targeted at cloth simulation (see below)~\cite{li_subspace-preconditioned_2023,LLL+24}.

For parameter estimation, motion planning and inverse problems, Du et al.~\cite{DWM+21} introduced DiffPD which speeds up backpropagation to facilitate efficient differentiable simulations based on PD.
Li et al.~\cite{LDW+22} draw inspiration from this to make cloth simulation with frictional contact differentiable.

\subsubsection{Multiphysics in Projective Dynamics}

Projective Dynamics is an established approach for interactive simulations especially involving deformables, cloth and multibody systems including skinning, coupled by frictional contact.
There are also a few works on multiphysics effects such as plasticity, granular media and fluids.
However, their coupling to deformables within PD remains an open problem.

\paragraph*{Multibody Systems, Contact \& Damping}

The original PD method does not directly support rigid bodies and hard constraints.
To simulate articulated multibodies coupled with deformables more accurately while maintaining the efficiency of PD, Li et al.~\cite{li_fast_2019,li_soft_2022} propose to represent rigid bodies using affine transformations that are locally projected to the rigid transformation manifold SE(3) while taking hard joint constraints into account in an iterative scheme.

For contact, many PD-based works either use dynamically introduced soft unilateral constraints or springs.
To avoid refactorizations of the global matrix and to support dry friction, Ly et al.~\cite{ly_projective_2020} reformulate Projective Dynamics in terms of velocity degrees of freedom and introduce semi-implicitly computed local per-vertex contact forces that satisfy the Signorini-Coulomb law at every iteration.
To facilitate real-time physically-based skinning of animated skeletons, Komaritzan and Botsch~\cite{KB18,KB19} propose Projective Skinning and a corresponding GPU implementation based on iterative solvers which supports global self-collisions.
For the simulation of very high resolution meshes with locally restricted collisions in interactive simulators, Wang et al.~\cite{wang_optimized_2021} present a highly optimized method based on partial Cholesky factorization with sparse updates and penalty-based collision constraints between selected vertices.

Based on the barrier potentials used by IPC~\cite{LFS+20}, Lan et al.~\cite{lan_penetration-free_2022} introduced a PD framework with non-penetration guarantees.
They propose a local projection for the barriers and a two-level iteration scheme where barriers are only added or removed by CCD after several PD iterations to prevent sticking and oscillations during iterations.
Furthermore, they present an aggregated Jacobi solver specifically tailored for better performance on the GPU.

To improve upon typical naive ``ether drag'' damping models, more accurate models such as Rayleigh damping~\cite{BOFN18} and Laplacian damping~\cite{LLK18} were proposed with adaptions for PD.
Dinev et al.~\cite{dinev_stabilizing_2018} investigated the numerical damping introduced through the backward Euler integration of PD and propose a blending technique with the midpoint rule.
However, this approach is not always successful in conserving energy~\cite{DLL+18}.
To improve on this, Kee et al.~\cite{KUJH21} proposed per vertex constraints to conserve energy and momentum.

\paragraph*{Rods and Cloth}
For the simulation of rods, Soler et al.~\cite{soler_cosserat_2018} introduce Cosserat rods into the PD framework by deriving the local projections and weights for the corresponding stretch, shear bend and twist constraints.

Simulating cloth with widely used geometrically motivated and example-based material models is already possible within the original PD framework.
Recent research mostly aims to improve performance and convergence for highly detailed simulations especially on the GPU.
Li et al.~\cite{li_subspace-preconditioned_2023} propose to accelerate the global solve using B-spline subspace simulation for low-frequency motion as preconditioning for a full-order iterative solver.
To handle penetration-free contact, they apply a time-splitting and contact proxy approach inspired by the work of Xie et al.~\cite{xieContactProxySplitting2023}.
As collision detection and line search filtering remained a large contributor to the cost of previous methods, Lan et al.~\cite{LLL+24} propose non-distance barriers that are dependent on the number of iterations they are active instead of exact distances and combine this with a partial CCD scheme.
They also use a much smaller subspace based on a small number of eigenvectors that further improves performance of the global solve.

\paragraph*{Elastoplasticity and Granular Materials}
Based on the Projective Dynamics local/global solver approach, He et al.~\cite{he_projective_2018} introduced the Projective Peridynamics framework for the meshless simulation of elastoplastic bodies, shells, rods and viscolestic fluid and granular flows.

\paragraph*{Fluids}
To simulate fluids within the PD framework, Weiler et al.~\cite{weiler_projective_2016} formulate a density constraint based on an SPH discretization similar to the Position Based Fluids~\cite{macklin_position_2013} approach (see Equation~\eqref{eq:fluid-constraint}) and derive the corresponding local constraint projections.
To avoid costly refactorizations due to changing particle neighborhoods, they propose to solve the global system with a matrix-free CG solver instead.

For real-time simulation of fluid and elastic solid coupling, Brandt et al.~\cite{brandt_fluid_2019} proposed the Reduced Immersed Method framework, coupling PIC/FLIP fluids~\cite{zhu_animating_2005} and deformables simulated with model reduction.
For the coupling they employed the \emph{immersed boundary method}~\cite{Peskin2002} where both fluids and solids are treated as a single incompressible medium.
This approach eliminates the need for fluid-solid surface tracking and explicit contact constraints in the PD solver, but also introduces no-slip conditions between deformables themselves and to the fluid, which can lead to sticking. The immersed boundary method can additionally lead to leakage of fluid through the boundary of thin solids.

\section{Constraint-Based Multiphysics Modeling}
\label{sec:unified:constraint-based}
\begin{figure*}
    \includegraphics[trim={0.085cm 0 0.085cm 0}, clip, height=6.35cm]{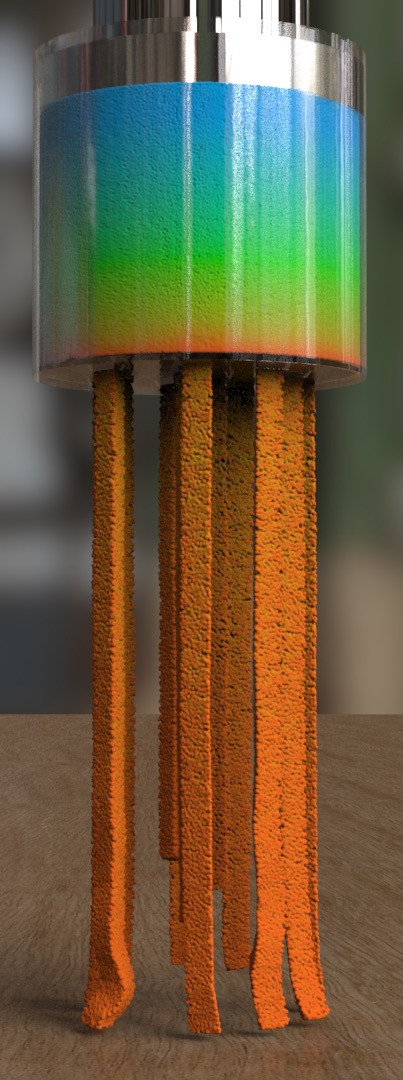}\hspace{0.04cm}%
    \includegraphics[trim={0 0 0.005cm 0}, clip, height=6.35cm]{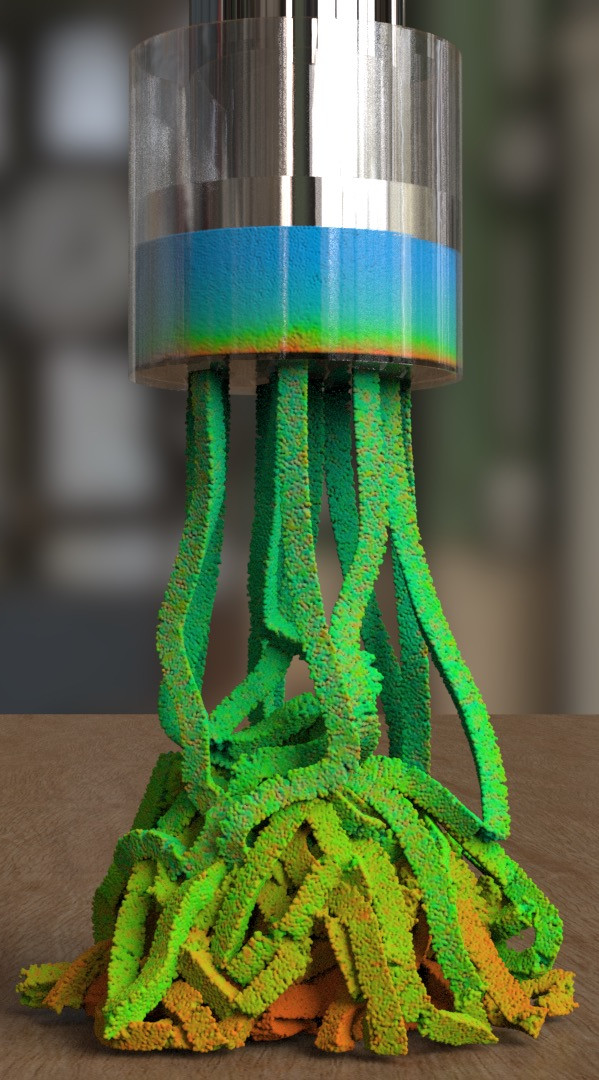}
    \hfill
    \includegraphics[trim={9.6cm 0 9.5cm 0}, clip, height=6.35cm]{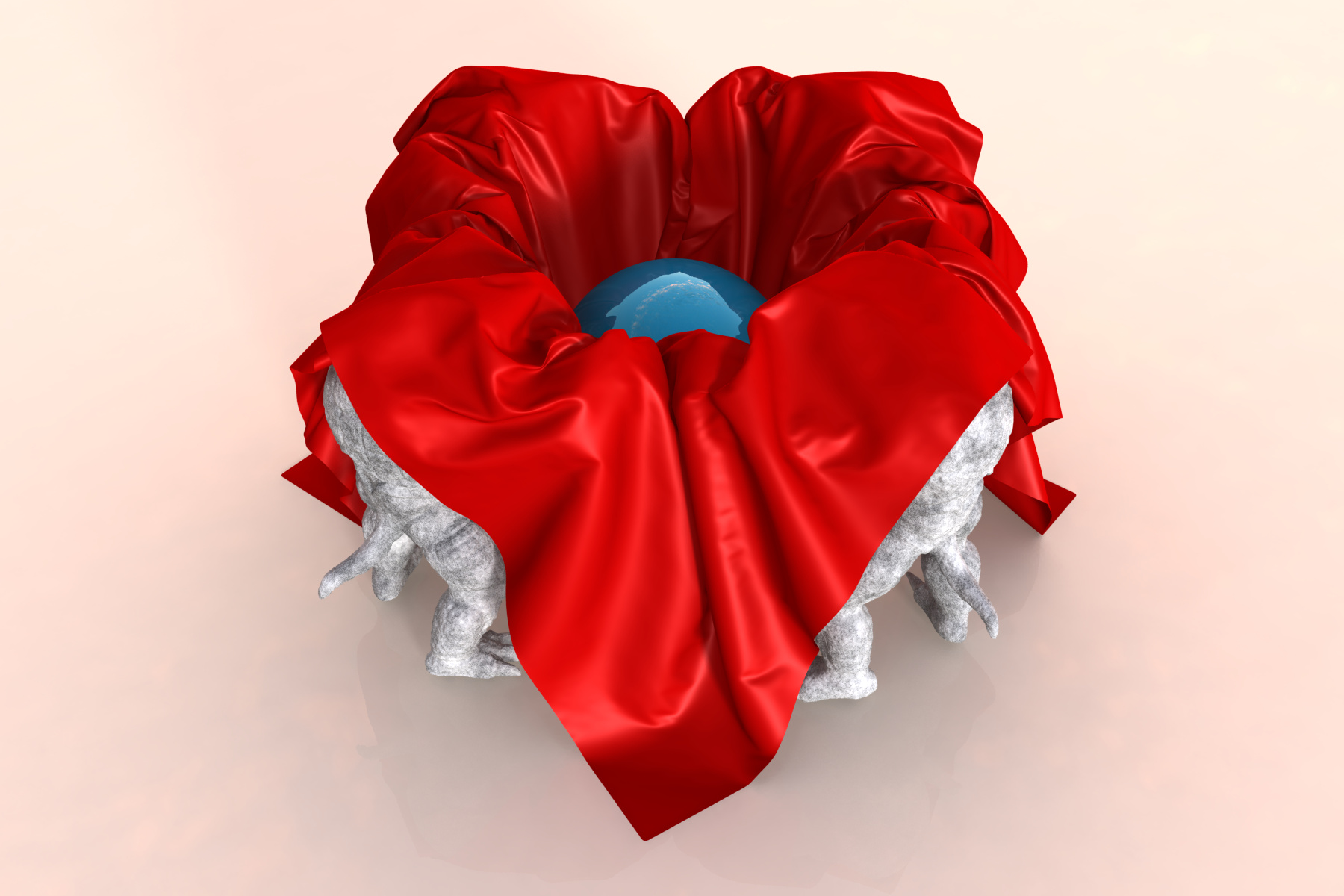}
    \hfill
    \includegraphics[height=6.35cm]{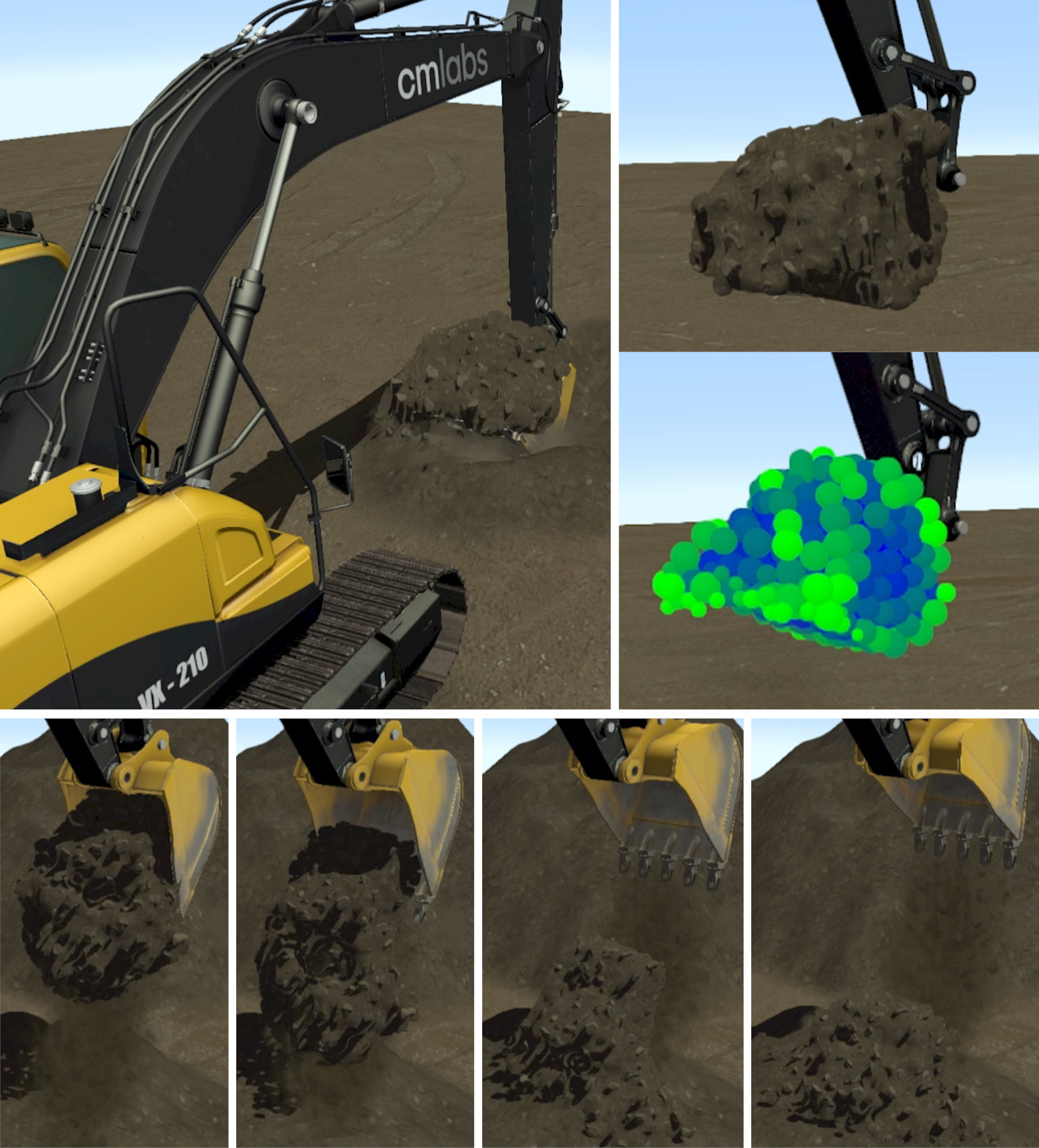}
    \centering
    \caption{\label{fig:constraint-based} Various examples of constraint-based material simulations. \textbf{Left:} A plastic material resembling modeling clay is pressed into strings and piles up on the ground~\cite{yu_xpbi_2024}. \textbf{Middle:} Realistic wrinkles form on a position-based elastic cloth being pushed down by a heavy object~\cite{bender_position-based_2014}. \textbf{Right:} Coupled real-time simulation of an articulated excavator, modeled as a stiff multibody system, interacting with position-based cohesive soil~\cite{holz_real-time_2018}.}
\end{figure*}

In constraint-based simulation methods, the simulated multiphysics system is discretized using a Lagrangian description. A set of constraint functions is then used to restrict the motion of the system's degrees of freedom and combined with the equations of motion. 
This principle modeling approach makes this group of methods commonly quite robust under coarser material discretizations. 
This is specifically beneficial for real-time use cases as it allows reducing the number of degrees of freedom in the system and can thus lead to significant workload reductions. 
In contrast to most approaches from the category of energy-based models in Section~\ref{sec:unified:energy-based}, constraint-based methods are also typically designed to handle hard constraints well.
Furthermore, as opposed to explicit penalty-based methods which struggle with stability when simulating stiff systems~\cite{servin_interactive_2006}, the constraint-based methods presented in this section can simulate such systems robustly even under large time steps.

In the most general terms, both approaches employ different strategies to solve the following non-linear system of equations
\begin{equation}
\begin{split}
\mathbf{M}\ddot{\mathbf{x}} &= \mathbf{f}(\mathbf{x}) \\
\mathbf{C}_b(\mathbf{x}) &= \mathbf{0} \\
\mathbf{C}_u(\mathbf{x}) &\geq \mathbf{0}
\end{split}
\label{eq:constraint-based}
\end{equation}
which combines Newton's second law with $n_b$ equality (or \emph{bilateral}) and $n_u$ inequality (or \emph{unilateral}) constraints $\mathbf{C}_k = [C_{k_1}(\mathbf{x}), \dots, C_{k_{n_k}}(\mathbf{x})]^T$, where $k\in\{b,u\}$, for the system's generalized coordinates $\mathbf{x} = [x_1, \dots, x_n]^T$, a system mass matrix $\mathbf{M}$, and a vector of internal and external generalized forces $\mathbf{f}(\mathbf{x})$ acting on the system's degrees of freedom.
This formulation provides a flexible modeling framework for multiphysics simulations. Through combination of different constraints tailored to the desired behaviors and interactions in the same system, it enables modeling various materials and physical phenomena in a strongly coupled, monolithic manner. Figure~\ref{fig:constraint-based} depicts several examples of constraint-based simulations in this framework.

\subsection{Simulation Methods}

In the following we introduce two popular constraint-based simulation methods, namely Position Based Dynamics (PBD) and eXtended Position Based Dynamics (XPBD).

\subsubsection{Position Based Dynamics}
\label{sec:PBD}
Introduced to the computer graphics community in 2006 by M\"uller et al.~\cite{muller_position_2006} for simulating soft bodies and cloth, Position Based Dynamics has emerged as a unified multiphysics simulation model for interactive and real-time simulation over the last two decades through a number of extensions that allows the approach to be used for the simulation of other materials and physical phenomena.

In its original form, the method employs a local optimization scheme to minimize the errors in a set of constraints that restrict the motion of a collection of particles representing the discretized material~\cite{bouaziz_projective_2014}.
These constraints can take different forms, making it possible to flexibly model different physical phenomena, materials and their interaction.
The method operates directly on the position-level. Particle positions can only move to admissible positions, making the scheme very stable and allowing for use of large time steps~\cite{muller_position_2006}. These qualities, paired with its simplicity, efficiency and versatility, have made PBD one of the most popular real-time physics simulation methods for games and VR applications of recent times. Among others, it is used in the real-time game multiphysics framework NVIDIA FleX~\cite{nvidiaflex_2024}. PBD also finds application in many popular visual effects tools, such as in Houdini~\cite{houdini_2024}, and is also leveraged in other areas, such as medical simulations~\cite{fang_state---art_2023}.
A position-based modeling scheme similar to PBD was previously proposed by Jakobsen~\cite{jakobsen_advanced_2001}.
For a more detailed introduction and overview of PBD, we refer to the survey of Bender et al.~\cite{bender_survey_2017}.

\paragraph*{Foundations of Position Based Dynamics}
PBD discretizes the simulated solids into $n$ particles defined by positions $\mathbf{x}_i$, velocities $\mathbf{v}_i$ and scalar masses $m_i$ and defines a set of $l$ position-based scalar constraints, each restricting the relative positions of a particle subset. These constraints can be both equality and inequality constraints, where equality constraints are used for modeling \emph{bilateral} interactions such as the resistance to stretching between two material points in a cloth, and inequality constraints are used for modeling \emph{unilateral} interactions as for example experienced in a collision.

In a predictor-corrector type fashion the method first produces a prediction of the particle positions at the beginning of each simulation step, denoted by $\mathbf{p}_i$, before subsequently correcting the resultant constraint violations. Assuming the particle motion is unrestricted by any constraints and only affected by external forces, the particles are moved forward in time using symplectic Euler integration, yielding
\begin{equation}
\label{eq:pbd-prediction}
\mathbf{p}_i = \mathbf{x}_i + \Delta t \mathbf{v}_i + \Delta t^2\frac{\mathbf{F}_{\mathrm{ext}}(\mathbf{x}_i)}{m_i},
\end{equation}
where $\Delta{t}$ denotes the simulation time step and $\mathbf{F}_{ext}(\mathbf{x}_i)$ corresponds to the external forces applied to particle $i$. With the predicted particle positions in hand, temporary unilateral collision constraints are generated based on the particles' predicted trajectories, and added to the set of permanent constraints, which are then iteratively corrected via constraint projections in Gauss-Seidel style. Given the vector of concatenated predicted positions $\mathbf{p} \coloneq [\mathbf{p}_{i_1}^T,\dots,\mathbf{p}_{i_{n_j}}^T]^T$ for a subset of particles $\{i_1,\dots, i_{n_j}\}$ constrained by a single constraint $C$, PBD finds a position correction $\mathbf{p} + \Delta \mathbf{p}$ that satisfies the constraint through the following linearization approach.

Without loss of generality, linearizing a bilateral, scalar constraint of the form $C(\mathbf{p}) = 0$ via a first-order Taylor expansion yields the approximation
\begin{equation}
\label{eq:pbd-taylor}
C(\mathbf{p}+\Delta\mathbf{p}) \approx C(\mathbf{p}) + \nabla C(\mathbf{p})^T \Delta \mathbf{p} = 0\,.
\end{equation}
The above equation corresponds to an under-determined system for the unknown $\Delta \mathbf{p}$. To address this issue, PBD introduces the scalar Lagrange multiplier $\lambda$ and restricts the position correction to lie in the constraint gradient direction, such that
\begin{equation}
\label{eq:pbd-deltax}
\Delta \mathbf{p} = \lambda\mathbf{M}^{-1}\nabla C(\mathbf{p})\,,
\end{equation}
with $\mathbf{M}$ denoting a diagonal matrix with dimensions $3n_j \times 3n_j$ containing the scalar particle masses.

The above condition ensures existence of a unique solution and also conveniently prevents any linear or angular momentum gain within the constrained system~\cite{bender_survey_2017}. By inserting Equation~\eqref{eq:pbd-deltax} into Equation~\eqref{eq:pbd-taylor} we can derive the Lagrange multiplier
\begin{equation}
\label{eq:pbd-lambda}
\lambda = \frac{-C(\mathbf{p})}{\nabla C(\mathbf{p})^T\mathbf{M}^{-1}\nabla C(\mathbf{p})}\,,
\end{equation}
which is used in Equation~\eqref{eq:pbd-deltax} to compute the position correction and to update the particle positions. With each position correction becoming immediately visible to other constraints, this iterative scheme is non-linear in nature. The approach can easily be generalized to also handle unilateral constraints robustly by skipping the position update for a unilateral constraint in iterations in which the constraint is not violated.

Finally, after a given number of iterations, the particles' new velocities are recovered from the change in their positions as 
\begin{equation}
	\mathbf{v}_i = \frac{1}{\Delta t}(\mathbf{p}_i - \mathbf{x}_i), 
\end{equation}
and the particle positions $\mathbf{x}_i$ are set to the final predicted and corrected particle positions $\mathbf{p}_i$ for the next simulation step.
The fact that the velocities are obtained from a change of admissible particle positions rather than through extrapolation makes the method very stable.

\begin{algorithm}
\caption{Position-based dynamics}\label{alg:pbd}
\begin{algorithmic}[1]
\ForAll{vertices $i$}
	\State initialize $\vec x_i$, $\vec v_i$, $m_i$
\EndFor
\Loop
\State \textbf{for all} vertices $i$ \textbf{do} $\vec v_i \leftarrow \vec v_i + \Delta t \vec F_\text{ext}(\vec x_i)/m_i$
\State \textbf{for all} vertices $i$ \textbf{do} $\vec p_i \leftarrow \vec x_i + \Delta t \vec v_i$
\State \textbf{for all} vertices $i$ \textbf{do} genCollConstraints($\vec x_i \rightarrow \vec p_i$)
\Loop {\ solverIteration \textbf{times}}
	\ForAll{constraints}
		\State compute $\lambda$ using Eq.~\eqref{eq:pbd-lambda}
		\State compute $\Delta \vec p$ using Eq.~\eqref{eq:pbd-deltax}
		\State $\vec p \leftarrow \vec p + \Delta \vec p$
	\EndFor
\EndLoop
\ForAll{vertices $i$}
	\State $\vec v_i \leftarrow (\vec p_i - \vec x_i) / \Delta t$
	\State $\vec x_i \leftarrow \vec p_i$
\EndFor
\EndLoop
\end{algorithmic}
\end{algorithm}

The position based dynamics approach is outlined in Algorithm~\ref{alg:pbd}.
Note that there also exist open-source implementations of the method~\cite{PBD_Library,PSVC}.

One limitation with this approach is that Gauss-Seidel iterations may not converge in certain situations~\cite{bouaziz_projective_2014}, which can be circumvented by using Jacobi iterations instead~\cite{macklin_unified_2014}.
As an additional benefit, Jacobi iterations allow for easy parallelization of the iterative solver, on both GPUs and CPUs, as demonstrated by Macklin et al.~\cite{macklin_unified_2014} and Holz~\cite{holz_parallel_2014}, respectively. This can lead to improved performance compared to the original Gauss-Seidel approach as demonstrated by Frâncu et al.~\cite{francu_unified_2017}. As another alternative to Gauss-Seidel iterations, Goldenthal et al.~\cite{goldenthal_efficient_2007} compute all Lagrange multipliers combined via a linear solve in each iteration, yielding improved efficiency for inextensible cloth simulations.

Other limitations of the original PBD method are that solutions are dependent on material resolution, time-step and the number of solver iterations~\cite{bouaziz_projective_2014}, and that global angular momentum is not preserved~\cite{dahl_global_2019}. The latter was addressed by Dahl and Bargteil~\cite{dahl_global_2019} through particle velocity adjustments.

\subsubsection{XPBD} %
The time-step and iteration dependence issue was addressed by the \emph{eXtended} Position Based Dynamics (XPBD) method, proposed by Macklin et al.~\cite{macklin_xpbd_2016}.
In XPBD, the non-physical constraint relaxation approach introduced with the original PBD method~\cite{muller_position_2006}, is replaced with the physically-plausible constraint relaxation proposed by Servin et al.~\cite{servin_interactive_2006}.
In this method, the calculation of the Lagrange multiplier in PBD, given by Equation~\eqref{eq:pbd-lambda}, is replaced by a Lagrange multiplier update, which produces a change $\Delta \lambda$ applied to a total Lagrange multiplier $\lambda$ in each solver iteration. For a purely elastic constraint $C(\mathbf{p})$, this change is calculated as
\begin{equation} \label{eq:xpbd-lambda}
\Delta \lambda = \frac{-C(\mathbf{p}) - \frac{\alpha}{\Delta t^2}\lambda}{\nabla C(\mathbf{p})^T\mathbf{M}^{-1}\nabla C(\mathbf{p}) + \frac{\alpha}{\Delta t^2}}\,,
\end{equation}
where $\alpha$ corresponds to the constraint's compliance (or inverse stiffness). Viscous or visco-elastic constraint behavior can also be modeled~\cite{macklin_xpbd_2016}.
Note that, for XPBD, the position correction from Equation~\eqref{eq:pbd-deltax} uses $\Delta \lambda$ instead of $\lambda$ as well.
We can see that the original PBD update is recovered from Equation~\eqref{eq:xpbd-lambda} for zero compliance with $\alpha=0$ (i.e. for hard constraints).

\begin{algorithm}
\caption{eXtended Position-Based Dynamics (XPBD)}\label{alg:xpbd}
\begin{algorithmic}[1]
\ForAll{vertices $i$}
	\State initialize $\vec x_i$, $\vec v_i$, $m_i$
\EndFor
\Loop
\State \textbf{for all} vertices $i$ \textbf{do} $\vec v_i \leftarrow \vec v_i + \Delta t \vec F_\text{ext}(\vec x_i)/m_i$
\State \textbf{for all} vertices $i$ \textbf{do} $\vec p_i \leftarrow \vec x_i + \Delta t \vec v_i$
\State \textbf{for all} vertices $i$ \textbf{do} genCollConstraints($\vec x_i \rightarrow \vec p_i$)
\Loop {\ solverIteration \textbf{times}}
	\ForAll{constraints}
		\State compute $\Delta \lambda$ using Eq.~\eqref{eq:xpbd-lambda}
		\State compute $\Delta \vec p$ using Eq.~\eqref{eq:pbd-deltax}
		\State $\lambda \leftarrow \lambda + \Delta \lambda$
		\State $\vec p \leftarrow \vec p + \Delta \vec p$
	\EndFor
\EndLoop
\ForAll{vertices $i$}
	\State $\vec v_i \leftarrow (\vec p_i - \vec x_i) / \Delta t$
	\State $\vec x_i \leftarrow \vec p_i$
\EndFor
\EndLoop
\end{algorithmic}
\end{algorithm}

Algorithm~\ref{alg:xpbd} outlines the XPBD approach. 
Note that only the lines 10 and 12 differ from the PBD algorithm~\ref{alg:pbd}.
XPBD is also available as open-source implementation~\cite{PBD_Library,PSVC}.

The XPBD solver is derived from a Quasi-Newton method which approximates the system's Hessian with its mass matrix, ignoring second derivatives of the constraint functions~\cite{macklin_xpbd_2016}.
Specifically for interactive real-time applications, this has several advantages, as calculating the Hessian is a time consuming process, and its use requires strategies for dealing with indefinite or singular Hessians~\cite{bouaziz_projective_2014}, creating challenges for robustness.
While this approximation in XPBD reduces the convergence rate (compared to Newton's method), it does not significantly impact the end result of the XPBD solver and makes it more competitive for interactive applications.
In some cases however, a second simplification employed in the derivation of XPBD, which always assumes the non-linear residual (corresponding to the right hand side in Newton's method) to be zero, can prevent it from reaching force balance with respect to the equations of motion.
For example, this entails that performing more solver iterations does not necessarily minimize the inertia potential of the system.
This was recently addressed by Chen et al.~\cite{chen_primal_2024}.

The high stability of (X)PBD is related to its derivation from implicit Euler integration~\cite{macklin_xpbd_2016, francu_unified_2017}, a characteristic also found in Projective Dynamics (PD) (see Section~\ref{sec:projective-dynamics}).
As a consequence, both methods suffer from artificial damping introduced by this numerical integration scheme, which causes strictly dissipating energy in the system depending on the time step size.
In order to reduce the artificial damping in PBD, higher-order integrators could be used~\cite{bender_survey_2017}. For inverse problems including parameter, pose and shape optimization involving deformables and contacts, XPBD can also be made differentiable as demonstrated by Stuyck and Chen~\cite{diff_xpbd}.

\subsubsection{Performance Improvements}
A multigrid approach for improving the convergence rate of the PBD solver was proposed by M\"uller~\cite{muller_hierarchical_2008}. With the same goal in mind, Wang~\cite{wang_chebyshev_2015} proposes the use of the Chebyshev method applied to the Jacobi-style PBD solver~\cite{macklin_unified_2014, holz_parallel_2014}. The authors' experiments show that while this approach works well in certain cases, it is unable to significantly improve the rate of convergence in all cases and can even lead to divergence~\cite{wang_chebyshev_2015}.
Mercier-Aubin and Kry~\cite{mercier-aubin_multi-layer_2024} employ model reduction techniques to adaptively rigidify portions of elastic solids simulated in XPBD, based on local strain rates within the material, significantly reducing the workload and increasing the convergence rate in XPBD.
Macklin et al.~\cite{macklin_small_2019} employ sub-stepping with reduced solver iterations, yielding improved convergence and higher accuracy, as well as reduced artifical damping through improved energy conservation.
Higher accuracy in XPBD can also be achieved by correcting energy residuals in constraints~\cite{cetinaslan_position-based_2019}.

Commonly, graph coloring algorithms~\cite{fratarcangeli_gpu-based_2013} are used for the parallelization of Gauss-Seidel-style solvers as featured in PBD.
For dense constraint networks this approach can yield large numbers of independent constraint partitions which reduces the level of solver parallelism. Thon-That et al.~\cite{ton-that_parallel_2023} address this issue by employing a block-iterative approach
and applying graph coloring on the resultant graph of sub-domains. This sub-structuring leads to faster solver convergence while also reducing the number of independent constraint partitions, consequently improving solver parallelism.

\subsection{Materials and Phenomena}

In the following we will discuss different materials and phenomena that can be modeled using the constraint-based simulation approach.

\subsubsection{Fluids}
For Position Based Fluids (PBF), Macklin and M\"uller~\cite{macklin_position_2013} employ the fluid model proposed by Bodin et al.~\cite{bodin_constraint_2012}. Here, an incompressible fluid is modeled using the constraint $C_i = \rho_i / \rho_0 - 1 = 0$, which enforces a uniform reference density $\rho_0$ at every position $\mathbf{x}_i$ in the medium.
The per-particle density $\rho_i$ is computed using SPH-based material interpolation (see Section~\ref{sec:SPH}), yielding the position-level fluid constraint
\begin{equation}
C_i = \frac{1}{\rho_0} \sum_{j=1}^n m_j W(\mathbf{x}_i-\mathbf{x}_j,h) - 1 = 0
\label{eq:fluid-constraint}
\end{equation}
which dynamically constrains each fluid particle $i$ to its neighboring particles $j$ within the compact support of the SPH kernel $W$.
As in SPH, here we also assume an entirely meshfree Lagrangian view of the medium, which makes the method ideal for modeling the large deformations observed in fluids.

Takahashi et al.~\cite{takahashi_fast_2014} extend the above method for the simulation of viscous fluids with elasticity, capturing buckling and coiling effects, and demonstrating phase transitions caused by temperature changes.
Xing et al.~\cite{xingPositionBasedSurfaceTension2022} introduce a strong surface tension model based on an area-minimization constraint. The constraint is derived from local meshes constructed at the fluid interface, thereby overcoming limitations in the original model which makes use of artificial pressure terms~\cite{macklin_position_2013}.
Abu Rumman et al.~\cite{aburumman_isphpbd_2020} propose a technique for the two-way coupling of deformable PBD solids and SPH fluids, which can handle challenging scenarios such as leakage-free interactions with thin deformable shells, and overcomes the issue of particle clumping and visible gaps at fluid/solid boundaries in the original PBF~\cite{macklin_position_2013}.
In their unified SPH-based multiphysics model, Shao et al.~\cite{shao_improving_2017} use position-based non-penetration constraints to improve fluid/solid coupling and enable small-scale turbulences through vorticity constraints that are solved using PBD.
Macklin et al.~\cite{macklin_unified_2014} propose a PBD framework for the unified simulation of position-based fluids, smoke, rigid bodies and deformable solids using a GPU-based Jacobi solver. Barreiro et al.~\cite{barreiroConformationConstraintsEfficient2017} demonstrate position-based simulations of viscous, viscoelastic, elastoplastic, and inviscid liquids, for which they employ a novel, doubly constrained PBD solver that can handle position-based constraints as well as the velocity-based constraints required by their proposed viscoelasticity model.

\subsubsection{Rigid Bodies}
Further extending the capabilities of the PBD framework for multiphysics modeling, Deul et al.~\cite{deul_position-based_2016} enable maximal coordinate rigid body dynamics simulation with joints in PBD by strongly coupling position-based rigid bodies with particles in a unified manner. This approach has stability benefits compared to the direct forcing method proposed by M\"uller et al.~\cite{muller_position_2006} for coupling PBD particles with rigid bodies.
The method was later also applied for modeling rigid bodies in XPBD, with further extensions for simulating restitution and friction~\cite{francu_unified_2017, muller_detailed_2020}.

\subsubsection{Continuous Materials}
To model continuous materials in PBD, various works derive position-based constraints from constitutive material models
for elastic, viscous and plastic deformable solids.

Bender et al.~\cite{bender_position-based_2014} were the first to demonstrate how to model continuous materials in the PBD framework independent of the resolution in the material discretization -- a limitation of the original PBD approach~\cite{bouaziz_projective_2014}.
The method uses an FEM-based material discretization and employs position-level constraints on the strain energy in deformed tetrahedral volume elements, and captures various complex physical phenomena such as lateral contraction, anisotropy, viscoelasticity and elastoplasticity.
M\"uller et al.~\cite{muller_strain_2015} propose a strain limiting approach by directly constraining the entries of the Green--Saint-Venant strain tensor via position-based constraints,
which was recently adapted for use in XPBD~\cite{cetinaslan_esbd_2024}.
Macklin et al.~\cite{macklin_xpbd_2016} and Francu et al.~\cite{francu_unified_2017} employ a constraint regularization method for resolution-independent simulation of elastic, continuous materials in XPBD, initially proposed by Servin et al.~\cite{servin_interactive_2006} for constraint-based modeling at the velocity-level. Rather than constraining strain energies directly as proposed by Bender et al.~\cite{bender_position-based_2014}, here, position-based constraints are derived from elastic energy potentials and included in the XPBD framework. While the method is limited to Hookean materials, a recent extension enables Neo-Hookean materials~\cite{macklin_constraint-based_2021}. Ton-That et al.~\cite{ton-that_parallel_2023} further improve the approach, demonstrating better convergence,
and Chen et al.~\cite{chen_primal_2024} later generalize XPBD to arbitrary hyperelasticity by reinterpreting the Lagrange multipliers as components of the stress tensor.
Their method also shows significantly improved accuracy and convergence behavior.
Saillant et al.~\cite{saillant_high-order_2024} introduce a constraint formulation of higher-order finite elements for XPBD, increasing simulation accuracy and performance. They also address that some continuous material constraints were not stable at rest in XPBD.

As an alternative to the physically accurate methods above, efficient simulation of visually plausible deformations can be achieved with shape matching, in which sets of material points are regrouped and their deformed state mapped to the undeformed rest state of a target shape without any need for connectivity information. This geometrically motivated approach was introduced by M\"uller et al.~\cite{muller_meshless_2005} and subsequently adapted to the PBD framework for deformation and fracture~\cite{choi_real-time_2014, chentanez_real-time_2016}. For more information about this family of techniques we recommend the survey on PBD by Bender et al.~\cite{bender_survey_2017}.

\subsubsection{Rods}
As a special type of continuous material, rods find application in various areas of computer graphics, in both real-time and offline scenarios. They are particularly well-suited for the representation of hair, fur, vegetation, cables and ropes, and several position-based approaches have been proposed for their simulation.

M\"uller et al.\cite{muller_fast_2012} use a \emph{follow the leader} approach for modeling simple, position-based rods in PBD designed for inextensible hair and fur. While the method guarantees zero stretch in a single iteration, making it very fast, it suffers from energy injection and results in implausible behavior, which is addressed through artificial damping.
Umetani et al.~\cite{umetani_position-based_2015} introduce a novel discretization of Cosserat theory for the position-based modeling of bending and twisting elastic rods by using \emph{ghost points} to define consistent rod material frames.
Kugelstadt and Sch\"omer~\cite{kugelstadt_position_2016} extend the approach through use of quaternions as material frames, allowing them to formulate strain measures directly as position and orientation constraints.
For the simulation of stiff, elastic rods and trees in XPBD, Deul et al.~\cite{deul_direct_2018} replace the standard Gauss-Seidel solver by a direct solver, which guarantees high material stiffness and yields significant speed-ups compared to \cite{umetani_position-based_2015} and \cite{kugelstadt_position_2016}.
Angles et al.~\cite{angles_viper_2019} add support for volumetric deformations, and demonstrate volume conservation of compressed or extended rods.

\subsubsection{Granular Materials}
We can identify two popular approaches for modeling granular materials in literature, both of which are compatible with the Position Based Dynamics framework. These are DEM-based approaches and approaches based on plasticity theory.

DEM-based simulation of cohesionless granular materials in PBD was demonstrated by both Holz~\cite{holz_parallel_2014} and Macklin et al.~\cite{macklin_unified_2014}, employing inter-particle Coulomb friction constraints modeled at the position-level for stable piling of granular materials. Holz~\cite{holz_parallel_2014} improved the behavior further by modeling constraints as spring-dampers, enabling viscoelastic material behavior and making the artificial damping proposed by M\"uller et al.~\cite{muller_position_2006} unnecessary for this use case. Viscoelastic constraint behavior was also proposed by Macklin et al.~\cite{macklin_xpbd_2016} in XPBD, and also by Francu et al.~\cite{francu_unified_2017} who demonstrated viscoelastic collisions among others for granular material simulation. Later, Holz and Galarneau~\cite{holz_real-time_2018} modeled position-based cohesive granular materials through use of a unilateral adhesion constraint derived from a capillary bridge model, details of which can be found in \cite{karpman_discrete_2020}.

In a variation of XPBD, Yu et al.~\cite{yu_xpbi_2024} employ plasticity theory for the simulation of granular and other inelastic materials.
Their proposed eXtended Position-Based Inelasticity (XPBI) framework can represent a variety of visco- and elastoplastic materials, such as snow, sand, metal and foam in a unified, strongly coupled manner.
Just like MPM (see Section~\ref{sec:MPM}), XPBI employs a hybrid grid-particle discretization, which facilitates capturing the large plastic deformations and fracture commonly observed in plastic and granular materials. However, without use of an Eulerian background grid, XPBI does not suffer from the grid artifacts common to MPM simulations.

\section{Lagrangian Point-Based Methods}
\label{sec:point-based}
\begin{figure*}
    \includegraphics[trim={280 0 280 0},clip,height=3.845cm]{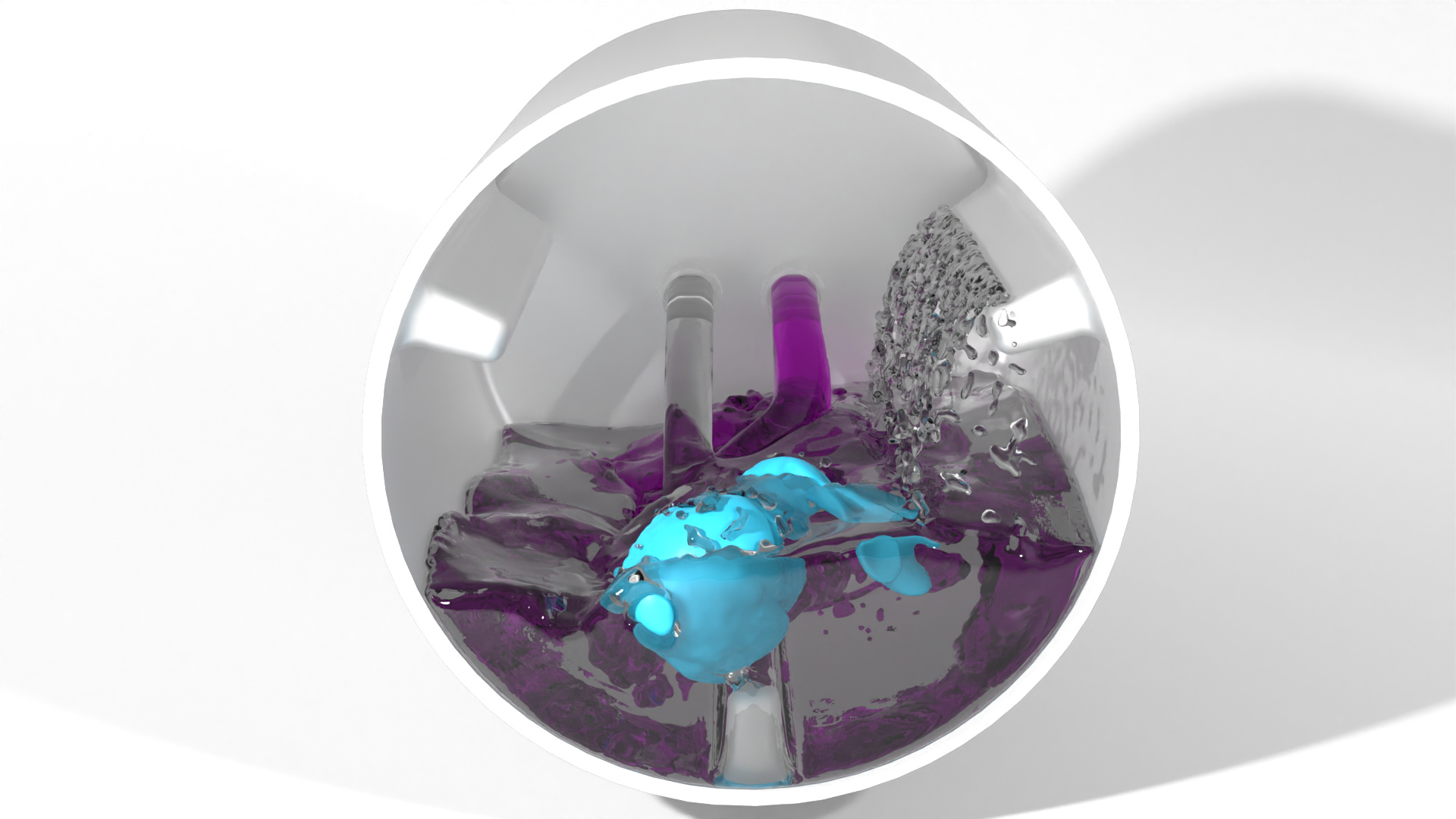}
    \hfill
    \includegraphics[trim={0 0 250 0},clip,height=3.845cm]{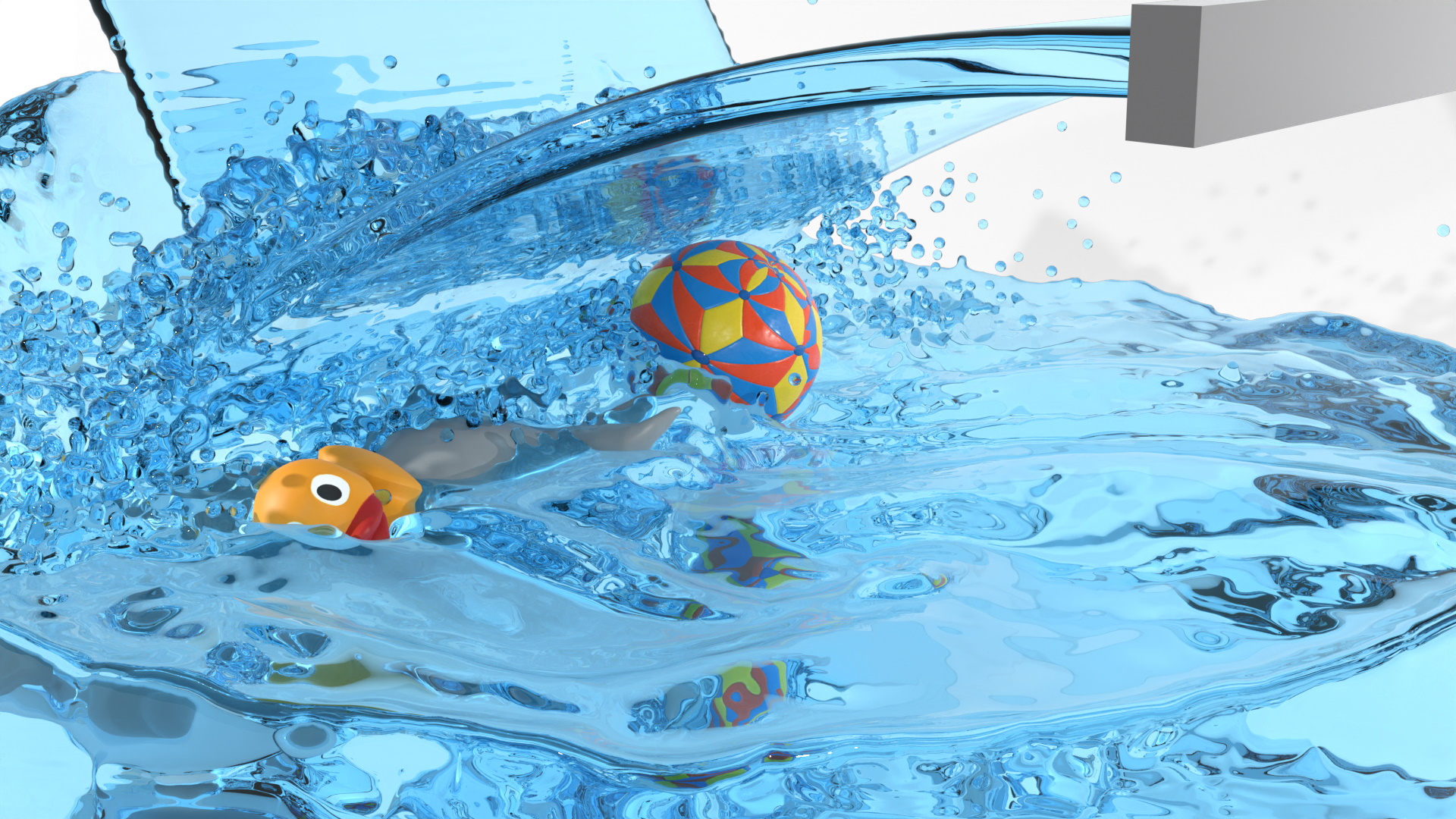}
    \hfill
    \includegraphics[trim={0 0 0 0},clip,height=3.845cm]{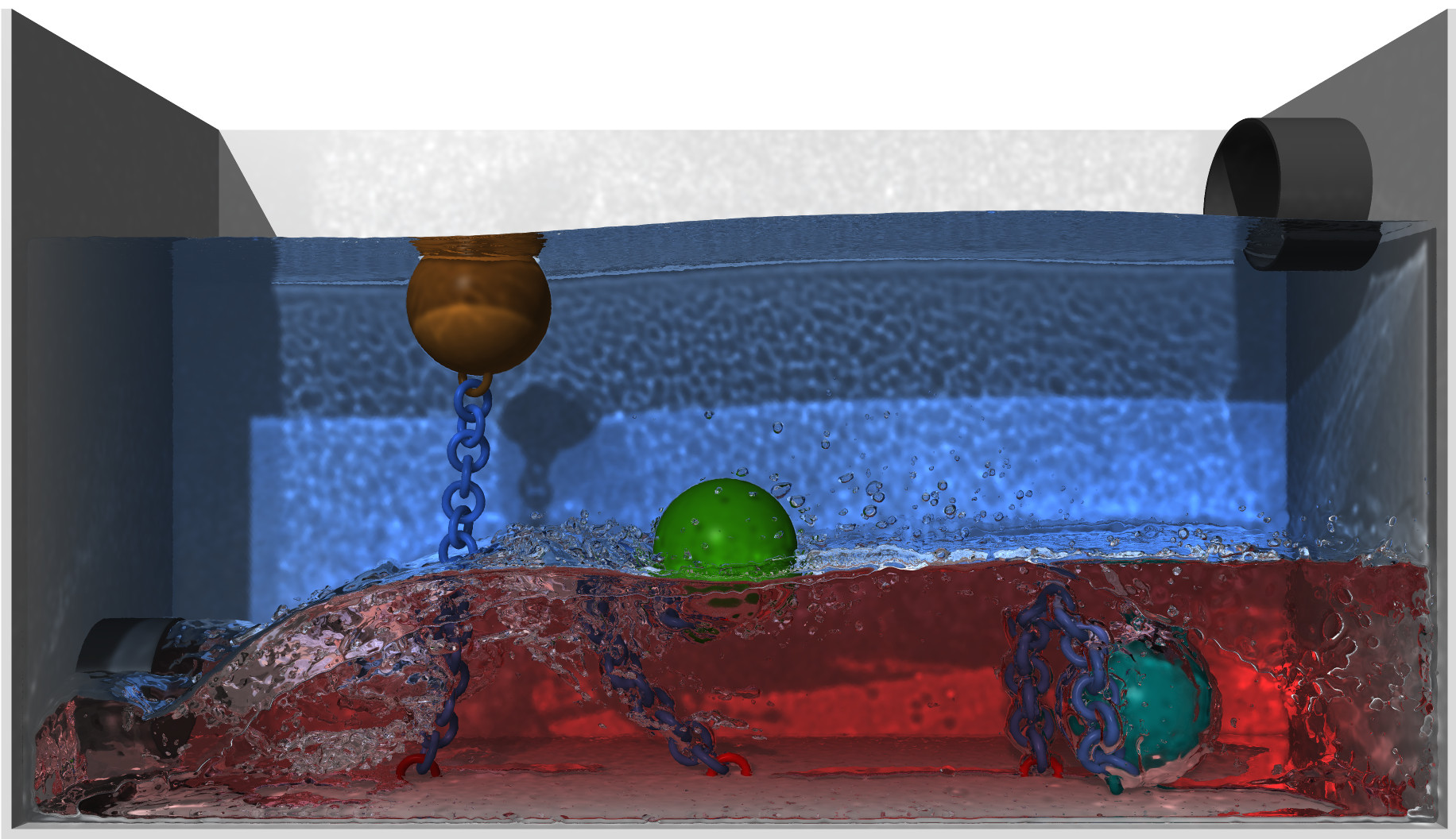}
    \centering
    \caption{\label{fig:point-based} \textbf{Left:} SPH simulation of 
    an elastic bunny which is coupled with a fluid, a highly viscous material and a rotating washing machine drum~\cite{westhofenComparisonLinearConsistent2023}. \textbf{Middle:} An SPH fluid interacts with a duck and a ball~\cite{LBJB23}. \textbf{Right:} A fluid and a highly viscous material are interacting with chains of rigid bodies~\cite{gisslerInterlinkedSPHPressure2019}.}
\end{figure*}

In this section we will look in more detail at current approaches using a point-based discretization of physical objects and phenomena.
Specifically, we will go into detail about how to solve problems defined in the continuum on a set of points.

In general, an unordered set of points is a versatile representation for a large variety of physical objects.
Particularly volumetric objects with changing topology, such as fluids and granular materials, are well suited to be represented in this way, due to their lack of inherent fixed structure.
The representation as points can also be very efficient in terms of computational efficiency and memory usage, since only regions of interest will be discretized.
This can be further improved by using spatial adaptivity, similar to grid- or mesh-based methods.
Objects where topological changes are less desirable, such as rigid bodies and deformable solids, can be represented by simply ``locking'' specific particle arrangements.
For most particle-based methods, this does not require special handling and fits directly into the respective paradigm.
More challenging are co-dimensional structures, such as shells, thin sheets, chains, or rods, which require special treatment in particle-based methods.
The difficulty stems from the fact, that the unordered particles do not natively carry any information about co-dimensionality and therefore have to be explicitly classified and transitioned.

For the case of fixed topology, the Finite Element Method (FEM) is also a popular choice in graphics, which differentiates itself from point-based methods by using a mesh.
While FEM is particularly useful for the Lagrangian simulation of deformable solids, it typically needs to be explicitly coupled to other physical effects through non-unified interaction terms.
In contrast, purely point-based methods more easily permit a unified discretization of multiple physical effects.
In our case of multiphysical simulation, FEM has become frequently used for the discretization of energy-based systems, which are outlined in more detail in Section~\ref{sec:energy-based-materials}.

With respect to Lagrangian point-based methods, we introduce the approach of Smoothed Particle Hydrodynamics (SPH) in Section~\ref{sec:SPH}.
Although SPH was initially proposed for astrophysics~\cite{lucy_numerical_1977,gingold_smoothed_1977} and is now mainly used for fluid dynamics, recent work has shown its remarkable ability for coupling of many different materials and physical effects (see Figure~\ref{fig:point-based}).
In Section~\ref{sec:mls}, we will briefly discuss the conceptual relationship to other purely particle-based methods, such as Moving Least Squares (MLS)~\cite{lancasterSurfacesGeneratedMoving1981} and the Reproducing Kernel Particle Method (RKPM)~\cite{liuReproducingKernelParticle1995}, both of which have also shown versatility in modeling multiphysics phenomena.
Finally, we will discuss how multi-material and multiphysics effects are typically incorporated into the point-based formulation.

\subsection{Simulation Methods}

\subsubsection{Smoothed Particle Hydrodynamics}
\label{sec:SPH}

The Smoothed Particle Hydrodynamics (SPH) method approximates a continuous spatial integral using unordered and unconnected point clouds with associated physical quantities such as mass $m$ and rest density $\rho_0$.
The core of SPH lies in computing material interactions of these finite points, using a (compactly-supported) kernel function $W$.
The kernel function in itself is an approximation of the $\delta$-distribution identity.
In the continuum, this can be written as
\begin{equation}
	\label{eq:sph_delta}
		f(\vec{x}) = \int f(\vec{y}) \delta(\vec{x} - \vec{y}) d\vec{y}      
		           \approx \int f(\vec{y}) W(\vec{x} - \vec{y}; h) d\vec{y} \,.
\end{equation}
In SPH, this continuous integral is approximated using a finite number of discrete points $\vec x_j$ within the support radius $\ensuremath{\hbar}$ which depends on the smoothing length $h$ in order to compute arbitrary field quantity $f(\vec x)$ as
\begin{equation}
	\label{eq:sph_function}
	f(\vec{x}) \approx \sum_j V_j f(\vec{x}_j) W(\vec{x} - \vec{x}_j; h) \,.
\end{equation}
This summation approximates the volume integral by assuming that each point has an associated finite volume $V_j$ in which the function value $f(\vec x_j)$ remains constant.
In order to satisfy this approximation, the SPH kernel $W$ has to fulfill the following conditions $\forall \; \vec{r} \in \Real^d, h \in \Real^+$~\cite{koschier_smoothed_2019}:
\begin{gather}
	\int_{\Real^d} W(\vec{r}^\prime; h) dv^\prime = 1 \; \quad \; \tag{\small normalization condition} \label{eq:normalization_condition} \\
	\lim_{h^\prime \to 0} W(\vec{r}; h^\prime) = \delta(\vec{r}) \vphantom{\int} \qquad \quad \; \tag{\small Dirac-$\delta$ condition} \\
	W(\vec{r}; h) \geq 0 \vphantom{\int} \quad \quad \; \tag{\small positivity condition}\\
	W(\vec{r}; h) = W(-\vec{r}; h) \; \quad \; \vphantom{\int} \tag{\small symmetry condition} \\
	W(\vec{r}; h) = 0 \ \text{for} \ \Vert \vec{r} \Vert \geq \ensuremath{\hbar}. \vphantom{\int} \tag{\small compact support condition}
\end{gather} 
The cubic-spline kernel is one of the most common choices, fulfilling all necessary conditions.

In the following we will use the common abbreviations $A_j = A(\vec{x}_j)$ for a field quantity $A$ and $W_{ij} = W(\vec{x}_i - \vec{x}_j; h)$ for the kernel function. 

Since we introduced the kernel $W$, derivatives can be computed by simply applying the derivative operator only to the kernel function. 
Therefore the gradient is determined as
\begin{equation}
	\label{eq:sph-sample-field-gradient}
	\nabla A_i \approx \sum_j V_j A_j \nabla W_{ij} \,.
\end{equation}
Over the years, multiple variations of computing derivatives using the SPH kernel have been proposed, each with their own properties and implications~\cite{koschier_smoothed_2019}.
The most most widely used formulations for first order derivatives are the difference formula: 
\begin{equation}
	\nabla A_i \approx \sum_j \frac{m_j}{\rho_j} \left (A_j - A_i) \right ) \nabla W_{ij},
\end{equation}
which provides an improved approximation, 
and the symmetric formula: 
\begin{equation}
	\nabla A_i \approx \rho_i \sum_j m_j \left ( \frac{A_i}{\rho_i^2} + \frac{A_j}{\rho_j^2} \right ) \nabla W_{ij},
\end{equation}
which is typically used to compute linear and angular momentum conserving forces.

In addition, it has become common to evaluate second derivatives without explicitly taking the second derivative of the kernel function~\cite{koschier_survey_2022}, but instead using
\begin{align}
	\label{eq:sph-laplacian}
	\nabla^2 A_i &\approx 2 \sum_j \frac{m_j}{\rho_j} \left(A_i - A_j\right) \frac{(\vec{x}_i - \vec{x}_j) \cdot \nabla W_{ij}}{||\vec x_i - \vec x_j||^2 + 0.01 h^2} 
	\\
	\label{eq:sph-laplacian2}
	\nabla^2 \vec A_i &\approx 2 (d+2) \sum_j \frac{m_j}{\rho_j} \frac{\left(A_i - A_j\right)  \cdot (\vec{x}_i - \vec{x}_j)}  {||\vec x_i - \vec x_j||^2 + 0.01 h^2} \nabla W_{ij},
\end{align}
for scalar and vector quantities, respectively, where $d$ is the spatial dimension.
The reason for this is that the Laplacian of the kernel function changes sign within its support radius and therefore the computation is prone to noise in the particle distribution~\cite{Pri12a}.

For more details, including details on implementation, the reader is referred to recent state-of-the-art reports~\cite{koschier_survey_2022,ihmsen_sph_2014,liu_smoothed_2010}.

\subsubsection{MLS and RKPM} 
\label{sec:mls}

In graphics, SPH is sometimes combined with two other methods due to their improved guarantees on the order of approximation or stability~\cite{westhofenComparisonLinearConsistent2023}.
These methods are Moving Least Squares (MLS) \cite{lancasterSurfacesGeneratedMoving1981} and the Reproducing-Kernel Particle Method (RKPM) \cite{liuReproducingKernelParticle1995} and can be considered variations of a kernel-based method.
Rewriting Equation~\eqref{eq:sph_function} using a generalized shape function $\phi(\vec x)$ and smoothing length $h$ yields
\begin{equation}
	\label{eq:sph_shape_function}
	f(\vec{x}) = \sum_j f(\vec{x}_j) \phi(\vec{x},\vec{x}_j; h).
\end{equation}
MLS and RKPM can now be represented by expressing $\phi(\vec x)$ in the following ways~\cite{adamsMeshlessApproximationMethods2009,westhofenComparisonLinearConsistent2023}:
\begin{equation}
	\label{eq:phi_sph}
	\phi^\mathrm{SPH}(\vec x, \vec{x}_j; \theta) = V_j W(\vec{x} - \vec{x}_j; h),
\end{equation}
\begin{equation}
	\label{eq:phi_mls}
	\begin{aligned}
		\phi^\mathrm{MLS}(\vec x, \vec{x}_j; \theta) = \vec{b}(\vec x)^T \matr{M}(\vec x) \vec{b}(\vec x_j) W(\vec{x} - \vec{x}_j; h) \\
		\matr{M}(\vec x) = \left[\sum_j W(\vec{x} - \vec{x}_j; h) \vec{b}(\vec x_j) \vec{b}(\vec x_j)^T \right]^{-1} \,,
	\end{aligned}
\end{equation}
\begin{equation}
	\label{eq:phi_rkpm}
	\begin{aligned}
		\phi^\mathrm{RKPM}(\vec x, \vec{x}_j; \theta) = V_j \vec{b}(\vec x)^T \matr{N}(\vec x) \vec{b}(\vec x_j) W(\vec{x} - \vec{x}_j; h) \\
		\matr{N}(\vec x) = \left[\sum_j V_j W(\vec{x} - \vec{x}_j; h) \vec{b}(\vec x_j) \vec{b}(\vec x_j)^T \right]^{-1} \,.
	\end{aligned}
\end{equation}
The vector $\vec{b}$ is typically chosen as a specific polynomial basis, such as the linear basis $\vec b(\vec x) = \left[ 1\,\, x\,\, y\,\, z \right]$.

In SPH simulations, first-order MLS derivatives can be used to guarantee linear consistency as demonstrated by Band et al.~\cite{bandMLSPressureBoundaries2018} and Westhofen et al.~\cite{westhofenComparisonLinearConsistent2023}.
MLS was also investigated to simulate deformable solids with elastic and plastic deformations~\cite{MKN+04}, fracturing~\cite{PKA+05} and as discretization for unified simulations of volumetric deformables, shells and rods~\cite{martin_unified_2010}.
Chen et al.~\cite{chenMovingLeastSquare2020} propose a moving least square reproducing-kernel method for multiphase continua simulation.
They also integrate a previous phase-field model for simulating mixtures, and show a multitude of coupled systems.

\subsection{Materials and Phenomena}

In this section we discuss SPH methods for the simulation of multiple materials and phenomena.

\subsubsection{Fluid Dynamics}

As described in the last section, SPH was proposed as a general function approximation framework.
However, as the name implies, Smoothed Particle Hydrodynamics mainly started out as a method for the simulation of incompressible fluids, commonly described by the incompressible Navier-Stokes equations
\begin{equation}
\label{eq:navier_stokes}
		\rho  \frac{D \vec{v}}{D t} = -\nabla p + \mu \nabla^2 \vec{v} + \vec{f}, \qquad \frac{D\rho}{D t} = -\rho (\nabla \cdot \vec v) = 0 \,,
\end{equation}
where $\rho$, $\vec v$, $p$, $\mu$ and $\vec f$ denote density, velocity, pressure, dynamic viscosity and external forces, respectively.
The first equation describes conservation of momentum, while the second describes conservation of mass. 
Both of these have to be fulfilled for any continuous physical medium, while the specific form in Equation~\eqref{eq:navier_stokes} describes the dynamic motion of incompressible fluids.

\paragraph*{Pressure}
Arguably the most important step in solving this equation is computing the pressure force density $-\nabla p$, which ensures constant density within incompressible fluids.
Mathematically speaking, it acts as a Lagrange multiplier enforcing the incompressibility constraint of the fluid.
It can also be interpreted as a kind of ``contact'' or ``collision'' resolution for point-based methods, as it prevents particles from ``intersecting'' each other.

\begin{algorithm}[tb]
	\caption{Simulation loop for SPH simulation of weakly compressible fluids.}
	\label{alg:WCSPH}
	\begin{algorithmic}[1]
	\ForAll{$particle~i$}
	\State compute density $\rho_i$ using using Eq.~\eqref{eq:SPH_density}
	\State determine pressure $p_i$ using state equation (Eq.~\eqref{eq:state_equation})
	\EndFor
	\ForAll{$particle~i$}
		\State compute pressure force $\vec f_i^\text{p} = -\nabla p_i$ using Eq.~\eqref{eq:SPH_pressure_force}
		\State compute viscosity force $\vec f_i^\text{v} = \nu \nabla^2 \vec v_i$, e.g, using Eq.~\eqref{eq:sph-laplacian2}
	\EndFor
	\ForAll{$particle~i$}
		\State $\vec v_i(t + \Delta t) = \vec v_i(t) + \frac{\Delta t}{\rho_i}\left (\vec f_i^\text{pressure} + \vec f_i^\text{viscosity} + \vec f_i^\text{ext} \right )$
		\State $\vec x_i(t + \Delta t) = \vec x_i(t) + \Delta t \vec v_i(t + \Delta t)$
	\EndFor
	\end{algorithmic}
\end{algorithm}

Early approaches, such as weakly compressible SPH (WCSPH) \cite{becker_weakly_2007}, used an equation of state (EOS) to explicitly relate density to pressure.
The density can be determined by applying the standard SPH approach (see Eq.~\eqref{eq:sph_function}) to the density field: 
\begin{equation}
\label{eq:SPH_density}
	\rho_i = \sum_j m_j W_{ij}.
\end{equation}
An example for an equation of state is 
\begin{equation}
\label{eq:state_equation}
 p_i = \frac{\kappa \rho_0}{\gamma} \left ( \left (\frac{\rho_i}{\rho_0} \right )^\gamma - 1 \right),
\end{equation}
where $\gamma$ and $\kappa$ are stiffness parameters. 
The equation of state determines the pressure values for each particle which can be used to determine the final pressure forces using the symmetric formula as
\begin{equation}
\label{eq:SPH_pressure_force}
	\vec f^\text{p} = -\nabla p_i \approx \rho_i \sum_j m_j \left ( \frac{p_i}{\rho_i^2} + \frac{p_j}{\rho_j^2} \right ) \nabla W_{ij}.
\end{equation}
A WCSPH simulation step is outlined in Algorithm~\ref{alg:WCSPH}.

WCSPH requires the use of a stiffness constant, which when combined with explicit time integration, could easily result in compression artifacts for low stiffnesses or instabilities for large stiffness values.
To mitigate this, implicit methods were proposed, which enforce constant density by solving the pressure Poisson equation (PPE)
\begin{equation}
\label{eq:pressure_poisson}
\Delta t \nabla^2 p = \frac{D\rho}{D t}
\end{equation}
in various ways.
Solenthaler and Pajarola~\cite{solenthaler_predictive-corrective_2009} introduced a predictive-corrective incompressible SPH (PCISPH) approach for reducing density deviations.
This method showed significant improvement over explicit pressure solvers.
Bodin et al.~\cite{bodin_constraint_2012} introduced a constant density constraint for modeling incompressible fluids in a constraint-based simulation framework, subsequently adopted by Macklin and M\"uller~\cite{macklin_position_2013} for use in a position-based fluid (PBF) solver (cf. Section~\ref{sec:PBD}). 
Later, Weiler et al.~\cite{weiler_projective_2016} solved the same constraint using Projective Dynamics (see Section~\ref{sec:projective-dynamics}).
Ihmsen et al.~\cite{ihmsen_implicit_2014} proposed implicit incompressible SPH (IISPH) to directly solve the PPE by computing the Laplace operator on the left hand side of Equation~\eqref{eq:pressure_poisson} using two SPH derivatives -- 
the first one to compute the pressure gradient $\nabla p$, and the second one to determine the divergence of this gradient $\nabla \cdot (\nabla p) = \nabla^2 p$. 
Bender and Koschier~\cite{benderDivergenceFreeSPHIncompressible2017} and later Cornelis et al.~\cite{cornelisOptimizedSourceTerm2019} proposed to solve a second PPE $\Delta t \nabla^2 p = \rho \nabla \cdot \vec v$ to enforce a constant density and a divergence-free velocity field.
The former is commonly referred to as divergence-free SPH (DFSPH).
Enforcing both improves the stability of the simulation and the performance since it reduces the iteration count of the pressure solver.
Note that there exist open-source implementations of most of the aforementioned pressure solvers~\cite{SPlisHSPlasH_Library,PSVC}.

All aforementioned implicit pressure solvers are equivalent except for small differences~\cite{koschier_survey_2022}.
While PCISPH uses a prototype particle to avoid a division by zero when computing the diagonal matrix entries, later works avoid this due to stability issues. 
PBF updates the particle positions and therefore the matrix in each iteration step of the solver. 
However, IISPH and DFSPH solve the PPE on the velocity level to avoid the expensive matrix updates. 
Finally, in contrast to the other solvers, DFSPH also enforces a divergence-free velocity field to improve the stability and performance.

For more details on SPH-based pressure solvers, we refer the reader to the surveys of Ihmsen et al.~\cite{ihmsen_sph_2014} and Koschier et al.~\cite{koschier_smoothed_2019,koschier_survey_2022}.

\paragraph*{Viscosity}
The Laplacian of the velocity field in Equation~\eqref{eq:navier_stokes} denotes the viscosity forces.
These are traditionally decoupled from the pressure solvers by operator-splitting and computed in isolation.
Some approaches simply ``smooth'' the velocity field to simulate viscous behavior, for example using artificial viscosity \cite{becker_weakly_2007} or XSPH \cite{schechter_12}.
When using XSPH, the velocity field is updated in each time step by computing a smoothed velocity $\hat{\vec v}_i$ for each particle $i$ as:
\begin{equation}
\label{eq:visco_XSPH}
	\hat{\vec v}_i = \vec v_i + \alpha \sum_j \frac{m_j}{\rho_j} (\vec v_j - \vec v_i) W_{ij}, 
\end{equation}
where $0 \leq \alpha < 1$ is a user-defined parameter. 
Others approximate the viscous stress tensor by either taking two first-order SPH derivatives~\cite{takahashiImplicitFormulationSPHbased2015} or a combination of an SPH derivative with finite differences~\cite{Pri12a} (see Equations~\eqref{eq:sph-laplacian} and \eqref{eq:sph-laplacian2}).

While low viscosity fluids can be simulated using explicit time integration, implicit methods have been investigated in recent years to simulate highly viscous fluids. 
Some of these methods use the strain rate tensor for an implicit formulation~\cite{takahashiImplicitFormulationSPHbased2015,peerPrescribedVelocityGradients2017,barreiroConformationConstraintsEfficient2017}.
However, Weiler et al.~\cite{weilerPhysicallyConsistentImplicit2018} showed that this leads to visual artifacts at the surface due to a particle deficiency problem. 
Therefore, they propose an implicit method which directly determines the Laplacian of the velocity field.

These implicit solvers are weakly coupled with the pressure solver and enable the simulation of highly viscous Newtonian fluids.
Liu et al.~\cite{liuAdaptedSIMPLEAlgorithm2022} introduced a method based on the idea of the SIMPLE algorithm to implement a strong coupling of pressure and viscosity solver. 
The simulation of non-Newtonian viscous material behavior was demonstrated by Zhang et al.~\cite{zhangMultiphaseViscoelasticNonNewtonian2024}.

\paragraph*{Multi-Phase, Mixtures and Porous Flow}
After determining how to compute the main terms in Equation~\eqref{eq:navier_stokes}, it becomes interesting how to couple multiple fluids either in a multi-phase simulation, or through mixture models.
The most widely-used method of stably coupling multiple fluids in SPH, even with relatively large density ratios, was proposed by Solenthaler and Pajarola~\cite{solenthalerDensityContrastSPH2008}. 
In this method, different fluid materials, i.e., phases, are represented as their own distinct set of particles.
An alternative approach for multi-phase simulations uses only a single set of particles and represents individual phases by volume fractions of these particles~\cite{renMultipleFluidSPHSimulation2014,yangFastMultiplefluidSimulation2015,yanMultiphaseSPHSimulation2016,renUnifiedParticleSystem2021,renIncompressibilityEnforcementMultipleFluid2022,yanHighDensityRatio2023,xuImplicitlyStableMixture2023}. 
This approach can model effects such as mixing, diffusion and dissolution very well.
However, conservation of mass and and momentum across all mixture components becomes more challenging, as does reconstruction of individual phase boundaries.

Porous flow is a type of interaction within a mixture of different materials, which describes the flow of a fluid through another porous medium and is governed by Darcy's law.
Early approaches relied on shrinking and growing particles~\cite{lenaertsPorousFlowParticlebased2008,lenaerts_mixing_2009,patkarWettingPorousSolids2013} as they enter and exit porous material.
Movement within the porous body is tracked using a saturation term, which can also affect material properties.
Another approach for porous flow is adopted by Yan et al.~\cite{yanMultiphaseSPHSimulation2016}, in which porous flow is computed using their multi-phase fluid framework.
This avoids changing particle sizes by adjusting per-particle volume fractions of each phase.
This is beneficial, as changing particle size and mass can result in instabilities due to large mass ratios.
Newer works propose an alternate way to compute porous flow with SPH particles.
The fluid particles enter and overlap the porous body, but become decoupled from the pressure solver~\cite{renUnifiedParticleSystem2021}.
Instead, their motion is entirely governed by Darcy's law when overlapping solid particles.
A similar approach is also adopted by Wang et al.~\cite{wangVisualSimulationSoilStructure2021} for simulating wet sand, where the sand is simulated using the Discrete Element Method (DEM).
Böttcher et al.~\cite{BWJB25} instead take local porosity into account when computing pressure forces to limit particle overlap and use coupling forces to simulate drag and capillary action.

\paragraph*{Further Fluid Phenomena}
There are a variety of further fluid-based effects that can be captured using SPH, which are typically modeled by an additional interaction force in Equation~\eqref{eq:navier_stokes}.
One such effect, which has garnered increased interest in recent years, is fluid turbulence.
SPH fluid simulations can suffer from numerical diffusion which leads to a loss in turbulent details.
Therefore, different methods were proposed to maintain these details, generally by reducing the amount of kinetic energy lost through dissipation.
Bender et al.~\cite{benderTurbulentMicropolarSPH2019} introduced rotational degrees of freedom to the SPH particles.
Using the equation for conservation of angular momentum and an additional momentum exchange term, it was possible to obtain better conservation of energy and turbulent details.
This was further improved by using vorticity refinement~\cite{liuTurbulentDetailsSimulation2021}, which uses a stream function to relate vorticity to velocity, instead of a momentum-exchange term.
An advantage of this approach is, that it allows for better control of turbulence and even allows for amplification of vortices.
Finally, where Liu et al.~\cite{liuTurbulentDetailsSimulation2021} used the stream function to avoid explicit computation of the Biot-Savart integral, Ye et al.~\cite{yeMonteCarloVortical2024} proposed an efficient computation of this integral using the Monte Carlo method.

Surface tension is another phenomenon that is important for realistic fluid simulations.
It is caused by pressure differences at fluid interfaces, which is formalized by the Young-Laplace equation, relating the force to the local curvature of the interface and the surface normal.
Initial attempts at computing surface tension explicitly, by means of the continuum-surface force (CSF)~\cite{mullerParticlebasedFluidSimulation2003}, have found the surface particle deficiency to result in unstable and erroneous forces.
Using a cohesion-force-based approach instead exhibited more stability \cite{becker_weakly_2007}.
This has been further improved on by using a combination of cohesion forces and approximate curvature forces with custom kernel functions~\cite{akinciVersatileSurfaceTension2013}.
Furthermore, there were attempts to introduce an energy for minimizing the surface area of the fluid~\cite{heRobustSimulationSparsely2015}.
It was found by Yang et al.~\cite{yangPairwiseForceSPH2017} that increasing the support radius of the SPH kernel and a pair-wise force formulation increases the stability and accuracy of interface-dominated phenomena, such as surface tension and adhesion.
However, this comes at the cost of significantly increased computational effort.
Zorilla et al.~\cite{zorillaAcceleratingSurfaceTension2020} presented an accurate computation of surface curvature using Monte Carlo integration.
Wang et al.~\cite{wangThinfilmSmoothedParticle2021} used the SPH discretization method to introduce a method for the simulation of thin films. 
To do this, they simulate a height field on bubbles consisting of a single layer of SPH particles.
Local surface reconstruction was also used to more accurately compute surface normals and coupled with a position-based fluid approach~\cite{xingPositionBasedSurfaceTension2022}.
Jeske et al.~\cite{jeskeImplicitSurfaceTension2023} instead proposed an implicitly integrated model based purely on cohesion forces, in order to simulate very large surface tension coefficients.
Finally, Wang et al.~\cite{wangCodimensionalSurfaceTension2020} pursued an MLS-based co-dimensional approach, where thin-films and filaments of fluids are treated individually.
This allows them to achieve greater accuracy on surface-driven phenomena, such as surface tension. 

Finally, solutions have been proposed to use SPH, partially in combination with other methods, to solve the shallow water equations (SWE).
Solenthaler et al.~\cite{solenthalerSPHBasedShallow2011} discretize 2D shallow water equations on particles instead of uniform grid cells, offering benefits common to particle discretization like simplification of sparsely filled domains and easy interaction with arbitrary boundary geometry. 
Chentanez et al.~\cite{chentanezRealtimeSimulationLarge2010} develop a shallow-water solver that spawns particles in regions difficult to capture with a height field, such as breaking waves and waterfalls, where these particles represent spray or foam independently before merging back into bulk fluid. 
By coupling a particle level-set method (FLIP) with an SPH solver, Losasso et al.~\cite{losassoTwoWayCoupledSPH2008} show simulations of diffuse phenomena like mixtures of water spray and air. 
Finally, Chentanez et al.~\cite{chentanezCoupling3DEulerian2015} improve on this approach by using a multi-resolution fully-adaptive approach, enabling the simulation of larger domains.
They combine an Eulerian solver with PBF or SPH, while allowing for different time step sizes among particles and the grid.

\subsubsection{Rigid Bodies and Frictional Contact}

Coupling fluids or deformable solids with static or dynamic rigid bodies is an essential topic in order to enable dynamic and expressive multiphysics simulations.
A common approach is to sample the surface of a rigid body using particles.
These particles can then be used to compute explicit interaction forces \cite{becker_direct_2009} or, more commonly, be used as additional sampling points in the density and pressure force computations~\cite{akinci_versatile_2012}.
Akinci et al.~\cite{akinci_coupling_2013} expand on this approach by dynamic resampling to enable coupling with thin structures such as cloth.
Gissler et al.~\cite{gisslerInterlinkedSPHPressure2019} extend the particle-based approach to also resolve contacts with friction between rigid bodies using the surface particles.
This enables a strong rigid-fluid coupling since fluids and rigid bodies are handled in one global SPH solver.
Probst and Teschner~\cite{probst_monolithic_2023} further improve this strong coupling by introducing a more realistic friction handling.

An alternative to the particle-based approach is to use an implicit boundary representation~\cite{fujisawaEfficientBoundaryHandling2015,koschier_density_2017,benderImplicitFrictionalBoundary2020,winchenbachSemianalyticBoundaryHandling2020}.
Fujisawa and Miura~\cite{fujisawaEfficientBoundaryHandling2015} propose a fast approach to compute the integral of the kernel within the boundary for triangular meshes and integrate this into a PBF solver.
Koschier and Bender~\cite{koschier_density_2017} use a fixed grid and precompute the density contribution of the boundary in the support radius of each grid point.
In the simulation the boundary contribution for each particle can be determined easily by interpolating the grid values of the cell which contains the particle. 
This approach was later improved by precomputing the volume intersection of the particle domain and the boundary instead of the density~\cite{benderImplicitFrictionalBoundary2020}.
Finally, Winchenbach et al.~\cite{winchenbachSemianalyticBoundaryHandling2020} compute boundary kernel contributions by defining a locally representative planar boundary for arbitrary geometries defined by signed distance functions and deriving an analytic solution.

When using additional sampling points, the standard SPH formulation commonly used in pressure solvers requires a pressure value at each sampling point. 
Different strategies were developed to determine these values: mirroring the pressure values from the corresponding fluid particles~\cite{akinci_versatile_2012}, considering the additional sampling points in the linear system solver~\cite{BGI+18} or extrapolating the pressure values from the fluid~\cite{bandMLSPressureBoundaries2018}.
However, recently it was demonstrated that an additional boundary pressure computation can be avoided entirely by a reformulation of the pressure solver~\cite{benderConsistentSPHRigidFluid2023}.

\subsubsection{Elastic and Elastoplastic Materials}

Deformable solids are often simulated using mesh-based methods since they can be more efficient for this application than meshless approaches.
However, point-based methods like SPH have the advantage that a unified representation for fluids and solids facilitates coupling between different materials, the simulation of state transitions like solidification or melting, as well as topology changes.

\paragraph*{Elasticity}

A common way to simulate a deformable solid using SPH is to first determine the particle displacement $\vec u = \vec x - \vec x^0$ as the difference of its current and its initial position~\cite{solenthaler_unified_2007}. 
Then, the SPH formulation is used to compute the gradient of the displacement field $\nabla \vec u$ and the deformation gradient $\vec F = \nabla \vec u + \identity$, where $\identity$ denotes the identity matrix.
This typically uses a fixed particle neighborhood from the rest state.
Becker et al.~\cite{beckerCorotatedSPHDeformable2009} propose a corotational approach to determine the gradients, which ensures a proper handling of rotations.
The strain can be calculated using the deformation gradient, and the stress is obtained by applying a constitutive model. 
Finally, symmetric forces are determined from the stress to simulate elastic behavior. 

The methods of Solenthaler et al.~\cite{solenthaler_unified_2007} and Becker et al.~\cite{beckerCorotatedSPHDeformable2009} are based on a conditionally stable explicit time integration. 
To improve the stability, Peer et al.~\cite{peerImplicitSPHFormulation2018} propose to apply a corotated linear elasticity model in combination with an implicit Euler time integration. 
Kugelstadt et al.~\cite{kugelstadtFastCorotatedElastic2021} improved the performance significantly by splitting the model in a stretch and a volume term. 
The separate consideration of the stretching term leads to a linear system with a constant system matrix which can be solved very efficiently using a precomputed Cholesky factorization.
While the aforementioned implicit methods are based on linear elasticity models, Kee et al.~\cite{keeOptimizationbasedSPHSolver2023} propose an implicit time integration for non-linear models using a quasi-Newton method.

\paragraph*{Elastoplasticity}

Inspired by the work of Zhu and Bridson~\cite{zhu_animating_2005}, elastoplastic material models were investigated in the field of SPH to simulate granular materials.
The core idea is that elastic material behavior is used to simulate static friction, and if a yield condition is met, the material can start to flow, deforming plastically. 

Lenaerts and Dutré~\cite{lenaerts_mixing_2009} also use an elastoplastic model to simulate granular materials. 
In their explicit solver, plasticity is modeled using the Mohr-Coulomb yield criterion.
Alduán and Otaduy~\cite{alduan_sph_2011} propose a similar approach but use the Drucker-Prager yield criterion and introduce an implicit solver.
Ihmsen et al.~\cite{ihmsenLagrangianFrameworkSimulating2013a} extend this method by an upsampling technique to increase the detail. 
Later, Gissler et al.~\cite{gisslerImplicitCompressibleSPH2020} introduced an SPH snow simulation based on an elastoplastic material model in combination with an implicit pressure solver for compressible fluids.

\subsubsection{Multiphysics Systems}
Besides the previously presented methods for modeling individual effects, which can readily be coupled in unified particle-based discretizations, there have been recent contributions to modeling multi-physical systems in general.
These contributions generally outline improvements to the fundamental particle-based methodology which can benefit the simulation accuracy and stability of multiple physical phenomena in a unified discretization.

Solenthaler et al.~\cite{solenthaler_unified_2007} early on proposed a system that allowed for coupled simulations of fluids, rigid bodies and deformables.
This included a simple model for phase-change including heat-transfer within and across different phases, allowing for effects such as melting, joining and solidification.
Interaction between different materials was handled by the pressure solver.
An improved system is presented by Yang et al.~\cite{yang_unified_2017}, in which all of the aforementioned effects can be simulated, but can further be combined with granular materials and plasticity, as well as multiple phases including solubility with continuous interfaces.
Instead of using a unified discretization method, Xie et al.~\cite{xieContactProxySplitting2023} introduce an approach with contact proxies to couple SPH fluids and deformables (see Section~\ref{sec:energy-based-materials}).

\section{Eulerian and Hybrid Methods}
\label{sec:eulerian_hybrid}
Thus far, we have focused on simulations methods with a Lagrangian viewpoint. This class of approaches treats a physical continuum like a particle system, where each bit of matter in a fluid or solid is represented by a separate particle with position and velocity and other physical quantities.

Eulerian simulations, on the other hand, track a set of fixed points in space and how physical quantities at these points evolve over time, e.g., using a uniform grid. 
The Lagrangian and Eulerian viewpoints are connected through the \emph{material derivative}, which for some physical quantity $q$ can be expressed as
\begin{equation}
\label{eq:material_derivative}
		\frac{D q}{D t} = \frac{\partial q}{\partial t} + {\vec v}\cdot \nabla q \,.
\end{equation}
The first term on the right hand side is simply the rate at which $q$ is changing at a specific point in space, whereas the second term is an advective derivative that gives the change due to movement through 
a velocity flow field ${\vec v}$.

We have seen the material derivative before.
For instance, in Equation~\eqref{eq:navier_stokes} the material derivative of the velocity of a fluid flow field appears in the Navier-Stokes equations. In the Lagrangian view, since physical quantities move with particles in the simulation, the advective derivative is not needed, and the material derivative is simply the time derivative of the velocity, $\frac{D \vec{v}}{D t} = \frac{\partial \vec{v}}{\partial t}$. However, since the Eulerian view measures quantities at fixed points, the advective term of the Navier-Stokes equations $\vec{v} \cdot \nabla \vec{v}$ must be considered.

In practice, Eulerian formulations are attractive because they turn spatial operators into local stencil computations on a fixed grid.
Before discussing specific phenomena, we briefly summarize the geometric representation and discretization pipeline that underlies many such methods.

\subsection{Eulerian Discretization Basics}
\label{sec:eulerian-discretization}

Building on the Eulerian viewpoint above, we now briefly revisit how continuum equations are turned into computable algebraic systems on a fixed grid.
A standard Eulerian setup introduces a bounded simulation domain $\Omega \subset \mathbb{R}^d$ and partitions it into cells of size $\Delta x$ (and analogously $\Delta y, \Delta z$ in higher dimensions).
Using regular grids is typical, but irregularly sampled tetrahedral meshes can also be supported.
Scalar or vector fields are then sampled at predefined locations (cell centers, faces, or nodes).
For fluid simulation, a common choice is a staggered marker-and-cell (MAC) layout, where pressure is stored at cell centers and velocity components are stored at face centers~\cite{harlow1965numerical,bridson2015fluid}.

\paragraph*{Geometric representation.}
In the Eulerian view, a continuum body moving through space is represented implicitly through fields defined on the fixed background grid.
For example, one may use an indicator or level-set field
\begin{equation}
\label{eq:eulerian_level_set}
\phi(\vec{x},t) =
\begin{cases}
< 0, & \vec{x} \text{ inside the body},\\
= 0, & \vec{x} \text{ on the interface},\\
> 0, & \vec{x} \text{ outside the body}.
\end{cases}
\end{equation}
The interface is recovered as the zero isosurface of $\phi$, while transport and deformation are captured by the evolution of the field values.
This is a common way to implement free-surface flows in the Eulerian viewpoint and is referred to as level-set method (LSM).

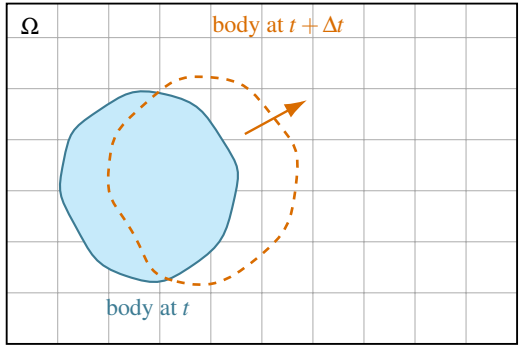
\begin{figure}[t]
    \centering
    \begin{tikzpicture}[scale=0.75]
        % Background Cartesian grid (Eulerian sampling locations are fixed).
        \draw[step=0.9, line width=0.25pt, gray!65] (0,0) grid (9,6);
        \draw[line width=0.7pt] (0,0) rectangle (9,6);

        % Simulation domain label.
        \node[anchor=north west] at (0.1,5.9) {$\Omega$};

        % Body at time t (filled shape).
        \filldraw[fill=cyan!20, draw=cyan!50!black, line width=0.8pt]
            plot[smooth cycle, tension=0.95] coordinates {
                (1.2,2.1) (2.1,1.2) (3.3,1.4) (4.0,2.5) (3.8,3.6) (2.8,4.4) (1.6,4.1) (1.0,3.2)
            };

        % Body at time t + dt (translated/deformed outline).
        \draw[orange!85!black, line width=1.0pt, dashed]
            plot[smooth cycle, tension=0.95] coordinates {
                (2.2,2.0) (3.0,1.1) (4.4,1.5) (5.1,2.8) (4.7,4.0) (3.6,4.7) (2.4,4.2) (1.8,3.1)
            };

        % Motion arrow.
        \draw[-{Latex[length=3mm,width=2mm]}, line width=0.9pt, orange!85!black]
            (4.2,3.7) -- (5.3,4.3);

        % Labels for time instances and fixed grid notion.
        \node[cyan!50!black] at (2.5,0.6) {body at $t$};
        \node[orange!85!black] at (4.8,5.6) {body at $t+\Delta t$};
        % \node[align=left, anchor=west] at (9.2,3.0) {fixed\\sampling\\locations};
    \end{tikzpicture}
    \caption{Conceptual illustration of a moving and deforming body within an Eulerian geometric representation on a fixed grid.}
    \label{fig:eulerian-grid-concept}
\end{figure}

\paragraph*{Finite differences and relation to FVM/FEM.}
A direct way to discretize PDEs on structured grids is the finite difference method (FDM), which approximates derivatives by local stencil operations.
For a sufficiently smooth scalar field $q(x)$ sampled at $x_i = i\Delta x$, a first derivative can be approximated as
\begin{equation}
\label{eq:fd_first_derivative}
\left.\frac{\partial q}{\partial x}\right|_{i}
\approx \frac{q_{i+1} - q_{i-1}}{2\Delta x} + \mathcal{O}(\Delta x^2).
\end{equation}
Compared with FDM, finite volume methods (FVM) emphasize conservation through flux balances over control volumes, while finite element methods (FEM) use weak formulations and basis functions on elements~\cite{Teran2003,bathe2006finite}.
While different in fundamental approach, on regular grids, these viewpoints are closely related and often lead to similar sparse linear systems.

\paragraph*{Canonical stencils in 1D, 2D, and 3D.}
In one dimension, the second derivative is approximated by
\begin{equation}
\label{eq:fd_1d_laplacian}
\left.\frac{\partial^2 q}{\partial x^2}\right|_{i}
\approx \frac{q_{i-1} - 2q_i + q_{i+1}}{\Delta x^2}.
\end{equation}
In two dimensions, the standard 5-point Laplacian stencil is
\begin{equation}
\label{eq:fd_2d_laplacian}
\left.\nabla^2 q\right|_{i,j}
\approx
\frac{q_{i+1,j} - 2q_{i,j} + q_{i-1,j}}{\Delta x^2}
+
\frac{q_{i,j+1} - 2q_{i,j} + q_{i,j-1}}{\Delta y^2}.
\end{equation}
In three dimensions, this extends to the 7-point stencil
\begin{align}
\label{eq:fd_3d_laplacian}
\left.\nabla^2 q\right|_{i,j,k}
\approx
\frac{q_{i+1,j,k} - 2q_{i,j,k} + q_{i-1,j,k}}{\Delta x^2}&\\
+
\frac{q_{i,j+1,k} - 2q_{i,j,k} + q_{i,j-1,k}}{\Delta y^2}&\\
+
\frac{q_{i,j,k+1} - 2q_{i,j,k} + q_{i,j,k-1}}{\Delta z^2}&.
\end{align}
Most stencils can be derived from Taylor expansion around a specific point, such that a variety of stencils exist for special cases.
For example, boundary nodes cannot utilize central finite difference methods.
The stencil may also have to be adjusted depending on the imposed boundary condition.

\begin{figure}[t]
    \centering
    \begin{tikzpicture}[scale=0.9]
        % 4x4 cell MAC grid.
        \draw[step=1.2, line width=0.35pt, gray!65] (0,0) grid (4.8,4.8);
        \draw[line width=0.8pt] (0,0) rectangle (4.8,4.8);

        % Pressure samples at cell centers.
        \foreach \i in {0.6,1.8,3.0,4.2} {
            \foreach \j in {0.6,1.8,3.0,4.2} {
                \fill[blue!70!black] (\i,\j) circle (1.4pt);
            }
        }

        % u-velocity on vertical faces.
        \foreach \x in {1.2,2.4,3.6,4.8} {
            \foreach \y in {0.6,1.8,3.0,4.2} {
                \fill[red!75!black] (\x,\y) circle (1.4pt);
            }
        }

        % v-velocity on horizontal faces.
        \foreach \x in {0.6,1.8,3.0,4.2} {
            \foreach \y in {1.2,2.4,3.6,4.8} {
                \fill[green!50!black] (\x,\y) circle (1.4pt);
            }
        }

        % Highlight a cell and annotate canonical stencil points.
        \draw[line width=1.0pt, orange!85!black] (1.2,1.2) rectangle (2.4,2.4);
        \fill[blue!70!black] (1.8,1.8) circle (1.7pt);
        \draw[-{Latex[length=2.8mm,width=1.9mm]}, line width=0.7pt] (1.8,1.8) -- (1.8,2.85);
        \draw[-{Latex[length=2.8mm,width=1.9mm]}, line width=0.7pt] (1.8,1.8) -- (1.8,0.75);
        \draw[-{Latex[length=2.8mm,width=1.9mm]}, line width=0.7pt] (1.8,1.8) -- (2.85,1.8);
        \draw[-{Latex[length=2.8mm,width=1.9mm]}, line width=0.7pt] (1.8,1.8) -- (0.75,1.8);

        % Labels and legend.
        \node[anchor=west] at (5.15,4.2) {\textcolor{blue!70!black}{$\bullet$} pressure $p_{i,j}$ (cell center)};
        \node[anchor=west] at (5.15,3.5) {\textcolor{red!75!black}{$\bullet$} $u_{i+\frac{1}{2},j}$ (vertical face)};
        \node[anchor=west] at (5.15,2.8) {\textcolor{green!50!black}{$\bullet$} $v_{i,j+\frac{1}{2}}$ (horizontal face)};
        \node[anchor=west] at (5.15,1.7) {$\nabla^2 p$: 5-point stencil};
    \end{tikzpicture}
    \caption{2D MAC grid with finite-difference stencil evaluation points.}
    \label{fig:fd-2d-stencil-concept}
\end{figure}
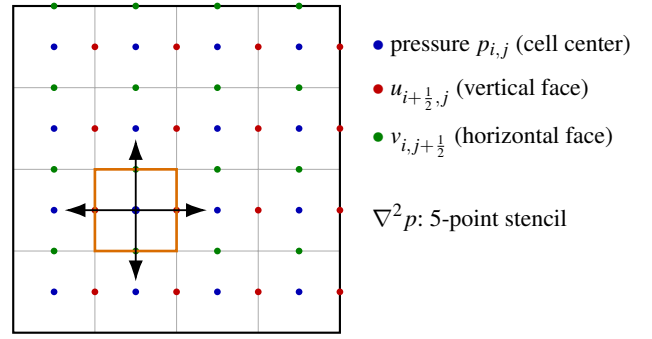

\paragraph*{Example PDE.}
As a representative model problem, consider advection-diffusion of a scalar quantity $q$:
\begin{equation}
\label{eq:advection_diffusion}
\frac{\partial q}{\partial t}
+
\vec{u}\cdot\nabla q
=
\kappa \nabla^2 q + s,
\end{equation}
where $\vec{u}$ is a prescribed velocity field, $\kappa$ is the diffusion coefficient, and $s$ is a source term.
This equation contains the two dominant operators that also appear in many fluid simulation substeps: transport and elliptic diffusion/projection.

\paragraph*{Semi-discretization via the method of lines.}
The most common approach to discretize space-time PDEs is to use the so-called method of lines.
With the method of lines, one first discretizes only space and keeps time continuous.
Using a grid vector $\mathbf{q}(t)$ containing all sampled values of $q$, Equation~\eqref{eq:advection_diffusion} becomes
\begin{equation}
\label{eq:mol_general}
\frac{d\mathbf{q}}{dt} = \mathbf{L}(\mathbf{q}) + \mathbf{s},
\end{equation}
where $\mathbf{L}(\cdot)$ is the discrete spatial operator assembled from advection and diffusion stencils.
For linearized advection and constant coefficients, one obtains
\begin{equation}
\label{eq:mol_linear}
\frac{d\mathbf{q}}{dt} = \mathbf{A}\mathbf{q} + \mathbf{s},
\end{equation}
with sparse matrix $\mathbf{A}$ induced by local stencil couplings.
A time integrator (e.g., forward Euler, Runge-Kutta, backward Euler) then advances this ODE system.

\paragraph*{Resulting algebraic systems.}
An explicit step has the form
\begin{equation}
\label{eq:explicit_update}
\mathbf{q}^{n+1} = \mathbf{q}^{n} + \Delta t\left(\mathbf{A}\mathbf{q}^{n} + \mathbf{s}^{n}\right),
\end{equation}
while an implicit backward Euler step gives
\begin{equation}
\label{eq:implicit_update}
\left(\mathbf{I} - \Delta t\,\mathbf{A}\right)\mathbf{q}^{n+1}
=
\mathbf{q}^{n} + \Delta t\,\mathbf{s}^{n+1}.
\end{equation}
Hence, the simulation loop repeatedly alternates between local stencil evaluations and the solution of sparse linear systems, exactly the structure encountered in pressure projection for incompressible flow~\cite{bridson2015fluid}.

\paragraph*{Strengths, limitations, and practical extensions.}
Structured-grid discretizations are simple, robust, and map incredibly well to parallel hardware because each stencil operation is local and regular.
This is one reason Eulerian solvers are highly effective on GPUs and many-core architectures.
At the same time, limitations include numerical dissipation (especially in low-order advection), CFL restrictions for explicit time stepping, and the need to predefine the simulated spatial extent~\cite{Stam99,Foster1996}.
This problem also causes free-surfaces reconstructed from a LSM to be afflicted by smearing artifacts, where the interface of the fluid becomes blurrier with passing time.
Practical extensions to Eulerian methods include adaptive grids, higher-order advection schemes, and hybrid particle-grid methods.

\subsection{Fluid Dynamics}
\label{sec:eulerian-fluids}
Eulerian methods are perhaps most well known in graphics for simulating fluid dynamics.
In this setting, a grid stores physical and flow field quantities that are typically updated with an operator splitting approach in a three step process:
\begin{enumerate}
    \item Update velocities due to external forces.
    \item Compute a pressure projection to make the fluid incompressible.
    \item Advect the velocity field and any physical quantities.
\end{enumerate}
Velocity updates are guided by the Navier Stokes equations as shown in Equation~\eqref{eq:navier_stokes}.
The second step is arguably the most computationally expensive, especially for higher resolution grids, since it requires solving a pressure Poisson equation (see Equation~\eqref{eq:pressure_poisson}).

The grid structure used for storing pressure and flow field quantities is an important consideration.
Staggered grids, where velocities are stored at the center of cell faces and all other physical quantities are stored at cell centers, are often preferred for their stability when using finite differences to evaluate spatial derivatives.  
The marker-and-cell (MAC) grid~\cite{harlow1965numerical} is a popular choice. Furthermore, work on efficient grid data structures and update algorithms is an active area of research~\cite{SPGrid,VDB2013,NeuralVDB,fVDB,BBT25}.

The reader is directed to the excellent reference material by Bridson~\cite{bridson2015fluid} for further background material on Eulerian fluid simulations. However, since our focus is on the multiphysics setting, we next consider how Eulerian fluid simulations can be coupled with other fluid phases, deformable objects and rigid bodies. 

\subsection{Solids}
Eulerian representations have proven effective for simulating elastic solids undergoing large deformations.  The regular grid structure used by many implementations avoids expensive remeshing steps, and further facilitates parallel implementations. Levin et al.~\cite{Levin2011} discretize the momentum equations on a grid and use the finite volume method~\cite{Teran2003} to compute the traction of each cell from a Cauchy stress field. Their approach uses an explicit scheme to model interactions between colliding bodies. However, Teng et al.~\cite{Teng2016} later extended the method with an implicit scheme allowing larger time steps, and further introduced fluid-solid coupling in a purely Eulerian regime.

\subsection{Multi-phase Fluids}
One of the advantages of using an encapsulating Eulerian grid, is that it becomes possible to capture multi-phase effects, particularly those between a liquid phase and an air phase.
This interaction is at the core of many real-world phenomena, such as water with bubbles.
Within the Eulerian regime of approaches, the ghost fluid method has been popular for modeling the jump in fluid densities~\cite{Hong2005, Boyd2012}.
Other approaches range from using stream functions~\cite{andoStreamFunctionSolver2015} and explicit bubble constraints~\cite{goldadeConstraintBubblesAffine2020}, to using the lattice Bolzmann method (LBM)~\cite{liEfficientKineticSimulation2022}.

\subsection{Fluid-Solid Coupling}
Many works in computer graphics have addressed the subject of fluid-solid coupling.
A central problem here is representing the solids and their boundaries in grid data structures. 
For instance, early work by Foster and Metaxas~\cite{Foster1996} demonstrated grid-aligned one-way coupling using a MAC grid. 
This approach requires rasterizing the solid onto the grid in order to determine the boundary interface, thus giving a one-way coupling. 
Takahashi et al.~\cite{Takahashi2002} achieve two-way coupling by assigning zero Neumann boundary conditions for pressures of any grid cell more than half filled with a solid. 
Cell velocities covered by a solid are then assigned velocities from the solid, although the approach cannot couple forces and torques due to fluid momentum. 

The immersed boundary method~\cite{Peskin2002} is another popular technique that gives a continuous forcing function that allows the fluid pressure field to change the solid's velocity by considering the solid to be part of the fluid. 
Similarly, Carlson et al.~\cite{Carlson2004} fixes the solid density to be the same as the fluid and enforces a rigidity constraint for velocities inside the solid using Lagrange multipliers. 
Guendelman et al.~\cite{guendelman_coupling_2005} proposed a solid-fluid coupling approach specialized for thin shells and cloth.
The immersed boundary method has also recently been applied to turbulent flow simulation using the lattice Boltzmann method (LBM) \cite{liFastScalableTurbulent2020}.

Cut-cell methods are another approach for computing interactions between Eulerian fluids and solids.
They are popular due to their simplicity and ability to encode fluxes due to geometric details smaller than the grid resolution, but without having to refine or change the grid structure.
The area of grid cell faces covered by the solid boundary is used to improve the accuracy of the divergence calculations. 
These approaches were first used in computer graphics by Roble et al.~\cite{Roble2005} and are subject of ongoing research~\cite{ZB17,TBFL19}.
Batty et al.~\cite{batty_fast_2007} accounts for partial grid cell overlap by approximated volume weights, which are used by a variational formulation of the pressure projection solve that improves accuracy of sub-grid fluid velocities near object boundaries.
Cut-cell methods are also used in combination with mesoscale approaches for more accurate coupling between fluids and solids~\cite{lyuFastVersatileFluidsolid2021}. 
Shi et al~\cite{Shi2024} proposed a coupling framework that target heterogeneous computing platforms combining a cut-cell method and adaptive grids for interactive simulation of rigid bodies and fluids.

Two-way coupling between fluids and solids often necessitates taking small time steps and using many coupling iterations to achieve stable simulations with correct pressure and momentum exchange between the two models~\cite{Carlson2004,guendelman_coupling_2005}.
Achieving strong two-way coupling requires accounting for both the solid and fluid dynamics during the pressure projection step. 
However, coupling techniques can sometimes result in non-physical momentum transfer, even though fluid-solid boundary conditions are satisfied. Robinson-Mosher et al.~\cite{RobinsonMosher2008} proposed a modification to the momentum update used by Batty et al.~\cite{batty_fast_2007} that improves fluid-to-solid momentum transfer for thin solids and shells by a mass-lumping strategy that combines equations for both fluids and solids in mixed cells. Follow-up work resolved problems with this approach, allowing for correct slip conditions of tangential velocities~\cite{RobinsonMosher2009}.
Klingner et al.~\cite{Klingner2006} proposed an implicit approach for rigid-smoke coupling, and later Chentanez et al.~\cite{Chentanez2007} extended the approach to liquids, that discretizes the domain using an adaptive tetrahedral mesh, which allows finer details to be preserved due to more accurate treatment of boundary conditions. However, the approach is computationally costly due to frequent remeshing for dynamic objects. 
Chentanez et al.~\cite{Chentanez2006} proposed a unified approach for strong coupling of fluids and deformable objects, although it results in a non-symmetric linear system that requires a more costly solver algorithm. 
Stable coupling can be achieved, however, without monolithic frameworks that solve for pressures and velocities across the fluid and solid domains. For instance, Akbay et al.~\cite{Akbay2018} showed that a partitioned impulse-based solver using a reduced model interface was effective for stable simulation using independent ``black-box'' fluid and solid solvers, requiring only Jacobian approximations collected from simulation data.

More recently, Takahashi and Batty~\cite{TB20} developed the Monolith solver for fluid-rigid coupling, and later extended the approach for elastic bodies~\cite{TB22}. While most work focuses on fluid-solid interactions of individual dynamical bodies, their approach also accounts for interactions between solids, for instance due to frictional contact, as well as viscosity. A hybrid approach, they combine an Eulerian fluid simulation with Lagrangian mesh-based elastic bodies, and rigid bodies using a strong coupling approach formulated as a unified constrained optimization problem. Although their fluid simulator uses a particle-in-cell method, coupling is solved using a grid, and hence can be considered an Eulerian approach.
However, hybrid approaches combining grid and particle representations have gained prominence in computer graphics, in particular the Material Point Method, which we consider next.

\subsection{Material Point Method}
\label{sec:MPM}

\begin{figure*}
    \includegraphics[trim={0 0 297 105},clip,height=3.98cm]{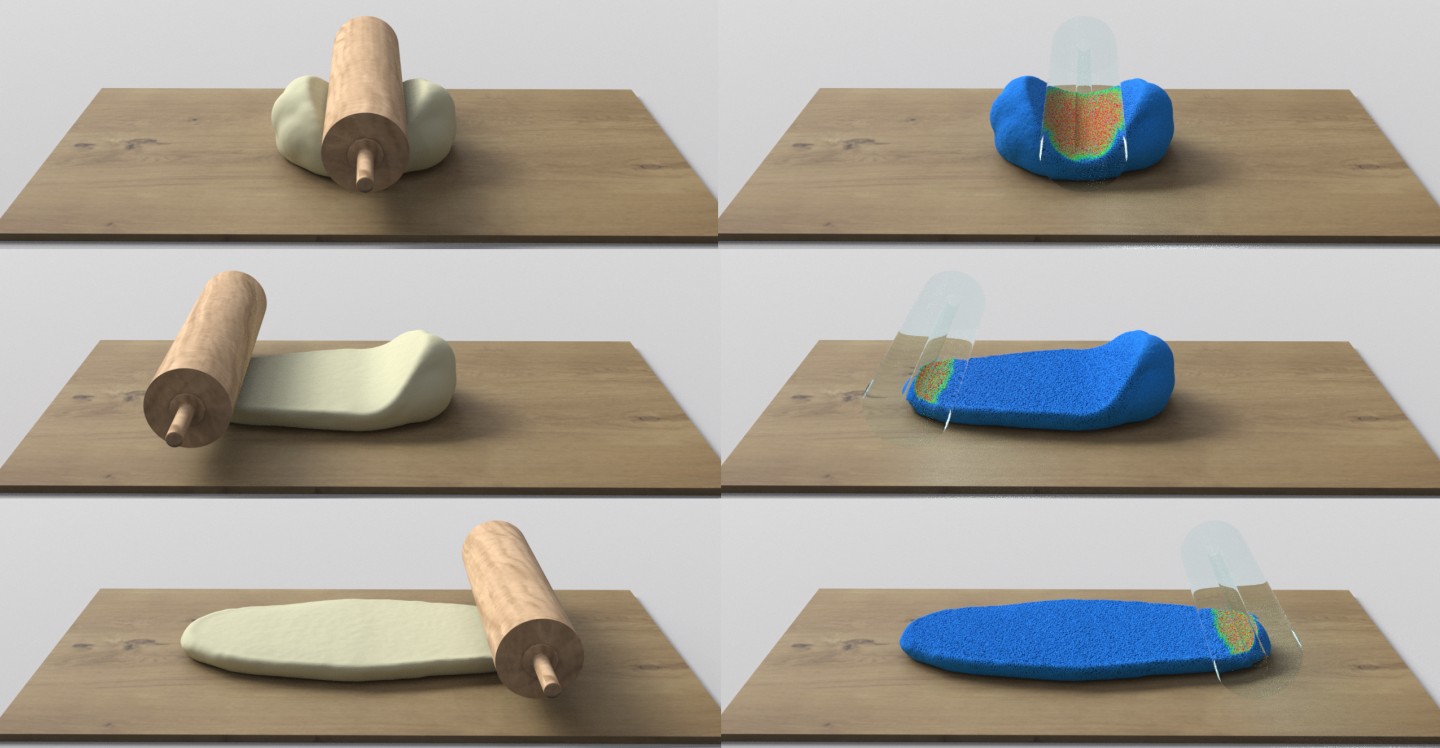}
    \hfill
    \includegraphics[trim={0 0 0 0},clip,height=3.98cm]{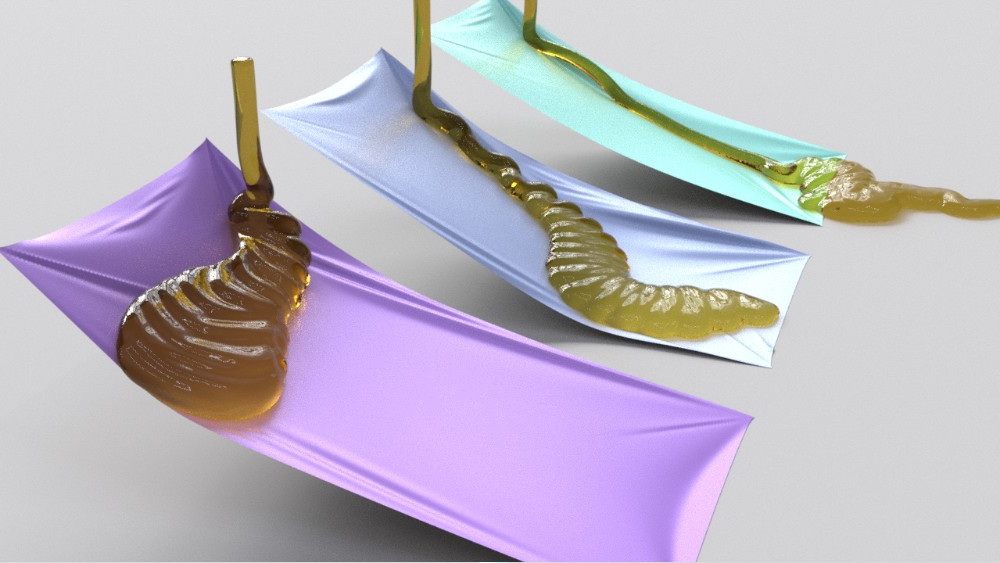}
    \hfill
    \includegraphics[trim={0 0 0 0},clip,height=3.98cm]{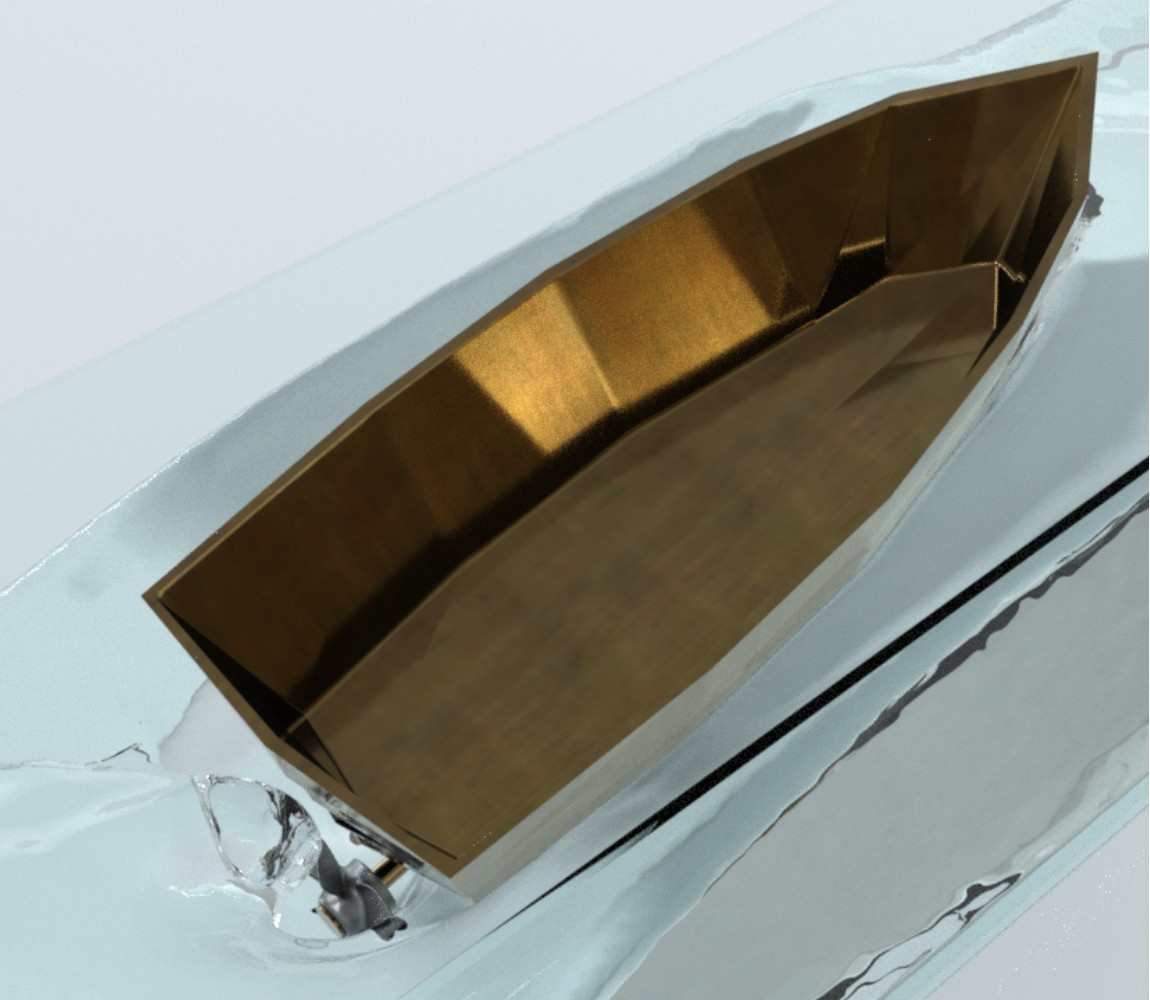}
    \centering
    \caption{\label{fig:mpm} MPM simulations of various interacting materials~\cite{li_dynamic_2024}. \textbf{Left:} Plastic dough material being flattened by a rigid, rolling pin. \textbf{Middle:} Highly viscous honey dropped on a cloth. \textbf{Right:} Two-way coupled FEM-MPM simulation of a motor boat interacting with water.}
\end{figure*}

The Material Point Method (MPM) is a hybrid Eulerian/Lagrangian method.
Hybrid methods inherit the strengths from both material descriptions: Lagrangian particles make advection trivial, while the underlying Eulerian grid eases the computation of spatial derivatives.
As one of the first hybrid methods, the Particle-In-Cell method (PIC) was used to simulate compressible fluid back in the 1960s~\cite{harlow_particle--cell_1962, harlow1965numerical}. 
Due to frequent grid-particle transfers, PIC yields excessive dissipation and the resulting simulation appears highly dampened.  
This issue was later fixed by Fluid-Implicit-Particle (FLIP)~\cite{brackbill_flip_1986}, which only averages the velocity and displacement increments during the grid-particle transfer. Zhu and Bridson~\cite{zhu_animating_2005} mixed PIC and FLIP with a blending weight to control the viscous damping and instability. The Material Point Method further generalizes PIC/FLIP~\cite{sulsky_application_1995} to solid mechanics by allowing particles to carry additional physical quantities such as the deformation gradient. MPM-based approaches have been firstly introduced to the graphics community for snow animation~\cite{stomakhin_material_2013}, and quickly generalized for a multitude of materials and phenomena (see Figure~\ref{fig:mpm}). After a brief summary of MPM in the following section, we will provide details on several of its applications for multiphysics modeling.

\subsubsection{Foundations of the Material Point Method}
Analogous to the Finite Element Method (FEM)~\cite{bathe2006finite}, the Material Point Method is derived from the weak form of the conservation of momentum, making it an accurate modeling framework for the discretization of constitutive laws~\cite{jiang_material_2016}.
Furthermore, its hybrid Lagrangian/Eulerian setting has various advantages, which makes it excel in handling phenomena involving topological changes and very large deformations. The Lagrangian representation is ideally suited for large deformations as the material and its quantities are effectively tracked by persistent particles~\cite{sulsky_application_1995}.
The Eulerian representation is only transient and is used for computing spatial derivatives and solving the material's governing equations. The resulting velocity field triggers the movements of particles.
Self-collision and fracture can thus be efficiently modeled on the Eulerian background grid, independent of the number of particles.

In order to make use of both representations, MPM needs to constantly transition information between Lagrangian particles and the Eulerian grid nodes. As such, while the exact details in different MPM-based method implementations vary, a basic symplectic Euler MPM algorithm in its simplest manifestation performs the following operations in each time step:
\begin{enumerate}
\item \emph{Particle-to-Grid (P2G) Transfer.} Transfer particle quantities to the grid, computing mass and momentum at grid nodes. Different P2G transfer schemes can be used in this step, such as Affine Particle-In-Cell (APIC)~\cite{jiang_affine_2015}.
\item \emph{Grid Velocity Computation.} Compute velocities at grid nodes from the mass and momentum computed above.
\item \emph{Explicit Grid Force Computation.} Compute forces acting on grid nodes due to nearby particles.
\item \emph{Grid Velocity Update.} Update grid node velocities based on the forces computed above and the grid node masses, while accounting for boundary conditions or specific collision handling.
\item \emph{Deformation Gradient Update.} Update the deformation gradient for each particle based on the grid velocity field.
\item \emph{Grid-to-Particle Transfer (G2P).} Transfer velocities from the grid to the particles using a specific transfer scheme. Depending on the scheme employed in the G2P and P2G steps, additional information needs to be transferred here.
\item \emph{Particle Advection:} Advect particles using their new velocities.
\end{enumerate}

For more information on the foundations of MPM we recommend the course by Jiang et al.~\cite{jiang_material_2016}.

\subsubsection{Accuracy, Stability and Performance Improvements}
Several notable improvements of MPM have made the framework more attractive over the years. With the development of the Affine Particle-In-Cell (APIC) method, proposed by Jiang et al.~\cite{jiang_affine_2015} and later generalized by Fu et al.~\cite{fu_polynomial_2017} with Polynomial Particle-In-Cell (PolyPIC), grid-particle transfer better conserves affine momentum and kinetic energy while maintaining the stability of the simulation.
The Power-Particle-In-Cell further takes particle volume into consideration~\cite{qu2022power}, which leads to better volume preservation.
Recently, Sancho et al.~\cite{STBA24} proposed the Impulse Particle-In-Cell Method (IPIC) to better conserve circulation and vortical details for fluids using flow maps.

Explicit time integration in MPM requires small time steps adhering to the CFL condition for feasible stability~\cite{sun2020effective}~\cite{wang2020hierarchical}.
For details, we refer to the stability analysis of Bai and Schroeder~\cite{BS22}.
Stomakhin et al.~\cite{stomakhin_material_2013} propose a time integration scheme to enable semi-implicit and implicit time integration, allowing the use of large time steps and thereby increasing efficiency of MPM. Hyde et al.~\cite{hyde2020implicit} adopt the implicit MPM formulation to simulate surface tension effects. Fang et al.~\cite{fang2019silly} use implicit MPM to handle viscoelastic and elastoplastic solids. Implicit MPM also contributes strong coupling between fluids and solids~\cite{fang2019silly}. 
Gast et al.~\cite{GSS+15} present an optimization-based formulation of implicit integration in MPM.
Combining the Moving Least Squares (MLS) material sampling method (see Section~\ref{sec:point-based}) with MPM, Hu et al.~\cite{hu_moving_2018} enable two times faster simulation rates, with a simpler implementation and with visually comparable results. Jiang et al.~\cite{jiang2017anisotropic} leverage MPM for processing frictional contacts. Gao et al.~\cite{gao2017adaptive} design an adaptive grid scheme to reduce the collision gap for MPM deformables. Ding and Schroeder~\cite{ding2019penalty} show the feasibility of combining MPM with Coulomb frictional contacts for accurate coupling with rigid bodies.
For further performance gains, prior works also explore GPU-based MPM implementations~\cite{gao_gpu_2018,wang2020massively,fei2021principles,qiu2023sparse}.

\subsubsection{Multiphysics Phenomena}
Multiphysics simulation is naturally supported in MPM by assigning different constitutive laws and material models to different particles~\cite{jiang_material_2016}.
Thanks to its hybrid discretization, MPM also exhibits a high versatility in robustly handling various phenomena ranging from snow~\cite{stomakhin_material_2013}, phase-changing materials~\cite{stomakhin_augmented_2014,tu_unified_2024,su2021unified}, foams~\cite{ram_material_2015,qu2023power,yue_continuum_2015} to granular materials~\cite{daviet_semi-implicit_2016,klar2016drucker,tampubolon2017multi}, and even magnetized materials~\cite{sun2021material}. The Lagrangian particles in MPM are insensitive to topological changes. Therefore, MPM has also been a popular choice for cutting~\cite{hu_moving_2018} and fracture~\cite{Wolper2019,wang2019simulation,fan2022simulating,CCL+22}.
In the following, we will provide brief descriptions of a representative selection of these methods.

Stomakhin et al.~\cite{stomakhin_material_2013} integrated an elasto-plastic constitutive model with MPM for modeling the dynamics of snow. A particle carries an elastic deformation gradient and a plastic deformation gradient. Out-of-range deformation accumulates in the plastic portion, while elastic deformation yields internal forces. Possible topology changes, caused by snow breaking or merging, are easily captured with the meshfree Lagrangian description, and the Eulerian background grid enables implicit handling of self-collisions and fracture. This feature inspires several follow-up works to use MPM to handle fracture~\cite{Wolper2019,wang2019simulation,fan2022simulating}.

Shortly after, Stomakhin et al.~\cite{stomakhin_augmented_2014} augmented MPM with heat transport on the grid so that particles with different temperatures are assigned different material properties. A deviatoric/dilational splitting of the constitutive model allows phase transitions during the simulation, such as melting and solidifying, which further enriches the MPM ecosystem.

Using MPM to simulate Non-Newtonian fluid has also been explored. Ram et al.~\cite{ram_material_2015} employed a modified Oldroyd-B model to approximate volume-preserving plastic flows. Yue et al.~\cite{yue_continuum_2015} demonstrate that Herschel–Bulkley fluid can also be handled within the MPM framework with particle resampling. Su et al.~\cite{su2021unified} proposed an extended POM-POM model~\cite{mcleish1998molecular,oishi2012numerical,verbeeten2001differential} that can uniformly handle a range of viscoelastic liquids with phase change.

Early MPM frameworks often chose explicit grid-level integrations for spatial differentiation of internal elasto-plastic forces. This simplification requires a highly conservative time step size for stiff simulation problems, e.g., granular materials with a high Young's modulus. Daviet et al.~\cite{daviet_semi-implicit_2016} further enhanced the stability of MPM for such cases by treating both elastic and plastic parts of the dynamics in a semi-implicit manner.
Other examples of MPM-based granular material simulation are the methods by Klar et al.~\cite{klar2016drucker} and Tampubolon et al.~\cite{tampubolon2017multi}.
To implicitly handle plasticity in optimization-based MPM frameworks, Li et al.~\cite{LLJ22} present strain energies for specific plasticity models (see also Section~\ref{sec:energy-based-materials}).

Hu et al.~\cite{hu_moving_2018} employ Moving Least Squares (MLS) interpolation in their MLS-MPM method for the unified simulation of fluids, elastic, plastic and granular materials with robust support for cutting. Any external rigid body dynamics simulator can be combined with their method, providing two-way coupling with rigid bodies.
Qu et al.~\cite{qu2023power} propose an extension of MLS-MPM which is capable of accurately capturing both macro- and mesoscale material geometries. They demonstrate the versatility of their approach with simulations of bubbles, sands, liquid and foams, and show that in contrast to traditional MPM, their method does not suffer from minimum particles-per-cell restrictions and is able to faithfully conserve material volume, all the while producing more uniform particle distributions.

Tu et al.~\cite{tu_unified_2024} combine MPM with a phase field solver for the unified simulation of elastic, plastic and viscous materials that undergo phase transitions. Their method is able to represent interacting fluids, deformable and rigid solids, and granular materials which can dissolve and melt, all in one unified framework. This is achieved through blending between different constitutive models in a unified particle representation. Collisions between different materials not undergoing phase changes was modeled more realistically by Gao et al.~\cite{gao_animating_2018} using MPM, although not in a unified manner, as we will see in the next section. The approach by Tu et al.~\cite{tu_unified_2024} also inherits a common limitation in MPM, which causes inaccuracies in contact behavior at multi-material interfaces -- a consequence of stickiness and numerical viscosity, leading to excessive cohesion and friction. This limitation was addressed in the modified Material Point Method introduced by Han et al.~\cite{han_hybrid_2019}, details of which are provided in the next section.
Chen et al.~\cite{CKM+21} propose an implicit model for spatially varying surface tension in MPM.
The surface tension coefficient can be dependent on local concentrations or temperatures and for the latter it is coupled with an implicit thermomechanical model to accurately simulate melting.

\subsubsection{Coupling Techniques}

In order to address the aforementioned shortcomings with collision modeling in traditional MPM manifesting in artifacts such as excessive cohesion and friction, Han et al.~\cite{han_hybrid_2019} introduce a modified Material Point Method which makes use of collision particles that are uniformly distributed over the surface of volumetric solids. This approach prevents information loss during particle-to-grid transfers and enables accurate Coulomb frictional contact modeling between volumetric solids. The method also effectively resolves the dependency issue found in traditional MPM between Eulerian grid size and Lagrangian particle density. Through use of two separate Eulerian background grids, the method supports coupling with traditional MPM and also provides a simple method for two-way coupling with rigid bodies.

Gao et al.~\cite{gao_animating_2018} propose an MPM-based method for simulating the interaction of granular materials with fluids through explicit two-way coupling. The authors use a density criterion to identify individual grains that become separated from the continuum body of the granular material which are treated as separate bodies, not subject to the elastoplastic Drucker-Prager constitutive law used to model still connected granular material clumps. The fluid and granular material domains are advanced in a sequential manner, with sub-stepping used for the granular material. The method captures intricate details in fluid/granular material interactions such as sediment transport in particle-laden fluid flows which can not be reproduced in the unified method proposed by Tu et al.~\cite{tu_unified_2024}.

Guo et al.~\cite{GHF+18} present a method for simulating elastoplastic thin shells based on an FEM subdivision surface discretization within MPM and demonstrate frictional contact and coupling to other MPM materials.
To simulate fracture due to shocks in compressible flow, Cao et al.~\cite{CCL+22} developed a monolithic solve for strongly coupled MPM and FEM.
This velocity-pressure system is discretized with a mixed-order finite element of B-spline shape functions.
More recently, for two-way coupling between MPM and FEM of possible co-dimensional geometry,
Li et al.~\cite{li_dynamic_2024} propose an asynchronous time-splitting scheme.
Specifically, an implicit integration scheme is used at the internal region of the FEM domain. The contact force is applied as a constant external Lagrangian force on MPM particles throughout substeps of explicit MPM integration.

\section{Multiphysics Frameworks}
\label{sec:frameworks}
This section presents a selection of frameworks for multiphysics simulation which make use of the mathematical models, coupling techniques and discretizations discussed in this report. These frameworks offer convenient out-of-the-box simulation capabilities via ready-to-use interfaces which allow easy deployment of simulations or extensions of the frameworks' capabilities. Some frameworks employ a single unified model while others use a modular approach and combine different formulations and discretizations through coupling techniques and expose them in a unified simulation suite.
We will briefly discuss several popular frameworks in this section.

\paragraph*{Chrono} %
Tasora et al.~\cite{kozubek_chrono_2016} propose the \emph{Chrono} framework which enables fully coupled multiphysics simulations of rigid and deformable bodies, fluids and granular materials.
It is designed for use in engineering applications, and, depending on the resolution and scale of the simulation, can achieve real-time frame rates, or can be used for accurate offline simulations. The framework is freely available with source code and documentation accessible at \url{https://projectchrono.org}.

The framework combines a number of simulation models and provides out-of-the-box coupling for their interactions. Constrained rigid bodies are represented using a Nonsmooth Multidomain Dynamics (NMD) formulation. Granular materials are represented using either a nonsmooth DEM approach within NMD or via a penalty-based method. Fluids are modeled using SPH and coupled with solids using boundary particles. Rods, sheets and thin beams are represented using the Absolute Nodal Coordinate Formulation (ANCF), and for volumetric flexible elements a corotational FEM approach is employed.
The advantage of using ANCF is that it can be embedded within the mathematical framework of NMD and thus provides strong coupling without need for any explicit coupling technique~\cite{kozubek_chrono_2016}.

\paragraph*{SOFA} %
With \emph{SOFA}, Faure et al.~\cite{payan_sofa_2012} propose a versatile, extensible framework based on multiple overlayed data models, providing flexible data flow and algorithm control that targets interactive and real-time multiphysics simulation for virtual surgery training. 
SOFA is an open-source framework with documentation and source code available at \url{https://www.sofa-framework.org}.

The framework provides a readily available multi-material simulation system. Fluids are modeled using SPH. Deformable bodies are simulated using corotational FEM, which interact with rigid bodies through unilateral contact constraints, solved using a projected Gauss-Seidel method. In addition, the framework offers various linear solvers and constraint solvers supporting both bilateral and unilateral constraints which can be employed to any simulation problem. 

\paragraph*{Loki} %
Entirely tailored to offline simulations for the film industry, the Loki multiphysics framework, proposed by Lesser et al.~\cite{lesser_loki_2022},
employs a generic coupling approach that uses interleaved Newton iterations of the individual methods at small time steps, which ensures a high level of solution accuracy. This approach also allows for the use of best-in-class methods for simulation of different material types, while avoiding combinatorial explosions from the use of an abundance of specialized coupling methods for the various pair-wise material type interactions. This is made possible through the use of generic and common material descriptions shared by all methods, such as Lagrangian particles and Eulerian grid nodes carrying common physical material quantities. For increased performance and scalability, Loki uses heterogeneous computing and distribution of workloads via MPI. Loki is a commercial framework that is currently not available to the public.

\paragraph*{STARK} was introduced by Fernández-Fernández et al.~\cite{FLL+24} as an open-source framework with state-of-the-art methods for energy-based simulations as outlined in Section~\ref{sec:unified:energy-based}.
In this original work, the framework is qualitatively validated in challenging scenarios of friction-dominated robot-cloth interactions.
STARK features implementations of widely used models for deformable solids, shells, rigid bodies with joints and motors as well as an IPC-based model for frictional contact.
For the numerical solution it uses Newton's method with line search.
The framework can be used as a black box simulator using high-level C++ and Python APIs.
However, it is also well suited for research and development of new models as it uses the symbolic differentiation engine \emph{SymX}~\cite{FLW25} to handle the differentiation and assembly components.
This makes it possible to add new models by only specifying a scalar energy potential in symbolic form.
STARK has already been used in several recent publications that demonstrate the advantages of energy-based strong coupling for multiphysics applications~\cite{LFJB24,WFJB24,LFJ+23}.
The source code is available at \url{https://github.com/InteractiveComputerGraphics/stark}.

\paragraph*{SPlisHSPlasH}~\cite{SPlisHSPlasH_Library} is a research-driven open-source SPH simulation framework.
The implementation contains many of the state-of-the-art methods discussed in Section \ref{sec:SPH}, including most pressure solvers and boundary handling approaches as well as implicit methods for elasticity, viscosity and surface tension.
Its architecture supports weak coupling in both a unified and non-unified way through interaction forces.
It is used by a multitude of recent works either as a data-generation tool for learning methods, or as a reference implementation for comparison purposes.
In recent years, it has also found application in the engineering community, e.g., for welding and thermally-sprayed coating simulations~\cite{JSS+22,JBB+22}.
It is actively being developed and provides a Python interface for easy access and usability, with source code available at \url{https://github.com/InteractiveComputerGraphics/SPlisHSPlasH}.

\paragraph*{NVIDIA Flex}~\cite{nvidiaflex_2024} is a freely available multiphysics framework for the real-time time simulation of various materials including rigid and deformable bodies and fluids on the GPU. The framework is tailored for use in video games and VR. Its implementation is based on the unified position-based dynamics method proposed by Macklin et al.~\cite{macklin_unified_2014}. Due to its lack of various functionalities required in gameplay mechanics such as collision queries and callbacks for integration with game applications, manual coupling of Flex with CPU-based rigid body dynamics engines commonly used in games might be required. 
NVIDIA Flex provides access to source code which is available at \url{https://github.com/NVIDIAGameWorks/FleX}.

\subsection*{Open-Source MPM Simulators}
Various open-source multiphysics frameworks based on MPM (see Sec.~\ref{sec:MPM}) are available.

\emph{Karamelo}~\cite{de_vaucorbeil_karamelo_2021} provides a lightweight C++ framework for unified CPU-based multiphysics simulations using MPM, designed to be easy to modify in order to facilitate research conducted on and with MPM. Parallelization is offered through use of MPI. Documentation and source code are available at \url{https://karamelo.org}.

With \emph{Taichi}, Hu et al.~\cite{hu_moving_2018} make the implementation of their method freely available at \url{https://github.com/yuanming-hu/taichi_mpm}.
Taichi provides a high performance CPU implementation of MLS-MPM with cutting support and two-way coupling with rigid bodies that offers a convenient Python API.

A GPU implementation of MPM using CUDA is provided by Gao et al.~\cite{gao_gpu_2018} that offers support for multiple particle-to-grid transfer schemes, FLIP, APIC and MLS, and enables both implicit and explicit time integration. The source code can be found at \url{https://github.com/kuiwuchn/GPUMPM}.

\section{Emerging Trends}
\label{sec:data-driven}
In this section, we briefly discuss trends and breakthroughs in multiphysics simulation that are emerging from the computer graphics community. 

Notably, there is increasing interest on developing simulation techniques
that leverage recent advances in machine learning, with the motivation being that increased levels of detail and realism can be achieved at reduced computational cost.
Much work on this topic focuses on deriving a reduced simulation model by sampling behavior obtained from traditional simulation techniques. Some approaches have focused on end-to-end learning of response to external interactions with learned physical objects~\cite{Tao2024,Holden2019}, whereas other approaches focus on improving performance by learning reduced latent sub-spaces~\cite{CROM, LiCROM, Shen2021}. Zong et al.\cite{Zong2023} used a common latent representation to learn neural fields embedding the deformation, stress and affine momentum for MPM simulations of fracture and elastoplasticity for a wide range of material behaviors.  The recently proposed Simplicits~\cite{Modi2024} addresses the problem of performing elastodynamic simulation of solids from arbitrary geometric representations, thereby foregoing the need for specific mesh-based or point-based discretizations. 
Models derived from the rendering community have also proven interesting for their ability to combine visual appearance and physical dynamics~\cite{PIENerf, PACNerf, PhysGaussian}. Point-based simulation frameworks in particular have recently been demonstrated to pair well with 3D Gaussian Splatting~\cite{VRGS}, and results demonstrating fluid-solid interactions have also be shown~\cite{Feng2024}.

Although many methods in this class of emerging approaches tend to focus on simulation of individual phenomena or a specific material behavior, there is increasing emphasis on integrating neural models in a multiphysics setting. Development of frameworks that facilitate research on this topic is therefore interesting~\cite{SimNet2021, Genesis}.

\section{Conclusion}
\label{sec:conclusion}

In this tutorial, we presented the mathematical foundations and key algorithms behind the most widely used multiphysics simulation methods in computer graphics.
We covered energy-based and constraint-based unified formulations, Lagrangian point-based methods, as well as Eulerian and hybrid approaches such as the Material Point Method.

For each approach, we discussed how different materials and physical phenomena -- from rigid and deformable bodies to fluids and granular materials -- can be represented within a common framework, and how unified models and coupling techniques each offer distinct trade-offs between simplicity, flexibility, and stability.
We also presented a selection of multiphysics software frameworks and discussed emerging trends such as machine learning-based simulation methods, which will likely have a significant influence on multiphysics simulation in the years to come.

We hope that these course notes serve as a useful reference for researchers and practitioners looking to implement or extend multiphysics simulation systems.

%------------------------------------------------------------------------

\section*{Acknowledgements}

We would like to thank Daniel Holz and Yin Yang for kindly providing images from their works for use in this tutorial.
This work is partially funded by the Deutsche Forschungsgemeinschaft (DFG, German Research Foundation) — 281466253; 310833819.

%-------------------------------------------------------------------------
% bibtex
\bibliographystyle{eg-alpha-doi} 
\bibliography{multiphysics}       

% biblatex with biber
% \printbibliography                

%\section{Appendix: Multiphysics Method Tables}
%\input{tables}

%-------------------------------------------------------------------------
%Color tables are no longer required for purely electronic publications.
% \newpage
% 
% 
% \begin{figure*}[tbp]
%   \centering
%   \mbox{} \hfill
%   % the following command controls the width of the embedded PS file
%   % (relative to the width of the current column)
%   \includegraphics[width=.3\linewidth]{sampleFig}
%   % replacing the above command with the one below will explicitly set
%   % the bounding box of the PS figure to the rectangle (xl,yl),(xh,yh).
%   % It will also prevent LaTeX from reading the PS file to determine
%   % the bounding box (i.e., it will speed up the compilation process)
%   % \includegraphics[width=.3\linewidth, bb=39 696 126 756]{sampleFig}
%   \hfill
%   \includegraphics[width=.3\linewidth]{sampleFig}
%   \hfill \mbox{}
%   \caption{\label{fig:ex3}%
%            For publications with color tables (i.e., publications not offering
%            color throughout the paper) please \textbf{observe}: 
%            for the printed version -- and ONLY for the printed
%            version -- color figures have to be placed in the last page.
%            \newline
%            For the electronic version, which will be converted to PDF before
%            making it available electronically, the color images should be
%            embedded within the document. Optionally, other multimedia
%            material may be attached to the electronic version. }
% \end{figure*}

\end{document}